\documentclass[12pt]{iopart}

\usepackage{graphics}
\usepackage[dvips]{graphicx}
\usepackage{bbm}
\usepackage{slashed}
\usepackage{color}
\usepackage[dvipsnames]{xcolor}

\newcommand{\beq}{\begin{eqnarray}}
\newcommand{\eeq}{\end{eqnarray}}
\def\r{{\boldsymbol r}}

\def\b{{\boldsymbol b}}

\def\z{{\boldsymbol z}}
\def\x{{\boldsymbol x}}

\def\y{{\boldsymbol y}}
\def\k{{\boldsymbol k}}
\def\q{{\boldsymbol q}}
\def\p{{\boldsymbol p}}
\def\u{{\boldsymbol u}}
\def\0{{\boldsymbol 0}}
\def\v{{\boldsymbol v}}

\def\E{{\boldsymbol E}}
\def\B{{\boldsymbol B}}

\def\cal{\mathcal}
\newcommand{\del}{\partial}

\def\bra#1{\langle#1\vert}
\def\ket#1{\vert#1\rangle}
\newcommand{\nn}{\nonumber\\ }

\usepackage{iopams}  
\begin{document}

\title[ High Gluon Density and  Heavy Ions]{High Gluon Densities in Heavy Ion Collisions}

\author{Jean-Paul Blaizot}

\address{Institut de  Physique Th\'eorique, CNRS/UMR 3681, CEA Saclay,  F-91191 Gif-sur-Yvette, France
}
\ead{jean-paul.blaizot@cea.fr}
\begin{abstract} 
The early stages of heavy ion collisions are dominated by high density systems of gluons that carry each a small fraction $x$ of the momenta of the colliding nucleons.  A distinguishing feature of such systems is the phenomenon of ``saturation" which tames the expected growth of the gluon density as the energy of the collision increases. The onset of saturation occurs at a particular transverse momentum scale, the ``saturation momentum", that emerges dynamically and that  marks  the onset of non-linear gluon interactions. At high energy, and for large nuclei, the saturation momentum is large compared to the typical hadronic scale, making high density gluons amenable to a description with weak coupling techniques. This paper reviews some of the challenges faced in the study of such dense systems of small $x$ gluons, and of the progress made in addressing them.  The focus is on conceptual issues, and the presentation is both pedagogical, and critical. Examples where high gluon density could play a visible role in heavy ion collisions are briefly discussed at the end, for illustration purpose. 

\end{abstract}

\maketitle

\section{Introduction}

By colliding nuclei at the highest possible energies, one creates new forms of matter, that have existed in our universe for only a few microseconds after the big bang.  These states of matter, where quarks and gluons are the main degrees of freedom, are generically called quark-gluon plasmas (QGP). Our understanding of such systems has evolved considerably over the years, from the early concept of a weakly interacting gas, as expected theoretically from the asymptotic freedom of Quantum Chromodynamics (QCD), to the strongly interacting plasma, the so-called ``perfect liquid", first revealed by experiments at the Relativistic Heavy Ion Collider (RHIC) and now intensively studied at the Large Hadron Collider (LHC). (For a recent review see \cite{Armesto:2015ioy} and references therein.)

Ultra-relativistic heavy ion collisions involve complex dynamics, that differ much at low and high energies. At low energy, it is sufficient to consider nuclei as being made of nucleons, and ignore the internal structure of the latter. Nucleons interact through meson exchanges  well accounted for by effective nucleon-nucleon interaction potentials. As energy increases  mesons start to be produced in abundance, as well as hadronic resonances, so that the hadronic matter produced in collisions becomes more and more complex. Some simplicity is recovered  at extremely high energy, when the hadron constituents, viz.  quarks and gluons, become the dominant degrees of freedom. 
This is expected to occur in the early stages of ultra-relativistic collisions, when quarks and gluons (mostly gluons) are freed from the nuclear wave functions.  In order to understand how, and how fast, quarks and gluons thermalize and form the quark-gluon plasma that exhibits the collective flow behavior seen in the experiments, we need  therefore to have a good understanding of  the gluon, and quark, content of a nucleus wave function at high energy. \\

When talking about  the ``wave functions" of nuclei at high energy, length and time scales are important, as well as the role of relativity and, of course, quantum mechanics. Because of Lorentz contraction, the two nuclei appear, in the center of mass frame of the collision, as two contracted pancakes, with a thickness of a small fraction of a femtometer ($\sim 10^{-3}$ fm at the LHC), while the diameter of a  Pb nucleus (spherical in its rest frame) is about 12 fm. The typical duration of a collision, as  measured by the time during which the two nuclei overlap is indeed quite short. However, this time is not the most relevant time scale. Recall that nucleons are  composite objects, made of three valence quarks bound together by a gluon field. This gluon field is accompanied by fluctuations, which, in low energy phenomena, are not directly visible because they are extremely short lived: they merely contribute to renormalization of observables, such as the nucleon mass for instance.  However, the lifetimes of these fluctuations get amplified by Lorentz dilation, to a point where they can be much larger than the time-scale that we referred to earlier. Such fluctuations play an essential role in the collisions: they make nucleons or nuclei look like clouds of virtual particles, generically called partons. During the collision time, these partons can be considered as quasi real and independent. One estimates that the partons involved in the production of the particles that are observed in a collision carry a tiny fraction (called $x$) of the longitudinal momentum of the colliding nucleons: of the order of $x\sim 10^{-2}-10^{-3}$ at RHIC, and $x\sim 10^{-4}-10^{-6}$ at the LHC, the highest values corresponding to particles produced at center of mass rapidity, the lowest ones to particles produced near the rapidity of one of the colliding beams. Such ``small $x$-partons" (mostly gluons) have relatively small longitudinal momentum, and accordingly they completely overlap longitudinally.  

Deep inelastic scattering of electrons on protons, in particular at the HERA facility in DESY, have provided  information about the distributions of low $x$ partons. Although information at very small $x$ is scarce, there is strong evidence, backed up by solid theoretical expectations based on QCD linear evolution equations, that the gluon density increases rapidly as one lowers $x$.  
This growth cannot go on forever though. It would indeed entail  a corresponding growth of the proton-proton cross-section with energy at a rate   incompatible with  bounds that result from the unitarity of the scattering matrix. In fact, it was recognized long ago that the mechanisms that lead to parton multiplication, namely radiation of softer and softer gluons, should be balanced, when the gluon density is large enough, by some kind of recombination processes, leading eventually to a so-called ``saturation" \cite{Gribov:1984tu}. From a naive statistical point of view, the phenomenon may look at first sight as a rather mundane effect. It turns out,  however that the microscopic  mechanisms that lead to it are rather subtle, and in fact  not yet fully understood. Also, while saturation is bound to occur at some point, finding direct and compelling experimental evidence for this phenomenon has remained a challenge.\\
 
Saturation is characterized by a  transverse momentum scale, referred to as the \emph{saturation momentum}, denoted by $Q_s$. That is, small $x$ partons may have different transverse momenta. Those with large transverse momenta occupy a small transverse size, and form a dilute system. Those with small transverse momenta, or large transverse wavelengths, will eventually overlap in the transverse plane as their number increases: saturation can in fact be pictured geometrically as occurring when the entire transverse plane is densely packed with partons. Thus the saturation momentum divides momentum space into two regions: a region of dilute partons with transverse momenta larger than $Q_s$, and a region of high density occupied by partons with small transverse momenta. The value of the saturation momentum grows with the energy of the collision and the size of the nuclei.  Typical values for large nuclei are $Q_s^2\simeq 2 {\rm GeV}^2$ at RHIC and $Q_s^2\simeq 5 {\rm GeV}^2$ at the LHC. 

At high energy, the saturation momentum is essentially the only relevant scale, and consequently it plays an important role in the phenomenology of heavy ion collisions: collisions of two nuclei involve partons with typical transverse momenta of the order of the saturation momentum. Thus the dependence of the bulk features of the data on the size of colliding nuclei or the energy of the collision can be expected to follow the corresponding dependence of $Q_s$. Furthermore,  being the only scale, $Q_s$ controls the magnitude of the QCD coupling constant. If it is large enough the coupling constant will be small, and the system amenable to a description with weak coupling techniques. This is not to say that the saturated system of gluons is ``weakly coupled" though, as collective behavior may emerge for instance from the large density (requiring resummations, i.e., going beyond strict perturbation theory). The ``strongly coupled'' character of the quark gluon plasmas seen in experiments may be caused by such a circumstance,  rather than be due to a genuinely large coupling constant, as assumed for instance in   approaches based on the AdS/CFT correspondence \cite{CasalderreySolana:2011us}. \\

Thus,  ultrarelativistic heavy ion collisions  offer us a unique opportunity to study dense, many-body systems of gluons, and hence explore QCD dynamics in unusual settings.  Such studies are difficult: on the experimental side because the complexity of the environment makes it difficult to probe fine details of the theory; on the theoretical side, the subject can become highly technical,  involving sophisticated formalisms whose mutual relations are not always visible. Focussing on theory, one can however identify several lines where important progress has been achieved in the recent years, and which have contributed to shape the field.  These will be discussed in this review. An important advance  was realized by McLerran and Venugopalan, whose model (the MV model) \cite{McLerran:1993ni,McLerran:1993ka} allows for an explicit, albeit approximate, calculation of the distribution of gluons in a nucleus.  This models builds on the idea that the field created by a large assembly of valence quarks (e.g. a large nucleus) is strong, and that its fluctuations during the collision can be ignored. This field   represents the small $x$ partons which, as we have seen, completely overlap and may be better treated by a field rather than by individual particles. This model provides a specific realization of the saturation phenomenon  in term on non linear interaction of the gauge field.  Another major part of the effort during the last two decades has been put in the derivation of non linear equations  for  the evolution with energy of the gluon density.  These equations allow in particular the determination of the energy dependence of the saturation momentum, and in addition they provide new, and sometime puzzling, perspectives on the phenomenon of saturation. The 
 Color Glass Condensate (CGC), prominently referred to in the field,  is often presented as an effective theory which encompasses features present  in the MV model, such as the averaging procedure over random classical fields, together with effects of radiative corrections that are described by the non linear evolution equations. In fact, it should be said that the reference to the CGC  has, to a large extent, lost over the years the strict connection to the original acronym significance: in practice,  ``CGC" has become synonymous of ``high density QCD",  ``saturation physics", etc., as well as of the set of techniques presently used to handle high gluon densities, and this is in this loose sense that we shall use it in this paper.   \\

Let me now explain what my goals are in writing this review. This is not meant to be a systematic review of the field.  Several excellent and well documented reviews, or lecture notes,  exist on the subject \cite{Iancu:2002xk,Iancu:2003xm,JalilianMarian:2005jf,Weigert:2005us,Triantafyllopoulos:2005cn,Gelis:2010nm,Lappi:2010ek,Gelis:2012ri,Albacete:2014fwa,Albacete:2013tpa}, with  complementary perspectives on small $x$ physics given for instance in \cite{Frankfurt:2005mc,Frankfurt:2011cs}.  Also the book by Levin and Kovchegov  \cite{Kovchegov:2012mbw} covers much of the material that will be discussed here.  However, as I already alluded to, the subject involves highly technical developments, and it is not always easy for the non expert to recognize the basic underlying physics principles, and also, within the formalisms, to distinguish what is solidly established from what are reasonable working assumptions or developments based on heuristic arguments. I feel therefore that there is a need for clarification, and this short review represents an effort to that goal. My intention is not to address experts,  but rather to provide non experts with  an elementary introduction to the physics of high gluon density systems,  focussing on the main conceptual issues,  and with a critical look at the recent developments. \\

This paper begins, in the next section, with a reminder about the parton picture of hadron wave functions. Most of the material presented there is well known, and the purpose of this section is to fix  notation and introduce some basic concepts, such as parton distributions, collinear factorization, linear and non linear evolution equations. The section ends with a first  introduction to the concept of saturation and the saturation momentum.

Much information concerning the nucleus wave function at high energy can be obtained by studying the collisions of elementary projectiles with a nucleus. The eikonal approximation plays a central role there, leading to a description of these collisions in terms of averages of Wilson lines, or of products of Wilson lines, the averages being taken over the color field of the nucleus. In Sect.~\ref{sec:eikonal} we focus on a particular combination of two Wilson lines in a color singlet state, referred to as a color dipole. Its interaction with a nucleus is analyzed, first in the  weak field regime, where the dipole interacts with nucleons via two-gluon exchange, as well as in the regime of the strong field of a nucleus, treated in terms of multiple scatterings. The section ends with a brief discussion of the phenomenological dipole model used in the analysis of deep inelastic scattering of leptons on protons or nuclei. An important insight obtained from this model is the  notion of geometrical scaling, which provides perhaps the best experimental evidence for the existence of a saturation momentum. 

In Sect.~\ref{WWfields}, I turn to the explicit calculation of the averages of Wilson lines, in particular of color dipoles. This is done using the McLerran-Venugopalan model. In this model, the color field of the nucleus is the non abelian generalization of the Weizs\" aker-Williams field that accompanies the valence quarks of the nucleus. The field is treated as a random field, the randomness coming from the changes in the configurations of the color charges from collision to collision. A simple Gaussian ansatz is used for the correlation function of this random field, so that all needed averages can be easily performed using Wick's theorem. Subtleties related to gauge choices are briefly discussed. The section ends with a calculation of the gluon distribution, and an analysis of its saturation properties which arise from  the non linear interactions of the gauge field. 

Sect.~\ref{sec:evolution} is an important section of this paper. It concerns the calculation of radiative corrections and the progress realized over the last two decades in developing non linear equations that account for the saturation phenomenon and  its evolution with energy.  I  give there a concise and unified perspective on these developments, indicating connections between various formalisms, and pointing out open issues.  
A particular subtle aspect of the discussion concerns the way the radiative corrections to a simple projectile, such as a color dipole, can be  interpreted as a modification of the properties of the target, such as the saturation momentum.

In section 6, I finally address collisions of complex nuclei, focussing on a few specific aspects of particle production. New features emerge as compared to the previous sections where the only collisions treated  involve an elementary projectile with a nucleus. The treatment of nucleus-nucleus collisions requires  further approximations.  This section naturally leads to a small phenomenological discussion, whose role is mostly illustrative: a limited number of selected examples are presented, where direct effects of high parton density could play a visible role. 

\section{The wave function of a hadron at high energy}\label{sec:partons}

In this section, I recall the parton interpretation of the hadron wave functions at high energy.  
I use here the words ``wave function" in a rather loose sense. The notion of a wave-function at high energy suffers indeed from well-known ambiguities: it depends on the frame where it is defined, on the gauge chosen,  with the parton picture emerging more naturally in the light-cone gauge and in the infinite momentum frame. A further ambiguity arises  in high order calculations of a given process in the separation of the constituents of the hadrons from the probe that is used to measure them. For  instance, when two colliding hadrons exchange gluons, one may choose to regard these gluons as being part of the wave function of the projectile, or of that of the target, or as part of the interaction mechanism. We shall have opportunities to return to this issue. 

The most accurate knowledge of these high energy wave functions has been acquired mostly from Deep Inelastic Scattering (DIS) of elementary particles (leptons, mostly electrons) on hadrons (mostly protons, with some data on nuclei). This will be briefly reviewed in this section,  where 
most of the material presented is well known \cite{Mueller:Cargese,Kovchegov:2012mbw,Collins:2011zzd,Ioffe:2010zz}. The section serves to fix the notation, introduce basic concepts, such as those of integrated or unintegrated gluon distributions, the linear evolution equations, as well as the important concept of collinear factorization. The section ends with a short discussion of non linearities  that lead to saturation,  and of the associated   saturation momentum.

\subsection{Some orientation from light-cone perturbation theory}

\begin{figure}[htbp]
\begin{center}
\includegraphics[scale=0.3]{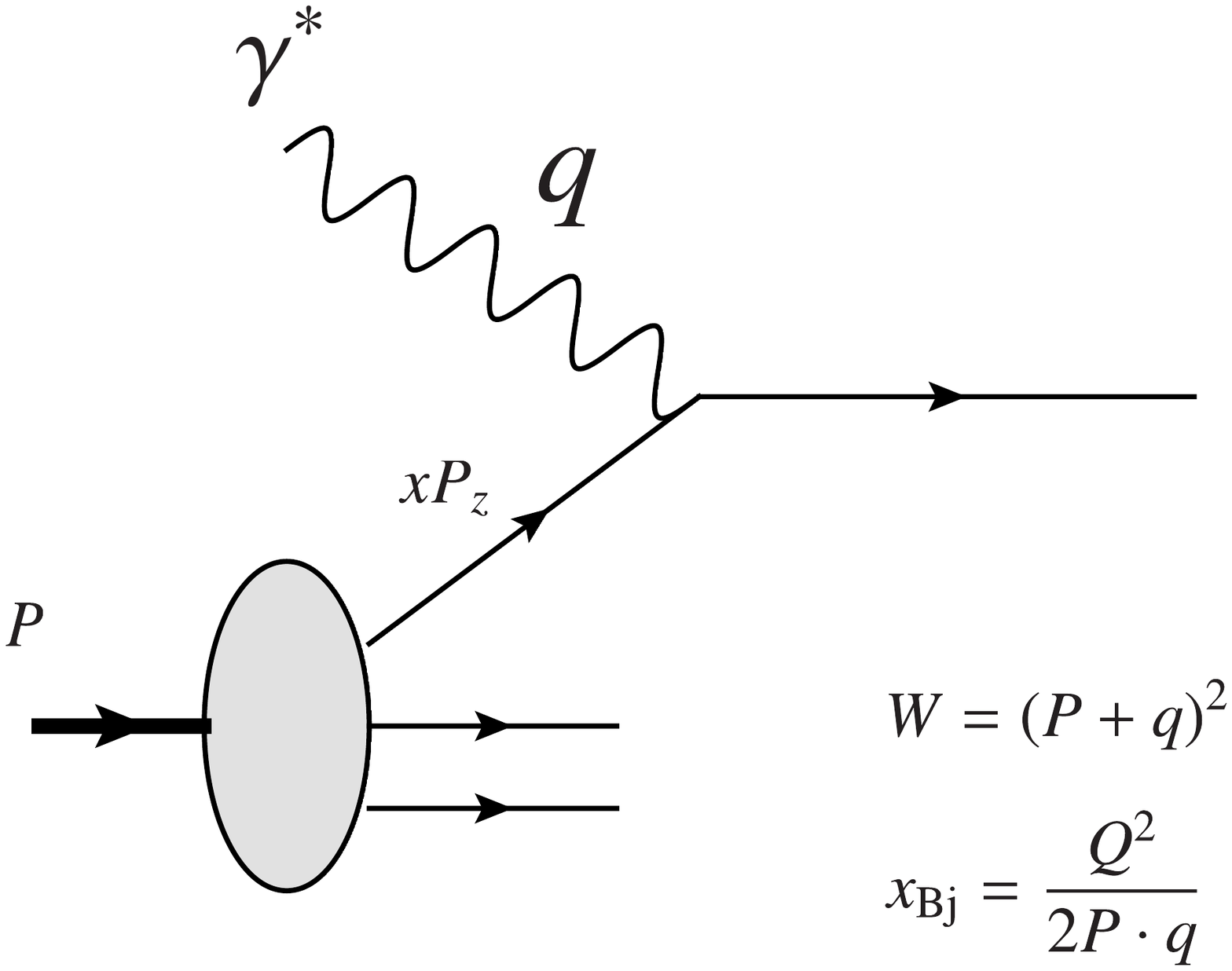}
\end{center}
\vspace{-1cm}
\caption{\label{fig:DIS0} Deep inelastic scattering of an electron on a proton, in  a frame where the proton is fast moving while the photon has little longitudinal momentum. The virtual photon $\gamma^*$ emitted by the electron (not shown)  is absorbed by one of the quarks present in the proton at the instant of the scattering. During the short interaction time, these quarks can be considered as independent particles.  The quark that absorbs the photon carries a fraction $x$ of the longitudinal momentum of the proton, with $x$ related to other kinematical variables as indicated by the formula. We have set $Q^2=-q^2>0$, with $q$ the four-momentum of the virtual photon, and $\hat s$ is the photon-proton  invariant mass. }
\end{figure}

The kinematics of a DIS process  is recalled in Fig.~\ref{fig:DIS0}.
For definitness, we consider the scattering of an electron on a proton, and focus first on the three valence quarks of the proton. The leading order picture corresponds to incoherent, and essentially elastic,  scattering of the electron on the individual valence quarks. The virtual photon $\gamma^*$, emitted from the electron with 4-momentum $q$, is absorbed by a quark of momentum $k$ which,  for the duration of the interaction,  can be considered as a real massless particle, i.e, $k^2=0=(k+q)^2$, or $q^2+2k\cdot q=0$.  Then,  setting $k=xP$, $Q^2=-q^2$,  we get  $x=x_{\rm Bj}=Q^2/(2P\cdot q)\approx Q^2/(\hat s+Q^2)$, where the approximate equality follows from ignoring the mass of the proton, and $\sqrt{\hat s}$ is the invariant mass of the photon-proton system. Deep inelastic occurs when  $Q$ is much larger that the typical momenta involved in the internal dynamics of the proton, of order $\Lambda_{QCD}$. This forces the interaction to take place on a small space-time scale, as we shall verify shortly. If in addition $\hat s\gg Q^2$, the partons that contribute to the process carry a small fraction of the longitudinal momentum of the proton, $x\approx Q^2/\hat s$. This limit of high energy and limited $Q^2$ is often referred to as the Gribov-Regge limit of QCD. 

Justification for this picture, often referred to as the ``naive" parton model,  is obtained  from simple order of magnitude estimates of basic time scales.  We  do this in the  ``infinite momentum frame", or Bjorken frame, in which the longitudinal momentum of the left moving  proton, $P_z=-p$,  is much larger than the proton mass $m$, while the photon has no longitudinal momentum $q^\mu=(q_0,\q,0)$, with $\q$ a vector\footnote{Throughout this paper, we denote vectors in the transverse plane by boldface symbols, e.g. $\u$, while the length of the vector $\u$ is denoted by $u_\perp$. } in the plane orthogonal to the momentum of the proton (the ``transverse plane"). A characteristic time scale for the interaction is given by $
\Delta t_{\rm int}\sim 1/q_0 \sim  {2xp }/{Q^2}$. 
This is to be compared to  the typical time scale of fluctuations within the proton.  In the rest frame of the proton, this time scale is $\sim 1/\Lambda_{QCD}$, but in the infinite momentum frame this is magnified by the Lorentz time dilation factor $p/m$, leading to $\Delta t_{\rm hadron}\sim p/(m\Lambda_{QCD})$. Since  $Q^2\gg m\Lambda_{QCD}$ and $x<1$, $\Delta t_{\rm int}\ll \Delta t_{\rm hadron}$. This large difference in time scales is the reason why one can consider the quark that absorbs the virtual photon as a free particle. 
\\

\begin{figure}[htbp]
\begin{center}
\vspace{-0.50cm}
\includegraphics[scale=0.3]{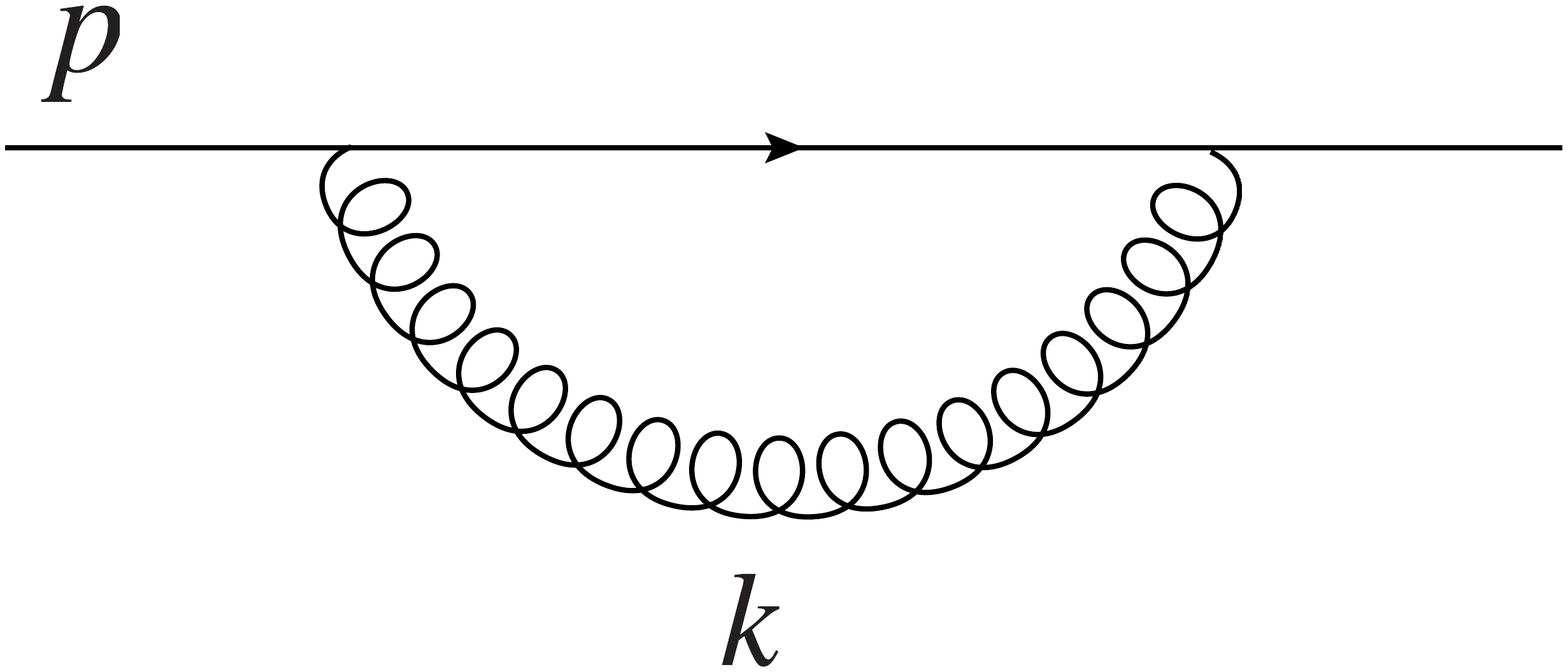}
\end{center}
\vspace{-2.5cm}
\caption{\label{fig:fluctuation0} An elementary gluonic fluctuation that accompanies a valence quark. The valence quark, represented by the straight line,  emits a gluon (the curly line) with  transverse momentum $\k$, and longitudinal momentum $k_z=xp_z$, with $p_z$ the longitudinal momentum of the valence quark. In the regime where $k_\perp\ll k_z$ and $k_z\ll p_z$, i.e., $x\ll 1$, the lifetime of the fluctuation is given by  $\Delta t\simeq 2xp_z/k_\perp^2$.
}
\end{figure}

\begin{figure}[htbp]
\begin{center}
\includegraphics[width=0.8\textwidth]{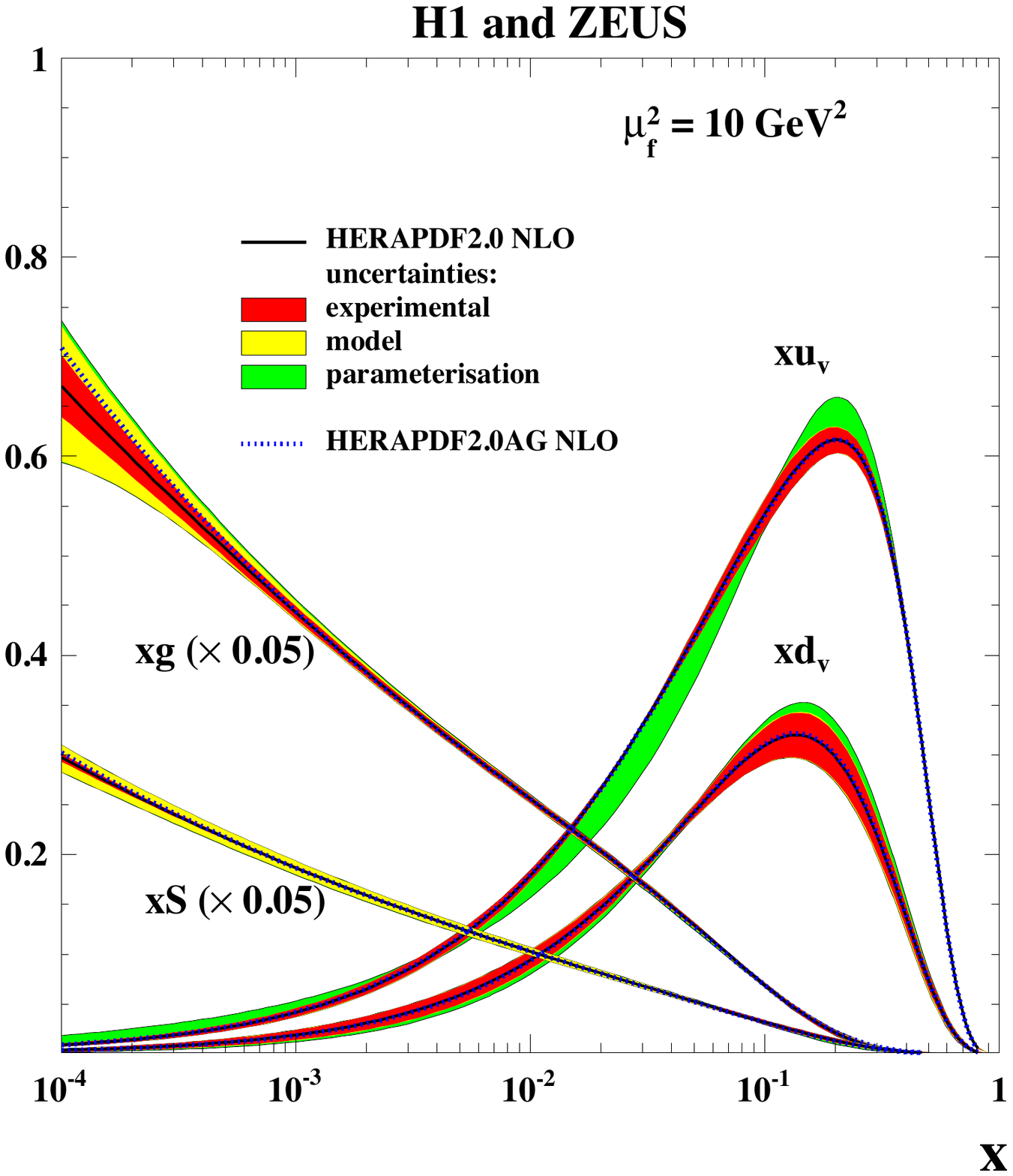}
\caption{Illustration of the $x$ dependence of the parton distribution functions of a proton,  as measured at HERA \cite{CooperSarkar:2012tx}.  At large values of $x$ (i.e., $x\gtrsim 10^{-1}$), the valence quarks dominate (their densities are denoted $xu(x)$ and $xd(x)$ for $u$ and $d$ quarks, respectively). However, at small $x$,  gluons ($xg(x)$)and sea quarks ($xS(x)$) completely dominate, with the gluon representing the largest parton density.  }\label{fig:HERA}
\end{center}
\end{figure}

So far we have been focussing on valence quarks, and these are indeed the dominant constituents of the proton when $x$ is not too small (see for instance Fig.~\ref{fig:HERA}). But we shall be eventually interested in partons that carry very small values of $x$. To understand where these partons come from, it is useful to recall the structure of  the proton wave function in the infinite momentum frame, as obtained in   light-cone perturbation theory \cite{Brodsky:1997de}. There, for instance, a valence quark may be viewed as a superposition  of the form $\ket{q}=a\ket{q}_0+b\ket{qg}_0+\cdots$, where $\ket{q}_0$, $\ket{qg}_0$, etc., are commonly referred to as Fock components, or Fock states (see  Fig.~\ref{fig:fluctuation0} for an illustration of the mixing of a ``bare" quark with a quark and a gluon). 
The calculation of the light cone  wave function, that is, the determination of the amplitudes ($a$, $b$, etc.) of the various Fock states,  requires a quantum mechanical calculation. However in the regimes relevant to the present discussion, such a calculation leads to  a probabilistic picture,  on which most of the forthcoming arguments will be based. Thus, the  elementary process displayed in Fig.~\ref{fig:fluctuation0} can be associated to gluon radiation, occurring with a   probability
\beq\label{eq:branching}
d{\cal P}\simeq \frac{\alpha_s C_R}{\pi^2}\,\frac{\rmd^2 \k}{k_\perp^2}\,\frac{dx}{x},
\eeq
where $\alpha_s\equiv g^2/(4\pi)$ is the strong coupling constant. 
This formula, analogous to that for QED bremsstrahlung,  is valid for the radiation from a quark (which belongs to the fundamental representation of the gauge groupe $SU(N_c)$), in which case $C_R=C_F=(N_c^2-1)/2N_c$, and for the radiation from a gluon (in the adjoint representation), where   $C_R=C_A=N_c$, with $N_c$ the number of colors ($N_c=3$ for QCD, but it is convenient to leave open the possibility to vary $N_c$, in particular when looking at simplifications that occur in the large $N_c$ limit). As revealed by Eq.~(\ref{eq:branching}), the radiation probability is enhanced when the emitted gluon carries a small transverse momentum  or a small energy fraction. These are commonly referred to, respectively, as ``collinear'' and  ``soft'' enhancements.   \\

Because gluons can themselves emit gluons, as well as quark-antiquark pairs, the structure of the wave function in terms of Fock states becomes gradually more and more complicated. Fortunately, we rarely need the details of this structure, but only a more inclusive information. Thus for instance, one may ask how many gluons are present in the wave function at a given energy. This is expressed in terms of a so-called integrated gluon distribution function, conventionally denoted $xG(x,Q^2)$.
 For just one valence quark, leading order perturbation theory yields 
 \beq\label{Gvalencequark}
xG(x,Q^2)=\frac{\alpha_s C_F}{\pi}\ln\left(  \frac{Q^2}{\Lambda_{QCD}^2} \right).
\eeq
This may be obtained simply by integrating the probability (\ref{eq:branching}) from some infrared cutoff $\sim \Lambda_{QCD}$ to the scale $Q^2$ at which the measurement is done.    
Roughly speaking,  $G(x,Q^2)\rmd x$ counts the number of gluons (of all colors and polarizations) in the hadron wave function (here the valence quark) with longitudinal momentum between $xP_z $ and $(x+dx)P_z$, and transverse momentum $k_\perp^2\lesssim Q^2$.  The infrared cut-off $\sim\Lambda_{QCD}$ accounts for the fact that the parton description ceases to make sense for partons that have wavelengths larger than the typical confinement scale $r_0\sim 1/\Lambda_{QCD}$. 
We shall also write  
\beq\label{undf}
xG(x,Q^2)=\int^{Q^2}_{\Lambda_{QCD}^2} \frac{\rmd k_\perp^2}{k_\perp^2} \varphi(x,k_\perp),
\eeq
where  $\varphi(x,k_\perp)=Q^2\frac{\del}{\del Q^2} xG(x,Q^2)$, is referred to as an  ``unintegrated parton distribution''. 
For a single valence quark,
\beq\label{varphiquark}
\varphi(x,k_\perp)=\frac{\alpha_s C_F}{\pi}\equiv \varphi_q.
\eeq
At this level of approximation, $\varphi_q$ is independent of $k_\perp$ and $x$.

\begin{figure}[htbp]
\begin{center}
\includegraphics[scale=0.3]{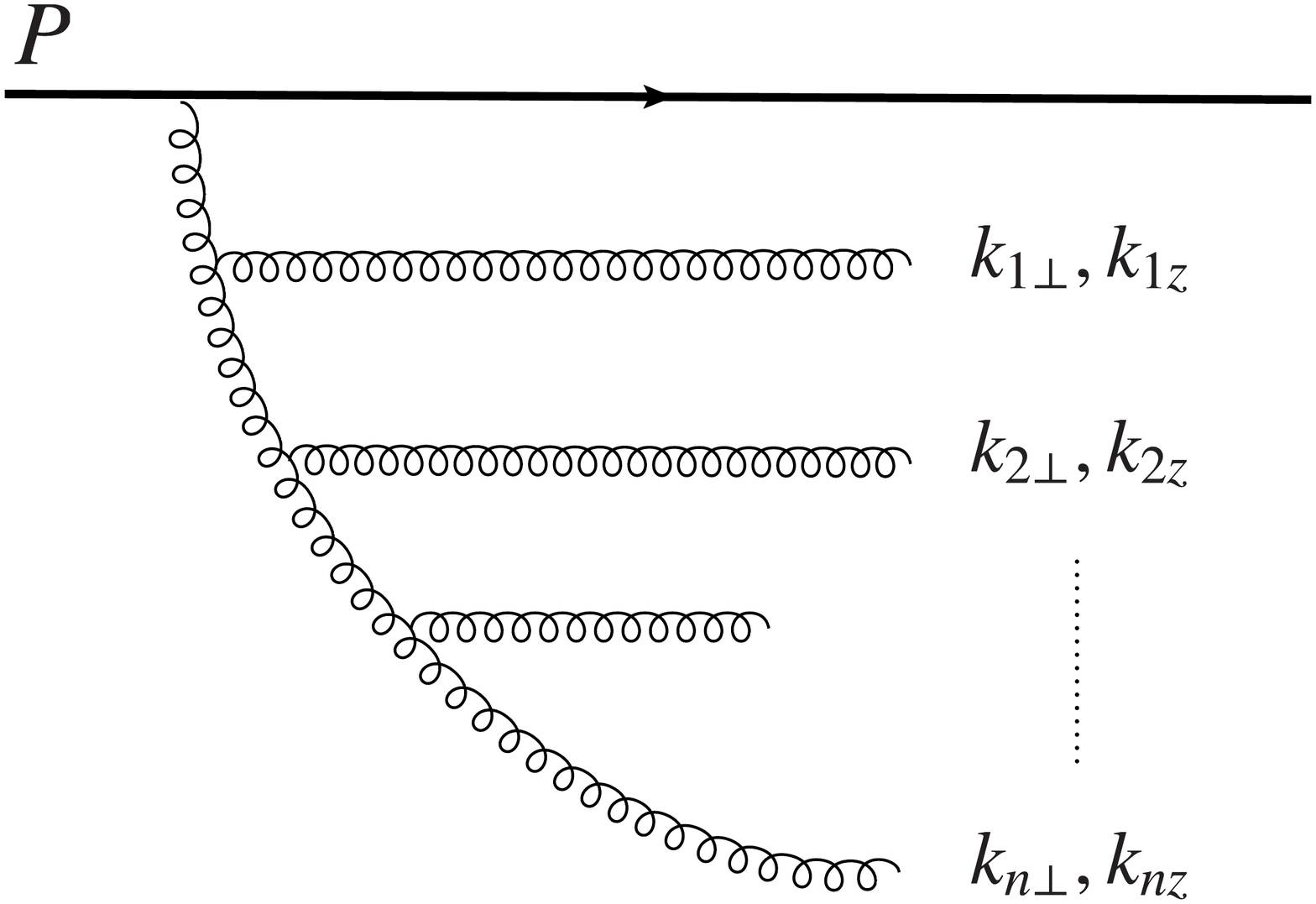}
\end{center}
\vspace{-1cm}
\caption{\label{fig:cascade} A typical cascade of gluon emissions. In DGLAP, the transverse momenta are ordered  $k_{1\perp}<k_{2\perp}<\cdots<k_{n\perp}$. In BFKL the longitudinal momenta are ordered $k_{1z}>k_{2z}>\cdots>k_{nz}$.   Note that the lifetimes of the successive fluctuations, $\Delta t_i=2k_{iz}/k_{i\perp}^2$, decrease as we move down the cascade.  }
\end{figure}

Leading order perturbation theory, which leads to Eq.~(\ref{Gvalencequark}),   is insufficient to determine the density of gluons which carry a large transverse momentum or a small longitudinal momentum.  When $\ln Q^2$ (or $\ln\frac{1}{x}$) increases, $\alpha_s \ln Q^2$ (and/or $\alpha_s\ln\frac{1}{x}$) may become of order unity, in which case higher orders have to be considered: there are indeed classes of diagrams where the smallness of the coupling is systematically compensated by such large logarithms. The relevant diagrams, those which maximize these logarithmic enhancements, correspond to relatively simple structures: they can be viewed as cascades of successive gluon emissions  such as that depicted in Fig.~\ref{fig:cascade}, where the transverse and/or the longitudinal momenta are strictly ordered. The first case (ordering in transverse momenta) corresponds to the Dokshitzer-Gribov-Lipatov-Altarelli-Parisi (DGLAP) cascade \cite{DGLAP}, the second (ordering in longitudinal momenta) to the Balitsky-Fadin-Kuraev-Lipatov (BFKL) cascade \cite{BFKL}. The properties of these cascades, and the linear equations that control their evolutions, will be recalled in the next subsections. 

\subsection{Linear evolution equations: DGLAP}

The gluon cascade that takes into account the corrections of order $\sim (\alpha \ln Q^2)^n$ is described by the DGLAP equation \cite{DGLAP}. For small (but not too small) $x$, and for gluons only  (we ignore here the quark contributions, subleading at high energy), this reads
\beq\label{eq:DGLAP}
Q^2\frac{\del }{\del Q^2}[xG(x,Q^2)]=\frac{\alpha_s(Q^2)C_A}{\pi}\int_x^1\frac{dz}{z} [zG(z,Q^2)],\eeq
where  $Q^2$ determines the scale of the running coupling constant. This equation has a probabilistic interpretation, its r.h.s. being proportional to the probability that a gluon carrying a momentum  fraction $z>x$ splits into two gluons (see Eq.~(\ref{eq:branching})), one of them carrying momentum fraction $x$. One should remember, however, that it results from  a quantum mechanical analysis;   in particular destructive interferences play an essential role in ordering successive transverse momenta along the cascade. 

Note an important feature of this equation, shared in fact by all the evolution equations that we shall meet in this paper: it does not allow the complete calculation of $xG(x,Q^2)$, but only its \emph{evolution} from some initial condition at a scale $Q_0\gg\Lambda_{QCD}$. 

The DGLAP evolution of parton distributions, because it is solidly  rooted in perturbation theory, which is under increasing control as $Q^2$ increases, plays an important role in the description of hadron structure, as probed for instance in DIS. The parton distributions have a universal character. Thus, once  measured in DIS for instance, they can be used to calculate other processes. This is due to  important factorization properties, to which we now turn.

\subsection{Collinear factorization} 

Factorization relies on a separation of time scales:  a long time scale associated with the ``preparation" of the wave function, and a short time scale during which the interaction takes place. The long time scale involves non perturbative physics, which in general one cannot calculate, and is buried into the initial condition for the parton distribution functions. The interaction, which occurs on a short time scale, involves a small coupling constant and  is calculable in perturbation theory. We have already seen an example of such factorization in our discussion of DIS. Another example, of direct relevance in the context of heavy ion physics, is that of the inclusive production of minijets, which are expected to be an important contribution to the total energy density produced in the collisions \cite{Kajantie:1987pd,Eskola:1996mb} (see also \cite{Niemi:2015voa}, and references therein). The corresponding  cross section is typically of the form
\beq\label{colfact}
\frac{\rmd \sigma}{\rmd p_\perp^2}= \int \rmd x_1  \int \rmd x_2\, x_1G(x_1,\mu^2)\,x_2 G(x_2,\mu^2)\, \frac{\rmd \hat\sigma_{gg\to gg}(\hat s,\mu^2)}{\rmd p_\perp^2},
\eeq
where $xG(x,Q^2)$ is a gluon distribution and $\hat\sigma_{gg\to gg}(\hat s,\mu^2)$ is an elementary parton-parton cross section  which can be calculated in perturbation theory. 
The quantity  $\mu$, the so-called factorization scale,  must be  sufficiently large to justify the factorization and the perturbative calculation of the parton-parton cross section. Corrections are expected to be power corrections in $1/\mu^2$, and are called higher twists.  The factorized formula (\ref{colfact}) can be read as a product of probabilities:  the probability to find in the colliding hadrons, gluons with fractions $x_1$ and $x_2$ of the  respective hadron momenta, times the probability  that these gluons, when colliding,  produce a jet with transverse momentum $p_\perp$. Such formulae are used in Monte Carlo codes describing particle production in nucleus-nucleus collisions (see e.g. \cite{Wang:1991hta}).

The scheme that underlies a formula such as Eq.~(\ref{colfact}) is commonly referred to as collinear factorization. In the context of heavy ion physics, collinear factorization is also used to determine the production of so-called hard probes (heavy quarks, $W$ and $Z$ bosons, jets, etc.). Hard probes are produced  over very short space-time scales through  parton-parton interactions which are not influenced by the surrounding medium. However, hard probes produced in nucleus-nucleus collisions may involve parton densities which  are different from those of a collection of independent nucleons. These are referred to as  nuclear parton distribution functions \cite{Eskola:2009uj}.

\subsection{Linear evolution: BFKL}

We now return to the evolution equations and consider increasing  the energy, keeping $Q^2$ bounded. Then $1/x$ decreases, and one eventually reaches a regime where $\alpha_s\ln (1/x)$ becomes of order unity, and the corresponding large logarithms need to be resummed. This new resummation is achieved by the BFKL equation \cite{BFKL}, which gives rise to a cascade similar to the DGLAP cascade (see Fig.~\ref{fig:cascade}), with however a different ordering of the various emissions, as we have indicated earlier.

Written as an equation for the unintegrated gluon density, the  BFKL equation takes the form \footnote{Throughout this paper we use the following shorthand notation for the integrals over two dimensional vectors in the transverse plane
$$
\int_\x=\int\rmd^2\x,\qquad \int_\k=\int\frac{\rmd^2\k}{(2\pi)^2},
$$
where the first integral is an integral over coordinates, while the second is over momenta. }
\beq\label{BFKL0}
\frac{\del \varphi(y,\k)}{\del y}=\frac{\bar\alpha_s}{2\pi} \int_{\p}\frac{\k^2}{\p^2(\k-\p)^2}\left[ 2 \varphi(y,\p)- \varphi(y,\k)  \right],
\eeq
where we have change variable from $x$ to the rapidity $ y=\ln (1/x)$, and we have set $\bar\alpha_s\equiv \alpha_s C_A/\pi$ (this conventional notation will be used throughout).
This equation  may be  given a probabilistic interpretation, with a loss term $\sim \varphi(y,\k)$ representing a process by which the gluon $\k$ disappears by splitting into gluons with momenta $\p$ and $\k-\p$, respectively, while the term $\sim \varphi(y,\p)$ summarizes the effect of gain terms corresponding to the splittings  $\p\to (\k,\p-\k)$ and $\p+\k\to (\k,\p)$. The kernel $\sim\bar\alpha_s \k^2\rmd^2\p/(\p^2 (\k-\p)^2)$ represents the probability for a gluon $\k$ to split into two gluons with momenta $\p$ and $\k-\p$.

The BFKL equation is linear, and local in rapidity. One of its  most remarkable feature, already alluded to,  is the exponential growth that it predicts for the gluon density (see Sect.~\ref{sec:solutionBK})
\beq
\varphi(y,\k)\sim {\rm e}^{\omega\bar\alpha_s y},
\eeq
with $\omega=4\ln 2$ (in leading order).  This explosive growth may be given a simple physical interpretation in terms of the color charge that keeps accumulating along the cascade, providing  a source of increasing strength for subsequent emissions \cite{Mueller:1994up}. \\

\begin{figure}[htbp]
\begin{center}
\includegraphics[scale=0.3]{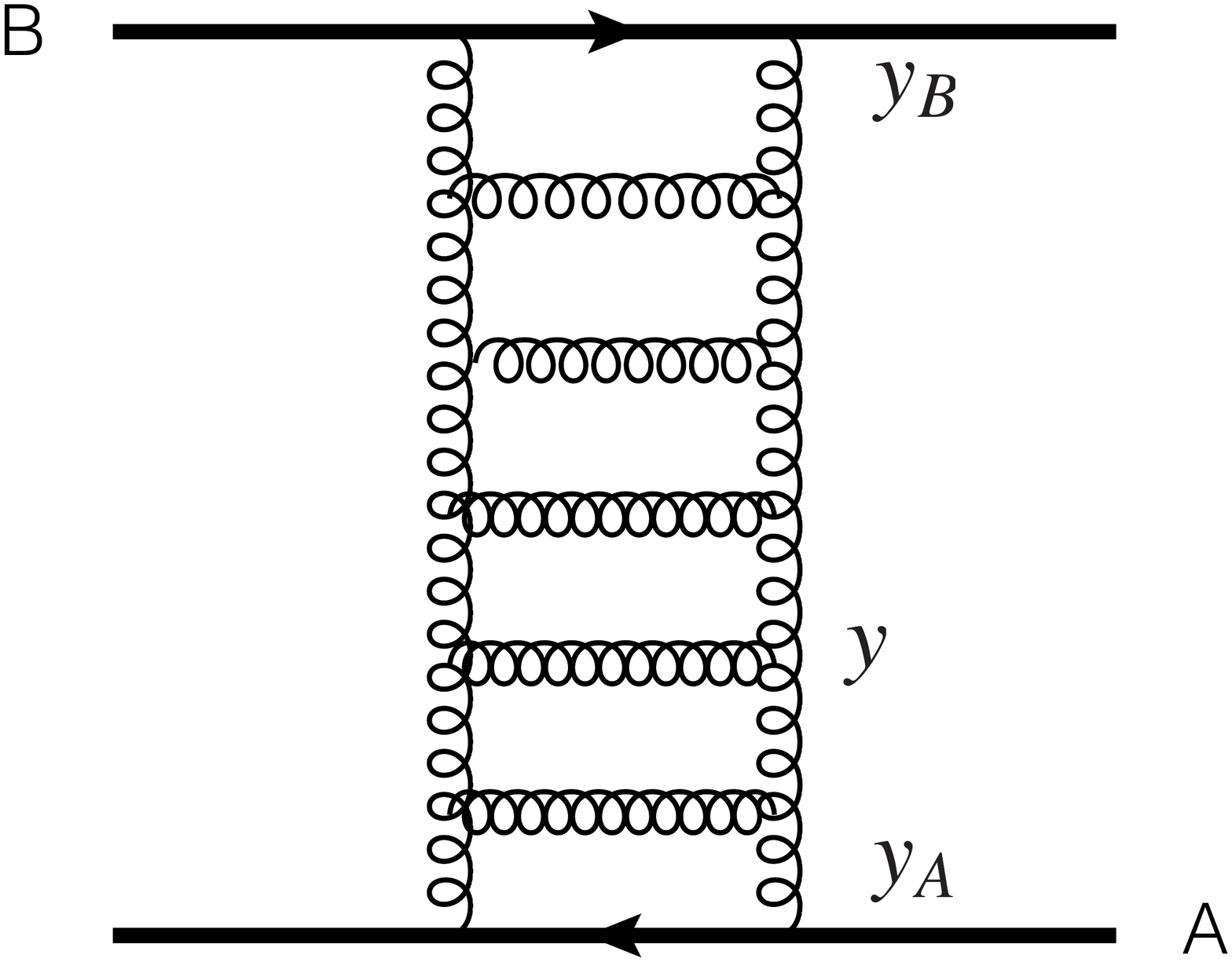}
\end{center}
\caption{\label{fig:gluon_ladder} A gluon ladder that contributes to the interaction between two hadrons (represented here by horizontal straight lines), and  often considered as a model for the (perturbative) Pomeron exchange  \cite{Forshaw:1997dc,Donnachie:2002en}
. Note that gluons propagators and vertices in such ladders receive non trivial corrections that will not be discussed here. This picture illustrates a remark made at the beginning of this section. The gluon ladder spans the entire rapidity gap between A and B. Unless we mark one of the gluons, as for instance we would do when calculating the inclusive particle production at a given rapidity $y$, it is difficult to attribute the gluons of the ladder to either A or B. One may view the process in different frames. In a frame with rapidity $y$ close to $y_A$  for instance, one would say that A suffers little evolution, while most evolution is put into B; in that case it could be natural to associate most gluons of the ladder to B. }
\end{figure}
 
The BFKL resummation has been also much studied in the context of hadron-hadron, in particular proton-proton, collisions. The focus in this case is not so much the  wave functions of the hadrons, but rather  the connection between QCD and Regge theory, on which many models of soft hadronic interactions are based. This emphasizes the exchange of particles, or more complex objects, in the ``$t$-channel''. Thus the BFKL ladder (see Fig.\ref{fig:gluon_ladder}) represents a contribution to the two gluon exchange process, dressed by gluon exchanges. This  point of view will not be discussed in this paper (see e.g.  \cite{Forshaw:1997dc,Donnachie:2002en} for pedagogical introductions), although it should be stressed that it has led to a very successful phenomenology of heavy ion collisions in a wide range of energies (see e.g. \cite{Werner:1993uh,Drescher:2000ha}). The ``$s$-channel'' point of view adopted in this paper, which focusses on the structure of the colliding objects,  lends itself more naturally to space-time descriptions of heavy ion collisions on which are based many theoretical approaches, such as hydrodynamics for instance.
  
\subsection{Gluon recombination and the saturation momentum}

Both the DGLAP and the BFKL equations lead to an increase of the parton density with increasing $Q^2$ or decreasing $x$, respectively.  In the case of DGLAP, the transverse size of the new partons shrinks as $1/Q^2$, while the number of new partons increases typically as $\ln Q^2$. It follows that  the area occupied by these new partons  in the transverse plane eventually decreases with increasing $Q^2$. The net result is that the system of partons produced by the DGLAP evolution is effectively more and more dilute, with the partons becoming weakly coupled (since the coupling strength $\alpha_s(Q^2)$ decreases with increasing $Q^2$): As $Q^2$ grows (with $x$ kept not too small) perturbation theory becomes more and more reliable in describing the changes in the hadron wave functions. This is to be contrasted with the BFKL evolution. The latter takes place with $Q^2$ being limited, so that the size of the partons remain roughly constant in the transverse plane. The exponential growth of the parton number predicted by the BFKL evolution then leads rapidly to a full occupation of the transverse plane. Since small $x$ partons are already overlapping in the longitudinal direction, one reaches a regime where partons fully overlap spatially. The system becomes then effectively strongly coupled, even though the coupling strength which controls parton-parton interactions may not be large. \\

\begin{figure}[htbp]
\begin{center}
\includegraphics[scale=0.4]{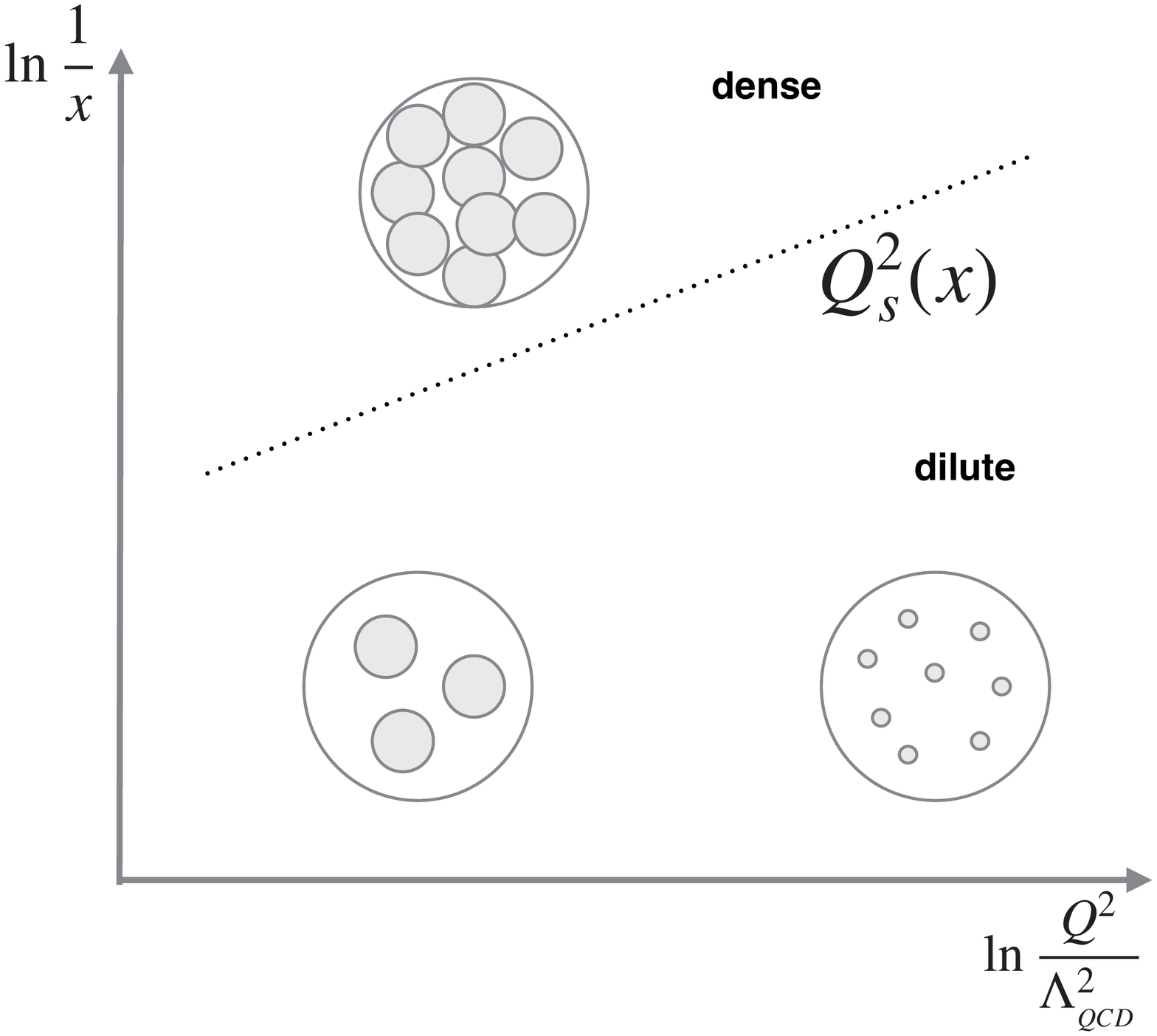}
\end{center}
\caption{\label{fig:xQplane} The various regimes in the plane $\ln (1/x), \ln Q^2$. At large $Q^2$, and moderate values of $x$, partons form a dilute system and are weakly coupled ($\alpha_s(Q^2)\ll 1$). At small values of $x$, partons form a dense (`saturated') system. Their high density makes them strongly coupled, although $\alpha_s\sim \alpha_s(Q_s)$ in this regime may still  be small. The divide between these two regions (dotted line) is the saturation momentum. Around this line, non linear effects strongly modify the evolution. The DGLAP evolution corresponds to increasing $Q^2$ with moderate variations of $x$ and leads to a more and more dilute system. The BFKL evolution corresponds to roughly constant $Q^2$ and decreasing $x$, and leads to a denser and denser system, eventually crossing the saturation boundary indicated by the dotted line \cite{Gribov:1984tu}. }
\vspace{0.5cm}
\end{figure}

When the partons start to occupy the same region in the transverse plane, one may expect corrections to the processes responsible for their multiplication due to gluon splitting: some form of gluon recombination may become important 
 \cite{Gribov:1984tu,Mueller:1985wy}. This effect was first estimated as a higher twist ($\sim 1/Q^2$) correction to the DGLAP equation:
 \beq
 Q^2\frac{\del}{\del Q^2}[xG(x,Q^2)]=\bar\alpha_s \int_x^1 \frac{\rmd z}{z}[zG(z,Q^2]-\frac{4\bar\alpha_s^2 \pi^3}{(N_c^2-1)Q^2}\int_x^1 \frac{\rmd z}{z} z^2 G^{(2)}(z,Q^2),\nn
 \eeq
 where $xG^{(2)}(x,Q^2)$ is a two-gluon density (at coincident values of its arguments).
 By using the simple approximation for the 2 gluon density, $x^2 G^{(2)}(x,Q^2)=(2/3) [x G(x,Q^2)]^2/(\pi R^2)$, and  taking a derivative with respect to $\ln x$ one obtains  \cite{Mueller:1985wy}
\beq\label{GLRMQ}
\frac{\partial^2\; xG(x,Q^2)}{\partial \ln(1/x) \,\partial \ln Q^2}=\bar\alpha_s \,xG(x,Q^2)-\frac{8}{3(N_c^2-1)}\bar\alpha_s^2\,\pi^2\frac{[xG(x,Q^2)]^2}{R^2Q^2} ,
\eeq
which differs from  Eq.~(\ref{eq:DGLAP}) by the second term accounting for ``gluon recombination".  Note that the terminology, suggestive of a kinetic interpretation, could be misleading. The recombination that is involved here is not quite the reverse of the gluon splitting (the first term of Eq.~(\ref{GLRMQ})), as indicated by the different powers of $\alpha_s$ in the two terms of Eq.~(\ref{GLRMQ}). In fact the gluons that are merging belong to two distinct cascades. 

 Saturation occurs when the two terms in the right hand side of Eq.~(\ref{GLRMQ}) balance each other.  A momentum scale naturally emerges from this condition. This scale, called the saturation momentum $Q_s$,  is (parametrically) given by
\begin{equation}\label{Qsaturation}
 Q_s^2
\sim \alpha_s(Q_s^2)\frac{xG(x,Q_s^2)}{\pi R^2}\; .
\end{equation}
As illustrated in Fig.~\ref{fig:xQplane}, the saturation momentum separates partons into dilute modes with $k_\perp\gg Q_s$ that are weakly coupled, and modes with $k_\perp\lesssim Q_s$ that are strongly coupled because densely populated. The onset of saturation coincides indeed with a breakdown of perturbation theory where 
gluon interaction energies become comparable to their transverse  kinetic energies, that is, loosely speaking,  $\partial ^2\sim \alpha_s
\langle A^2\rangle_Q$, where $\langle A^2 \rangle_Q\sim xG(x,Q^2)/\pi R^2$. This is the place where Eq.~(\ref{GLRMQ}) ceases to make sense.
Another important consequence of Eq.~(\ref{Qsaturation}) is that, at saturation, the phase space density of modes with $k^2_\perp\lesssim Q^2$ is large, of order  $xG/(Q_s^2R^2)\sim 1/\alpha_s$. 

We shall later give a more precise definition of $Q_s$, where the numerical coefficients will be fixed (see Eq.~(\ref{barQs})). But the parametric relation (\ref{Qsaturation}) already allows  for a number of important observations. 
Equation (\ref{Qsaturation}) shows that the square of the saturation momentum is directly proportional to the density of gluons in the transverse plane. Since this density grows with energy, so does the saturation momentum.  Also, one expects the gluon density of a nucleus $xG_A(x,Q_s^2)$ to be additive, in a first approximation, i.e,  $xG_A(x,Q_s^2)\sim A$, with $A$ the number of nucleons in the nucleus. Since $R^2\sim A^{2/3}$,  $Q_s^2$ grows with the nucleon number as $A^{1/3}$. Finally, as suggested by the running of the coupling explicitly indicated in Eq.~(\ref{Qsaturation}), if $Q_s$ is large, the coupling constant is small. Thus, although naive perturbation theory is inapplicable because of the large density of partons, weak coupling techniques may still be used to describe the saturated gluon systems. Finally, let us recall that  early estimates of effects of saturation in nucleus-nucleus collisions, based on Eq.~(\ref{GLRMQ}), 
 were made in \cite{Blaizot:1987nc}.

\section{Propagation of fast partons in dense QCD matter}\label{sec:eikonal}

The propagation of fast partons in nuclear matter is an essential ingredient in the description of hadronic collisions at high energy: A typical projectile can be  viewed as a collection of partons which, as they cross the target,  do not interact among themselves, and follow essentially straight lines trajectories (exceptions are rare, hard, collisions in which partons are deflected by large angles). These aspects of  high energy interactions are well captured by the eikonal approximation \cite{Nachtmann:1991ua,Nachtmann:1996kt,Buchmuller:1995mr,Kovner:2001vi}. 

In fact, we shall focus in this section on a simple projectile, a color dipole made of a quark-antiquark pair (or a pair of gluons) in a color singlet state.  Such dipoles are ubiquitous. It could be the dipole into which a virtual photon fluctuates during a deep inelastic scattering, or the quark antiquark pair inside a pion, or the constituents of charmonium, etc.  A dipole also emerges naturally when one studies the propagation of a single fast parton (quark or gluon) in a nucleus.  

In this section, we use the dipole as a test particle to probe the state of the target. We treat the interaction in the eikonal approximation, where parton propagators take the form of  Wilson lines. The average of the Wilson lines over the state of the target is done first in the weak field approximation, where the Wilson lines can be expanded, and then in the strong field regime via a  multiple scattering calculation. The latter calculation provides a simple mechanism for the ``unitarization" of the interaction between the dipole and the target, that is akin to saturation, and relates  the saturation momentum to momentum broadening. The section ends with a brief discussion of the phenomenological dipole model applied to deep inelastic scattering, and the experimental evidence it gives for  the existence of the saturation momentum. 

\begin{figure}[htbp]
\begin{center}
\includegraphics[scale=0.35]{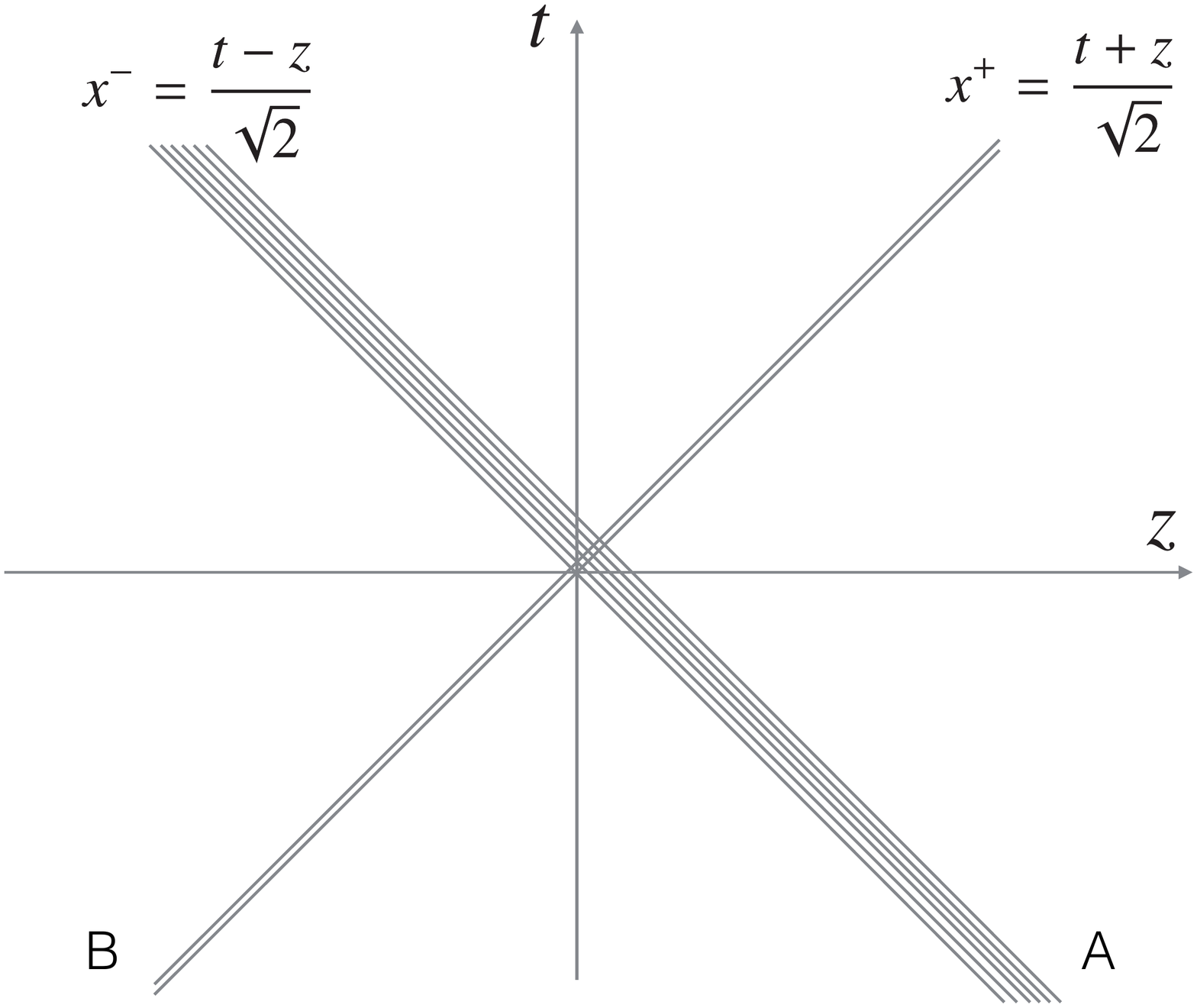}
\end{center}
\caption{\label{fig:light_cone} Space-time diagram for the collision of an elementary projectile  B (e.g. a color dipole) with a target A (e.g. a nucleus), with light cone coordinates.  The target  is Lorentz contracted so the spread in $x^+$ is nearly negligible.  The coordinate $x^+$ plays the role of a longitudinal coordinate for the target, while $x^-$ plays the role of time. Correlatively, for the target,  $p^+$, variable conjugate to $x^-$ in a Fourier transform, plays the role of an energy, and $p^-$ that of a longitudinal momentum. The roles of $x^+$ and $x^-$, as well as $p^+$ and $p^-$,  are interchanged for the projectile. }
\end{figure}

\subsection{Eikonal, Wilson lines }

As will be the case in most of this paper,  we take the projectile to be right moving, while the target, which may be a proton or a nucleus,  is left moving. We assume that the interaction does not involve large momentum transfer so that, in  a first approximation, the transverse coordinates of the projectile partons do not change as they propagate through the target. Under these conditions, the parton propagation  is described by an eikonal phase along a straight trajectory, which here takes the form of a path ordered exponential
\begin{equation}\label{Ueikonal}
U({x})\,\equiv\,{\rm T}\,{\rm exp}\left({\rm i}g\int_{-\infty}^{x^+} dz^+
A^-_a(z^+,{\x}) t^a\right),  
\end{equation}
where $A^-_a(z^+,{\x})$ is the color field of the target, and $t^a$ a color matrix. 
The path ordering, denoted here by the symbol T, plays indeed a role similar to the  time ordering (recall also that $x^+$ plays the role of a time for the right moving projectile -- see Fig.~\ref{fig:light_cone}). It is needed to take care of the non commutation of the color matrices: it places the operators with larger $x^+$ to the left of those with smaller $x^+$. Throughout, we use light-cone coordinates, $x^\mu=(x^+,x^-,{\x})$, with $x^\pm=(t\pm z)/\sqrt{2}$ and ${\x}$ is a 2-dimensional vector in the transverse plane. We shall refer to $U(x)$ in Eq.~(\ref{Ueikonal}) as a Wilson line. In cases where $x^+$ can be sent to infinity, this will be denoted simply as  $U_\x$. Equation~(\ref{Ueikonal}) is written for a quark, with $t^a$ a color matrix in  the fundamental representation. A similar Wilson line describes the eikonal propagation of a gluon with $t^a$ replaced by a matrix $T^a$ of the adjoint representation. 
We have  chosen a gauge where $A^+=0$, suited to the parton description of the projectile,  and such that the only relevant component of the gauge field of the nucleus is $A^-(x^+,\x)$.

\begin{figure}[htbp]
\begin{center}
\includegraphics[scale=0.2]{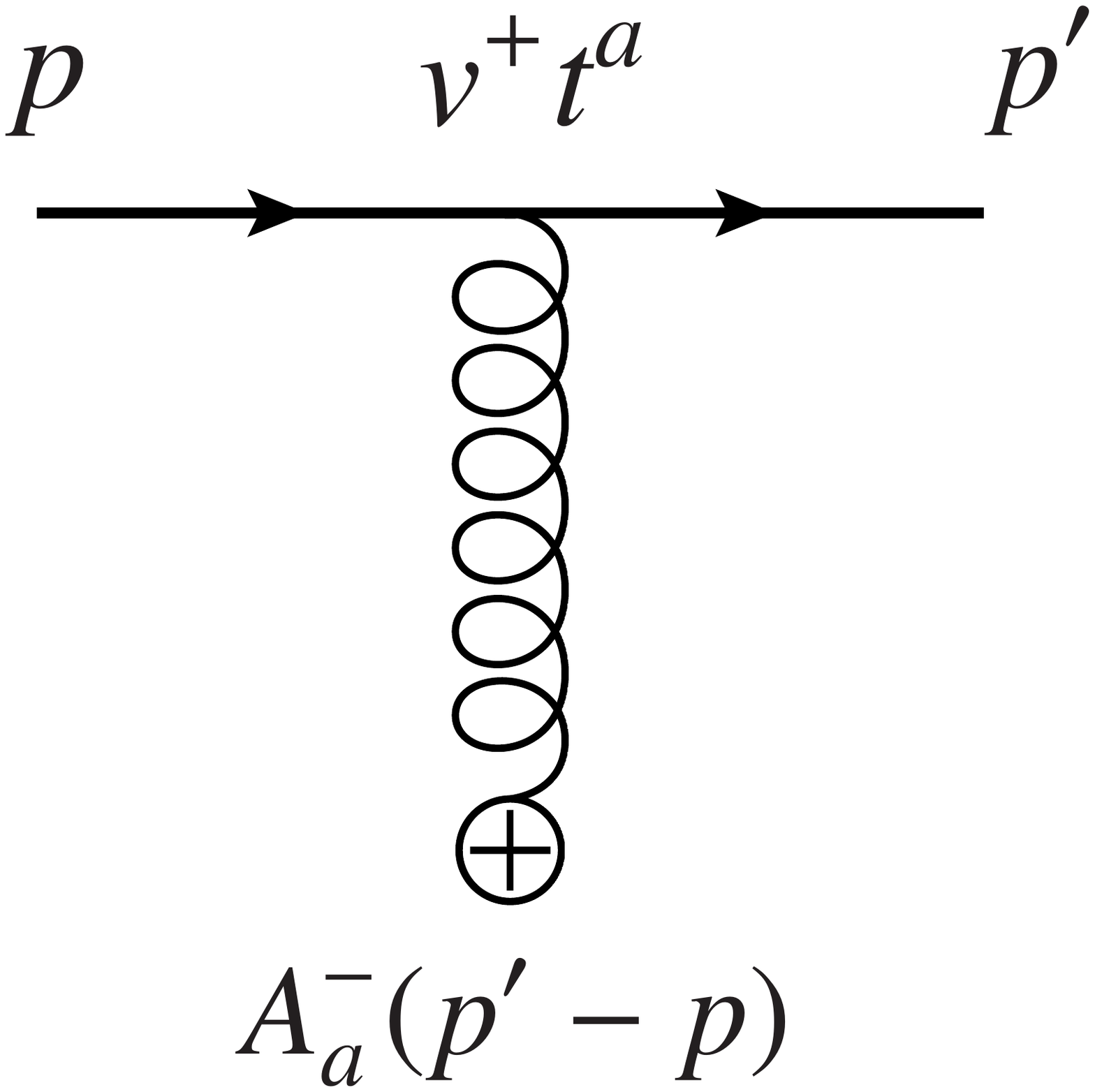}\includegraphics[scale=0.22]{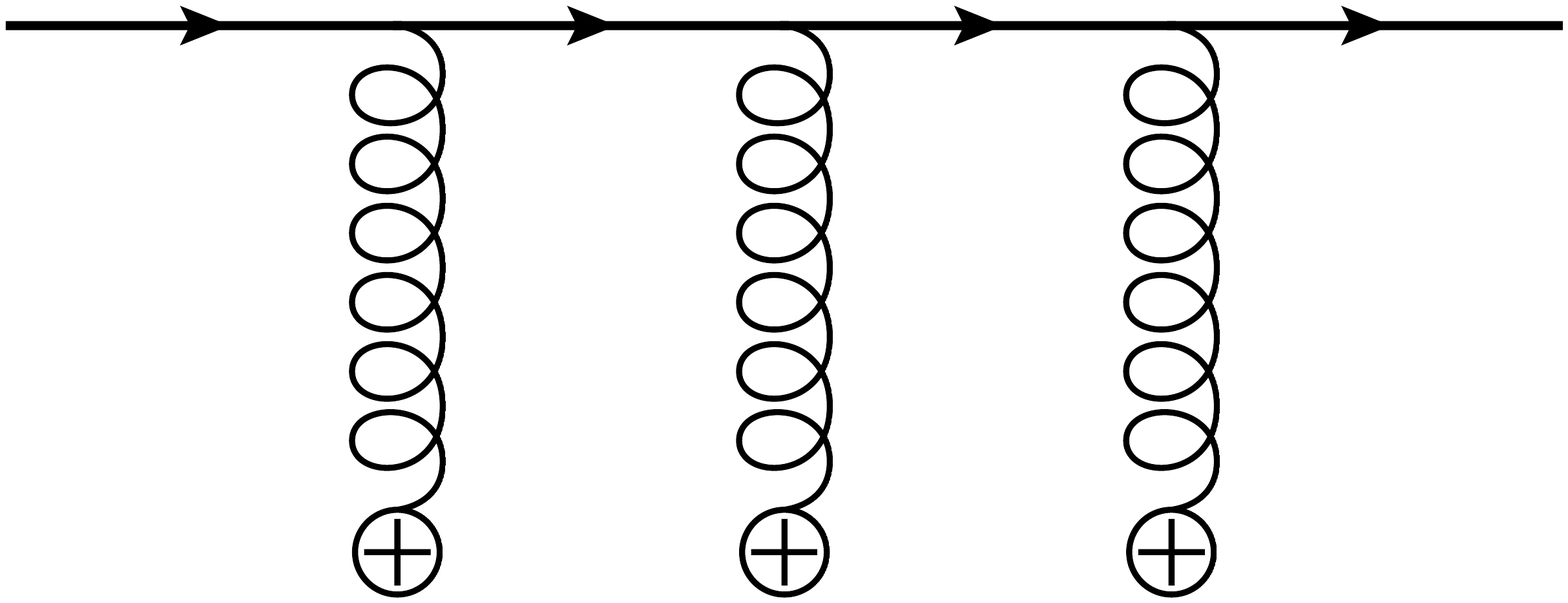}
\end{center}
\caption{\label{fig:eikonal} The eikonal vertex (left) for the propagation of a quark in the color field $A^-_a$ of a target. The quark momentum is along the + direction, i.e., its transverse momentum vanishes and $p_z\simeq p_0$.  The interaction with the field $A^-$ is soft, with negligible transverse momentum exchange. The + component of the momentum is preserved in the interaction (i.e., $p'^+=p^+$), as well as spin or polarization. The gluon coupling would be similar, with $v^+ t^a$ replaced by $p^+ T^a$. The diagram on the right illustrates the multiple insertions of the field which need to be resummed in a Wilson line when the field is strong, that is when $g\int_{x^+} A^-(x^+)\gtrsim 1$.}
\end{figure}

Diagrams illustrating the eikonal coupling and eikonal propagation are given in Fig~\ref{fig:eikonal}. The coupling between the parton, taken here to be a quark, and the gluon field is given by the interaction Hamiltonian $ g\bar q(x) \gamma_\mu A^\mu_a(x) t^a q(x)$ with,  in light cone variables, $\gamma_\mu A^\mu=\gamma^+ A^-+\gamma^- A^+-\gamma_\perp\cdot A_\perp$. With our present choice of gauge, and the fact that we ignore the transverse motion of the projectile partons, this reduces to $\gamma^+ A^-\to v^+ A^-$, from which Eq.~(\ref{Ueikonal}) follows.
 In the eikonal approximation the component $p^+$ of the momentum is conserved (there is no exchange of $p^+$ with the target), which is related to the fact that the field $A^-$ can be considered as independent of $x^-$  during the interaction.  Note also that the projectile parton only probes the field $A^-$ at (or near) $x^-=0$.

\subsection{Deep inelastic scattering in the dipole frame}

We now return to DIS, and view the process in a frame,  the ``dipole frame'',  where the photon has a large longitudinal momentum. In this frame, Lorentz time dilation renders the  fluctuations of the virtual photon into $q\bar q$ pairs  long lived, so that  the interaction of the photon with the nucleus is best  viewed as the interaction of the  $q\bar q$ Fock component of the photon with the target (see Fig.~\ref{fig:xQplane}). The picture in the dipole frame is complementary to that given earlier in the  Bjorken frame. The Bjorken frame is convenient to exhibit the parton structure of the target, while the dipole frame puts emphasis on its color field (see e.g. \cite{Mueller:Cargese}).

\begin{figure}[htbp]
\begin{center}
\includegraphics[scale=0.4]{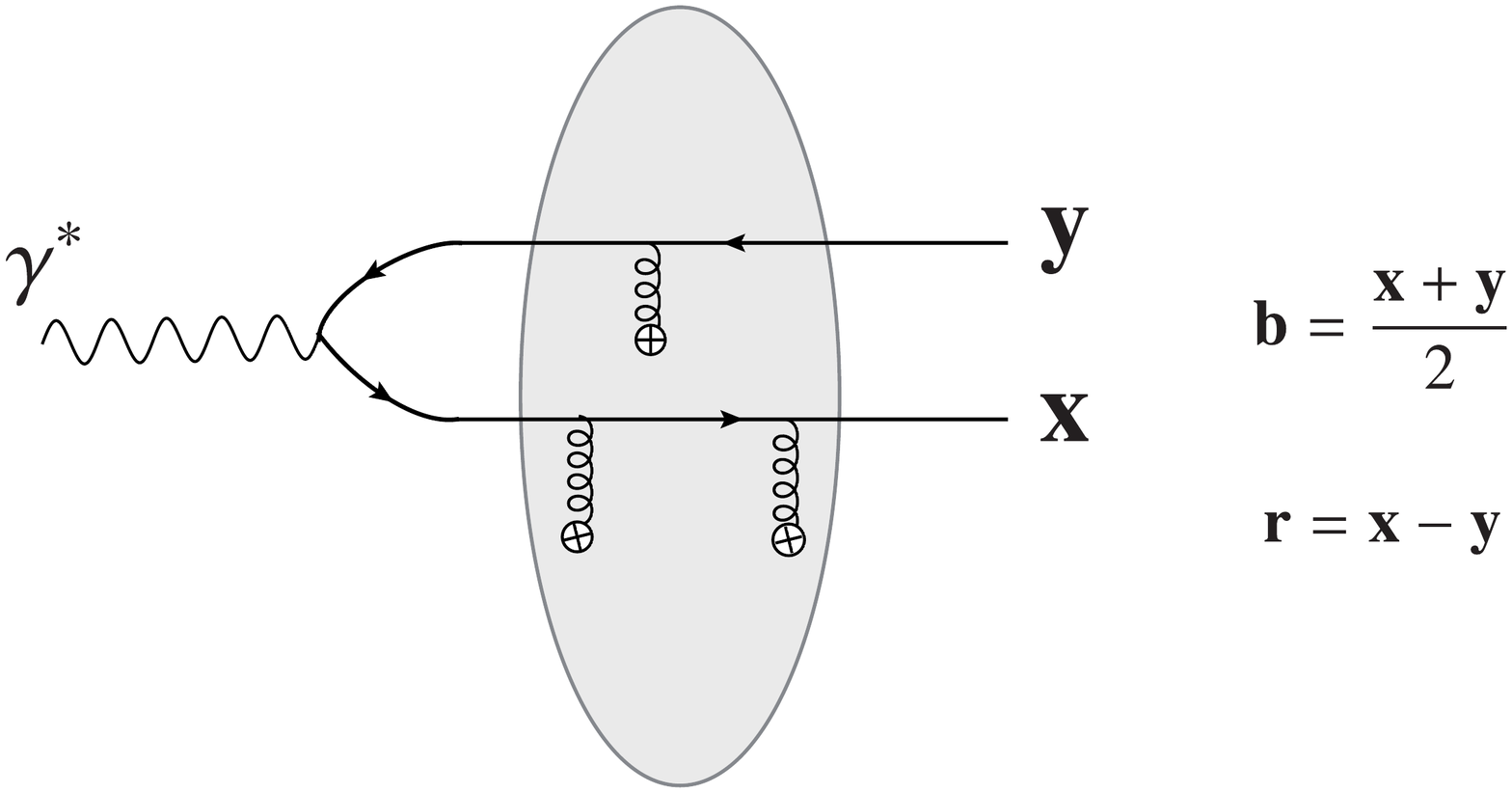}
\vspace{-3cm}
\end{center}
\caption{\label{fig:xQplane} DIS viewed in a frame where the lifetime of the fluctuation of the virtual photon into a $q\bar q$ pair is enhanced by Lorentz time dilation.  In this frame, the interaction of the virtual photon with the hadron is visualized as the interaction of the  $q\bar q$ pair, a color dipole, with the color field of the hadron. }
\end{figure}

In the dipole frame,  the total cross section 
$\sigma_{\gamma^* p}$  (at a given impact parameter)  is of the schematic  factorized  form \cite{Nikolaev:1990ja}
\begin{equation}
\sigma_{\gamma^* p}(x,Q^2)=\int\limits_0^1 dz\int_\r
\left|\psi(Q^2,z,\r)\right|^2
\sigma_{\rm dip}(\r),
\label{eq:fact-F2}
\end{equation}
where we leave aside here the distinction between transversely and longitudinally polarized virtual photons. 
In this formula, $\psi(Q^2,z,\r)$ is the Fock component of the
virtual photon light-cone wave function that corresponds to a
$q\bar{q}$ pair; it depends on the invariant mass
$Q^2$ of the virtual photon, on the transverse distance $\r$ between the quark and the antiquark, and on the fraction $z$ of the photon longitudinal
momentum taken by the quark.  The other factor in this formula, the dipole cross section
$\sigma_{\rm dip}(\r)$,  summarizes the QCD
interaction of the dipole with the hadronic target.  It can be calculated in terms of the (forward) $S$-matrix element
\beq\label{Stau}
S({\b,\r})\,\equiv\,\frac{1}{N_c}\,
\langle \tr\big( U_\x U^\dagger_\y\big)
\rangle\eeq
where the two Wilson lines $U_\x$ and $U^\dagger_\y$ describe  the propagation of the quark (at transverse position $\x$) and the antiquark (at $\y$) in the field of the target, and the brakets indicate an averaging over the color field of the target, which will be specified shortly. We have set $\b=(\x+\y)/2$, representing the impact parameter of the dipole and $\r=\x-\y$, and we shall use interchangeably the notations $S_{\x\y}$ and $S(\b,\r)$ for the $S$-matrix. The dipole cross section can then be obtained from the general relation, in the eikonal approximation, between the $S$-matrix and the total cross section, namely  
\beq\label{sigmatoteikonal}
\sigma_{\rm dip}(\r)=2\int\rmd^2\b\, (1- {\rm Re}\, S(\b,\r)).
\eeq  

The formula (\ref{eq:fact-F2}) exhibits a factorization into a product of probabilities:  the  square of the light-cone wave function of the virtual photon $\gamma^\ast$, giving  the probability that
$\gamma^\ast$ splits into  a $q\bar q$ pair of a given transverse size, and the probability that this color dipole interacts with the target, represented by the dipole cross section. This factorization  relies, as emphasized before, on the existence of well separated time scales: a long time scale associated with the fluctuation of the virtual photon into a $q\bar q$ pair (this time scale is magnified by Lorentz time dilation in the dipole frame), and the short time scale of  the interaction of the dipole with the target color field. During the short duration of the interaction, the dipole transverse coordinates may be considered as frozen.

\subsection{Momentum broadening of a fast quark in a medium}

The dipole $S$-matrix appears, albeit in a somewhat less direct way, in the calculation of the momentum broadening of a fast parton propagating in a thick target. 
We are interested in the probability ${\cal P}(\b,\p )$ that a parton, e.g. a quark, entering the target at impact parameter $\b$ with some transverse momentum $\p_0=0$ leaves it with some transverse momentum $\p$. The amplitude for this to happen is given by 
\beq
\int_\x\rme^{i(\p_0-\p)\cdot \x}\,\bra{\Psi_n,\beta} U_\x \ket{\Psi_0,\alpha},
\eeq
where the initial state of the target is denoted by $\Psi_0$, the final state by $\Psi_n$. The indices $\alpha$ and $\beta$ refer to the initial and final colors of the quark, respectively.  The probability ${\cal P}(\b,\p )$ is obtained by squaring the amplitude, averaging over the initial color, and summing over the final color, as well as summing over the final  states of the target.  One obtains
\beq
&&\int_\x\int_\y\,\rme^{-i\p\cdot (\x-\y)}\,\bra{\Psi_n,\beta} U(\x)\ket{\Psi_0,\alpha}\bra{\Psi_n,\p,\beta} U(\y)\ket{\Psi_0,\alpha}^*\nn
&&=\int_\b  \int_\r\, \rme^{-i\p\cdot \r}\,S(\b,\r)= \int_\b {\cal P}(\b,\p ),
\eeq
where  $S(\b,\r)$ is defined in Eq.~(\ref{Stau}), with the average made in the state $\Psi_0$. 
The probability  ${\cal P}(\b,\p )$ is normalized so that  $
\int_\p {\cal P}(\p,\b)=\frac{1}{N_c}{\rm Tr} \langle\left(U(\b)U^\dagger(\b)\right)\rangle=S(\b,\r=0)=1$, the last equality reflecting the property of  ``color transparency" (see below). Observe  that the difference $\r$  in the coordinates of the quark in the amplitude and in the complex conjugate amplitude, i.e.,  the ``size" $\r$ of the dipole, is conjugate to the momentum $\p$.

This connection between momentum broadening and the forward $S$-matrix element of a dipole has been obtained in a strict eikonal approximation (frozen transverse coordinates), an approximation that can be  easily relaxed (see e.g. \cite{Blaizot:2012fh}). It also holds when one includes soft gluon radiation (soft gluon exchanges between the dipole constituents) to next to leading order, but presumably breaks down in higher order (see \cite{Mueller:2012bn} for a thorough discussion of this point).

\subsection{The dipole-nucleon $S$-matrix in the two-gluon exchange approximation}\label{sec:twogluonexch}

In the preceding subsections, we have seen two examples of observables that can be expressed as matrix elements of a color dipole, a product of two Wilson lines in a color singlet configuration. To complete the calculation  we need to specify the state of the target and the fluctuations of its color field (the average color field vanishes). We shall here  relate these fluctuations to the integrated gluon distribution. We start by considering the scattering of a color dipole on a nucleon, assuming the color field of the nucleon to be weak so that we can proceed with a perturbative calculation. In leading order, this amounts to a two gluon exchange approximation. The perturbative calculation makes sense if  the dipole size is small enough, that is $r_\perp\ll r_0$ with $r_0$ the size of the nucleon. 

\begin{figure}[htbp]
\vspace{-0.4cm}
\begin{center}
\includegraphics[scale=0.5]{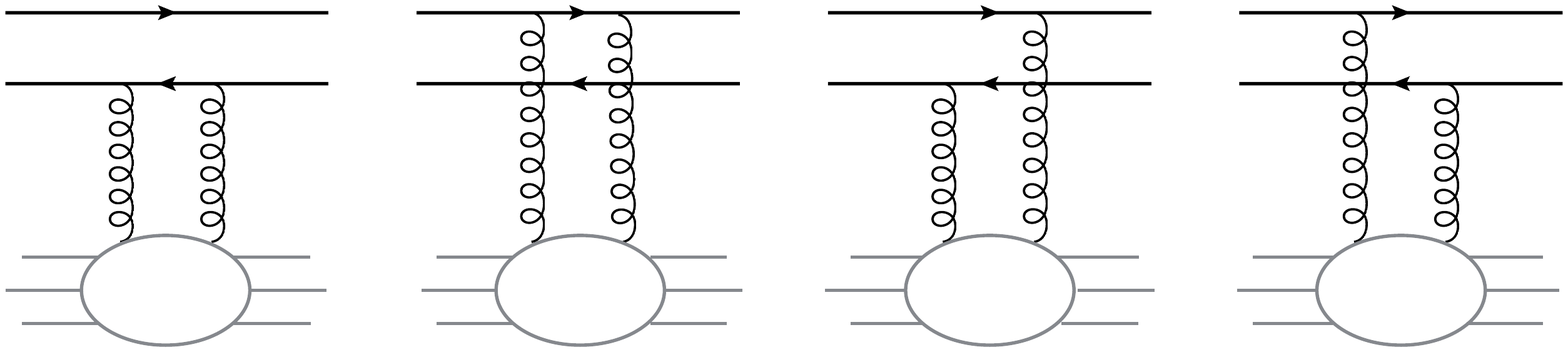}
\end{center}
\vspace{-6cm}
\caption{\label{fig:dipole_nucleon} The four diagrams that contribute to the dipole-nucleon interaction in the two gluon exchange approximation. The dipole is represented by the two oriented line at the top of the diagrams. The gluons are drawn as curly lines, as usual. }
\end{figure}

Let us then evaluate the matrix element (\ref{Stau}) in a nucleon state. Assuming the field to be weak,  we  expand the Wilson lines in powers of $gA$
\beq\label{expansionU}
U_\x
&\simeq& 1+ig \int \rmd x^+ \, A^-_a(x^+,\x)\, t^a\nn
&&-\frac{g^2}{2}\int \rmd x^+_1 \int\rmd x^+_2\, {\rm T}\left[A^-_a(x^+_1,\x)\, t^a\,A^-_b(x^+_2,\x)\, t^b\right]+\cdots,
\eeq
and similarly for $U^\dagger_\y$. The leading order  contribution to $S_{\x\y}$ is that of second order in $g$: the first order involves the average color field of the nucleon, which vanishes. The various contributions that appear  when calculating $U_\x U^\dagger_\y $ are illustrated by Fig.~\ref{fig:dipole_nucleon}. 
Since the dipole probes the field only very near $x^-=0$, the various operators  $A^-_a(x^+,\x)$ that appear in the expansion of $U_\x U^\dagger_\y $ are located at points that are separated by space-like intervals, and therefore commute. This allows us to rewrite $S_{\x\y}$ simply as 
\beq
S(\b,\r)&\approx&1+ \frac{g^2}{2N_c}\sum_a\int_{x_1^+, \,x_2^+ }\frac{1}{2} \bra{N} \left( A^-_a(x_1^+,\x) -A^-_a(x_1^+,\y)\right) \nn
&&\qquad\qquad\qquad\qquad\qquad\times \left( A^-_a(x_2^+,\y)- A^-_a(x_2^+,\x)\right)  \ket{N}  ,
\eeq
where we have used ${\rm Tr} (t^a t^b)=\delta^{ab}/2$. In  covariant gauge, $F^{i-}_a=\del^i A^-_a$, so that $A^-_a(x^+,\y) -A^-_a(x^+,\x)\approx r_i F^{i-}(\b)$, where $r^i=x^i-y^i$ and  we have assumed that  the field of the target varies smoothly over the size of the dipole. We get, after performing an angular average, 
\beq\label{Sonegluonexch}
S(\b,\r)&\approx& 1-\frac{g^2}{8N_c} r_\perp^2 \sum_{i,a}\int\rmd x_1^+\rmd x_2^+ \bra{N} F^{i-}_a(x^+_1,\b) F^{i-}_a(x^+_2,\b) \ket{N}\nn
&\approx&1-\frac{\alpha_s \pi^2}{2N_c} r_\perp^2 \frac{xG_N(x,1/r_\perp^2)}{\pi R^2}, 
\eeq
where we have introduced the integrated gluon distribution of the nucleon, $xG_N(x,1/r_\perp^2)$, and we have ignored the dependence on impact parameter\footnote{That is, we have assumed that the gluon density is constant over the transverse area occupied by the nucleon. This is a crude approximation, but sufficient for our present purpose, and we shall use it systematically in this paper.}. Leaving aside issues of gauge invariance (these will be addressed somewhat in the next section), one can indeed relate the integrated gluon density  to the correlator of the color electric field:
\beq\label{xGxFF}
xG(x,Q^2)=\int^{Q^2} \frac{\rmd k_\perp^2}{4\pi^2} \langle F^{i-}_a(k^-,\k) F^{i-}_a(-k^-,-\k)\rangle.
\eeq
where
\beq\label{FFweak}
&&\langle F^{i-}_a(k^-,\k) F^{i-}_b(-k^-,-\k)\rangle\nn
&&=\int \rmd x^+\rmd y^+\int_{\x,\y} \rme^{-ik^-(x^+-y^+)} \rme^{i\k_\perp\cdot(\x_\perp-\y_\perp)} \langle F^{i-}_a(x^+,\x) F^{i-}_b(y^+,\y)\rangle.
\eeq
In setting the coordinates of the fields to be both equal to $\b$ in Eq.~(\ref{Sonegluonexch}) we have assumed an upper cutoff in the integration over $\x-\y$, that is $|\x-\y|\lesssim  1/Q^2$, while the integration over $\b=(\x+\y)/2$ yields a factor $\pi R^2$ (assuming that $\langle F^{i-}_a(\x) F^{i-}_b(\y)\rangle $ is independent of $\b$). Similarly, in the integral over $x^+$ and $y^+$, we can first perform the integration over  $(x^++y^+)/2$ over the longitudinal size of the proton, assuming that   $\langle F^{i-}_a(x^+) F^{i-}_b(y^+)\rangle $  depends only on $x^+-y^+$, which leaves us with the integration over  $x^+-y^+$. The resulting value of $k^-$  will eventually fix the value of $x=k^-/p^-$ in the gluon distribution, with $p^-$ the minus component of the proton momentum.  This value of $k^-$  depends on the process.  If the dipole that one is considering  is that occurring in DIS,  the natural choice is $x=x_{\rm Bj}$. Arguments guiding the relevant choice of $x$ in the context of momentum broadening are given in \cite{Baier:1996sk}.  

One can then deduce from the expression (\ref{Sonegluonexch}) above, and the relation (\ref{sigmatoteikonal}), the dipole-nucleon cross section in the two-gluon exchange approximation \cite{Blaettel:1993rd,Frankfurt:1993it}
\beq\label{sigmadipole}
\sigma_{\rm dip}(r_\perp)=\frac{\pi^2\alpha}{N_c} r_\perp^2 xG_N(x,Q^2),\qquad Q^2\sim 1/r_\perp^2.
\eeq
  The quadratic dependence on the dipole size is characteristic of gauge theories. It reflects the phenomenon of color transparency \cite{Brodsky:1988xz,Farrar:1988me,Nikolaev:1990ja,Frankfurt:1993it} (see  \cite{Jain:1995dd}  for a review), already alluded to earlier : when the size of the dipole is small, it is seen as a color neutral object by the field of the target, and it passes through it without interacting, i.e., $S(0)=1$. 

Note that what is meant by a ``small' or a ``large" dipole depends on a scale. As can be seen on the expression (\ref{Sonegluonexch}) of the $S$-matrix element, this scale is proportional to  the gluon density per unit transverse area (in fact, we shall identify it shortly to the saturation momentum). In a collision of the dipole with a large nucleus, the gluon density seen by the dipole can become large, thereby invalidating the 2-gluon exchange approximation. However, if the collisions between the dipole and the nucleons can be treated as independent, we can handle this situation of large gluon density by  using a standard multiple scattering calculation. This is what we do in the next subsection.

\subsection{Multiple scattering, momentum broadening, saturation} \label{sec:momentumboradening1}

To proceed simply, we note that $|S( \b,\r)|^2$ can be interpreted as the survival probability, that is, the probability that the dipole crosses the target without suffering interaction. Knowing this probability for the case where the target is a nucleon, for which a perturbative calculation has been done in the previous subsection, one can easily obtain the corresponding result for a target nucleus via a  standard multiple scattering calculation \cite{Mueller:Cargese}. (A calculation that handles directly the average of Wilson lines in the strong field regime, as illustrated by the right panel of Fig.~\ref{fig:eikonal},  will be presented in the next section.) This calculation assumes that individual dipole-nucleon collisions can be treated as independent, i.e. are not overlapping.  One gets
 \beq\label{S2multiple1}
S^2(\b,\r)=\rme^{-L/\lambda},\qquad \frac{1}{\lambda(r_\perp)}=\rho\,\sigma_{\rm dip}(r_\perp),
\eeq
where $L$ is the length travelled by the dipole through the nucleus,  $\rho$ the density of nucleons per unit volume,  $\lambda$  the mean free path, and we have taken into account that $S$ is real in the two-gluon exchange approximation, i.e., $|S|^2=S^2$.  

One can then write the $S$-matrix for the dipole-nucleus interaction as follows
\beq\label{Sexpon}
S(\b,r_\perp)= \rme^{-L/2\lambda}=\rme^{-\frac{L\rho}{2}\sigma_{\rm dip}(r_\perp)}=\rme^{- Q_s^2 r_\perp^2/4},
\eeq
where 
\beq\label{barQs}
 Q_s^2= \frac{2\pi^2 \alpha}{N_c} L\rho\, xG_N(x,1/r_\perp^2)=\frac{2\pi^2 \alpha}{N_c} \frac{AxG_N(x,1/r_\perp^2)}{\pi R^2}
\eeq
is the saturation momentum for a quark dipole \cite{Mueller:Cargese} (the corresponding saturation momentum for a gluon dipole would be obtained by multiplying Eq.~(\ref{barQs}) by a factor $C_A/C_F$). The independence of the successive nucleon-nucleon collisions is reflected here in the additive character of the gluon density, which grows with the nucleon number.  
   The saturation momentum (\ref{barQs}), which  has the same parametric dependence as in Eq.~(\ref{Qsaturation}), also grows with $A$, as $A^{1/3}$.   It follows that at fixed dipole size,  the dipole-nucleus interaction can be large if the nucleus is large. Note however that in contrast to what would happen for the perturbative estimate (\ref{Sonegluonexch}) where the $S$-matrix could become negative for a too large gluon density, here $S\to 0$ as $Q_s^2 r_\perp^2$ becomes large. Multiple scattering provides a mechanism by which saturation of the interaction between the dipole and the nucleus is achieved, a feature of saturation often referred to as ``unitarization" of the dipole cross section. 
   
   Multiple scattering is also  essential for momentum broadening, which leads to another interpretation of the saturation momentum. 
To see that, let us take the Fourier transform of Eq.~(\ref{Sexpon}). One gets 
\beq\label{momentumbroadening1}
{\cal P}(p_\perp)=\int_\r ~\rme^{-i\p\cdot\r} \,\rme^{-\frac{L\rho}{2}\sigma_{\rm dip}(\r)}\approx \frac{4\pi}{Q_s^2}\rme^{-p_\perp^2/Q_s^2}  ,\eeq
 where the final result has been obtained by assuming a constant $Q_s$ (i.e., ignoring the weak dependence of the gluon distribution  on the scale $r_\perp$ -- see also end of Sect.~\ref{momentumbroadMV}). 
Equation~(\ref{momentumbroadening1}) has a natural interpretation in terms of diffusion in momentum space, with   $Q_s$ playing the role of a diffusion constant: $Q_s^2$ is the  average momentum squared acquired by the quark after traversing a length $L$ of the medium. This is  traditionally denoted  by $\hat q L$ \cite{Baier:1996sk},  so that  $Q_s^2=\hat q L$ (recall that $Q_s^2$ is indeed proportional to $L$, see Eq.~(\ref{barQs})). 

\subsection{Phenomenological dipole model}\label{sec:dipolemodel}

\begin{figure}[htbp]
\begin{center}
\includegraphics[scale=0.4]{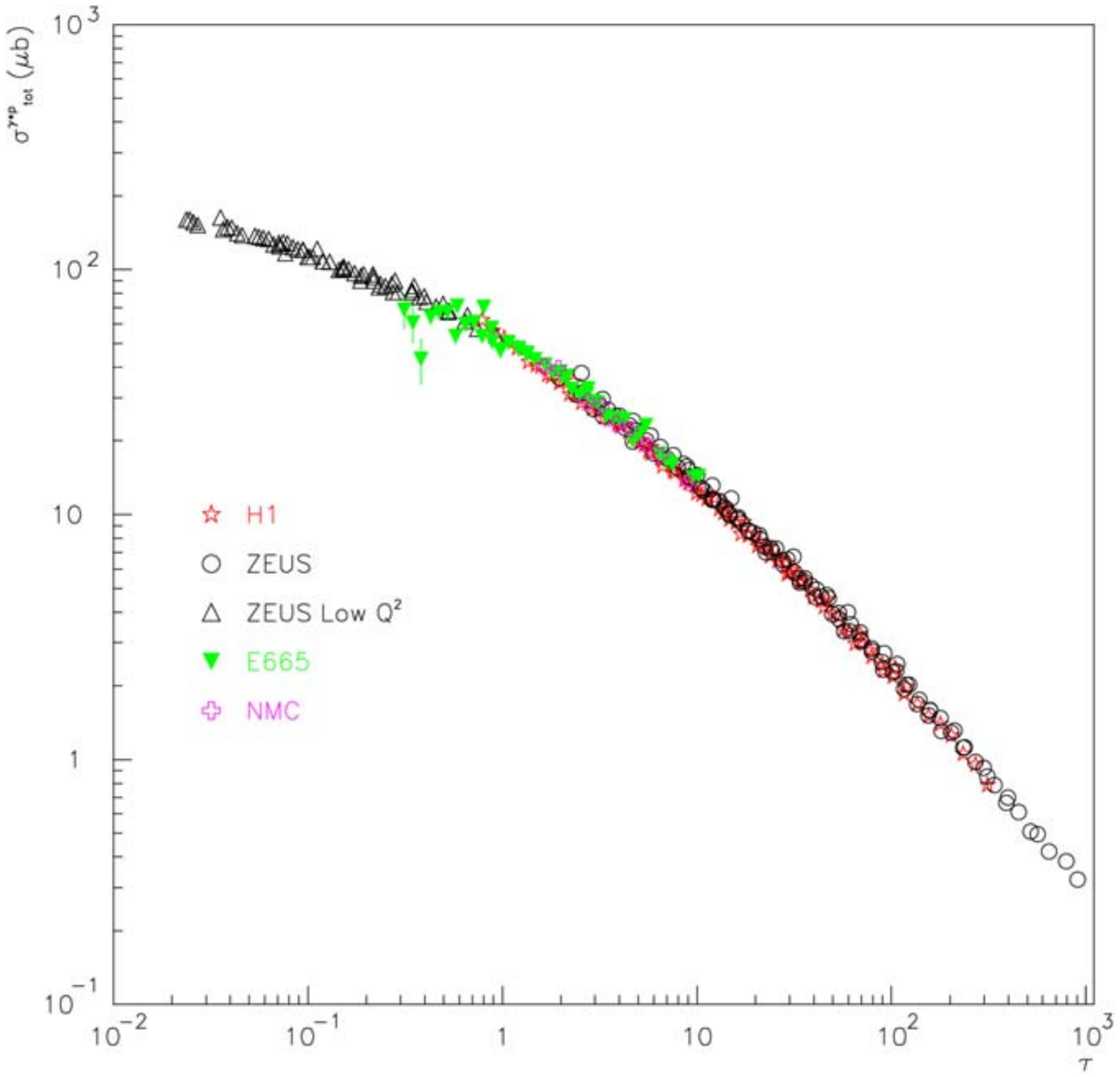}
\end{center}
\caption{\label{fig:geomscaling} Illustration of geometrical scaling:  the total cross section $\sigma_{\rm tot}^{\gamma^*p\to X}$, which depends a priori on the independent variables $x$ and $Q^2$ is seen to be a function only of $\tau\equiv \ln (Q^2/Q_s^2(x))$.  Figure taken from Ref.~\cite{Marquet:2006jb}.}
\end{figure}

The previous discussion has revealed that the strength of the interaction between a dipole  and a target is determined by the product of the dipole size and the saturation momentum. 
The saturation momentum characterizes the change of regime between the dilute regime ($r_\perp\ll 1/Q_s$) and the saturated one ($r_\perp\gg 1/Q_s$), and is proportional to the gluon density of the target.  Assuming that all the energy dependence  can be absorbed in the saturation momentum, via its dependence on the gluon density,   Golec-Biernat and W\"usthoff
(GBW)  \cite{GolecBiernat:1998js} have proposed the following parameterization for the DIS dipole cross section
\begin{equation}\label{GBW0}
\sigma_{\rm GBW}(x,r_\perp)=\sigma_0 \left[
1-e^{-\frac{1}{4}Q_s^2(x)r_\perp^2}
\right],\qquad Q_s^2(x)\equiv Q_0^2 \left({x_0}/{x}\right)^\lambda.
\end{equation}
Quite remarkably,  this led to a  good description of early HERA data at $x<10^{-2}$
and moderate $Q^2$, including diffraction data \cite{Stasto:2000er}. Typical values of parameters are $\sigma_0\simeq 23$ mb, $\lambda\simeq 0.288$, and $x_0\simeq 3\times 10^{-4}$, while $Q_0=1$ Gev fixes the dimension. The fact that the $x$ dependence of the cross section enters only through the saturation momentum  has been dubbed `geometrical scaling' \cite{Stasto:2000er}. This is illustrated in Fig.~\ref{fig:geomscaling} where the total cross section $\sigma_{\rm tot}^{\gamma^*p\to X}$ is plotted as a function of the scaling variable $\tau\equiv \ln (Q^2/Q_s^2(x))$. (For more recent fits of HERA data that take into account latest advances in non linear evolution equations, see \cite{Iancu:2003ge,Albacete:2010sy,Iancu:2015joa}. For a thorough and critical discussion of the dipole model see Ref.~\cite{Ewerz:2004vf}. See also \cite{Balitsky:2010ze} and \cite{Beuf:2011xd,Beuf:2016wdz} for calculations of NLO corrections. )

This phenomenon of geometrical scaling provides perhaps the best experimental evidence  for the existence of a saturation momentum, and therefore for saturation itself.  We shall see later that it emerges as a natural consequence of the non linear evolution equations that take saturation into account (see Sect.~\ref{sec:evolution}). A similar phenomenon has been looked for in d-Au collisions at RHIC and in p-Pb collisions at the LHC \cite{McLerran:2010ex}.

\subsection{One more remark}
 
Before leaving this section, and in anticipation of the next, I would like to make a very simple observation. Much of the previous discussion would apply as well to the scattering of an electric dipole   on a random electric field. Let us indeed consider an electric dipole  $d$ propagating with velocity $v$ through a region of spatial extent $L=vT$ in which exists a constant electric field $E$.  For simplicity,  I assume the dipole and the field to be parallel, and  orthogonal to the direction of propagation of the dipole.  At each passage of the dipole, the  field  changes randomly (equivalently, one may also consider that the dipole hits successively various regions where the electric field changes randomly from region to region  \cite{Hufner:1990uv}).   The $S$ matrix for one passage through the field is given by 
\beq
S= \rme^{i\int \rmd z \frac{E d}{v}} = \rme^{i\frac{L}{v}E d}.
\eeq
This is just a phase and the probability  that the dipole emerges from the field unaffected is unity, i.e., $|S|^2=1$.  However, when averaging over many passages through the field, phases interfere, leading to a very different result. Let us assume, for definiteness, that  the distribution of $E$ is a Gaussian, i.e.,  $P(E)\sim \rme^{-\frac{1}{2} \frac{E^2}{a^2}}$. Then the average $S$ matrix can be written as 
\beq\label{elecdipole}
\langle S\rangle={\rm e}^{-\frac{1}{4} {Q_s^2 d^2}},\qquad   Q_s^2=2\langle E^2 \rangle \frac{L^2}{v^2},
\eeq
 where the scale $Q_s$ characterizing the interaction with the random field is, not unexpectedly, related to the variance of the distribution ($\langle E^2\rangle=a^2$). After averaging over the random field, the $S$-matrix becomes real. The survival probability $|\langle S\rangle|^2$ has been written in the same way as in Eqs.~(\ref{S2multiple1}) and (\ref{Sexpon}): it shows that large size dipoles ($Q_s d\gtrsim 1$) have  a reduced probability to emerge intact after interacting with the random field.

\section{Propagation in random fields}\label{WWfields}

So far, we have seen that the eikonal propagation of high energy partons is described by Wilson lines in the field of the target.  In order to obtain, for instance, the  $S$-matrix for the dipole-nucleus interaction, an average over the field fluctuations need to be performed. In the previous section, this was done by relating the field fluctuations to the gluon density. Here, we shall proceed differently. We observe that during the short duration of the interaction, the color field of the target behaves as a ``classical''  background field, that is, a field whose fluctuations  can be ignored. Of course,  this field changes from event by event, and the values of observables   result therefore from an average over events, with some probability distribution  $W[A] $.   Schematically
\beq\label{observableaverage}
\left< {\cal O} \right> =\int {\cal D} A \;W[A] \;{\cal O}[A],
\eeq
where ${\cal O}[A]$ is, for instance, a color dipole $S$-matrix in the background field $A$, and $W[A]$ is  the probability to find the particular background $A$ in a given event. 
One may imagine $W[A]$ as given by the square  of the  light cone wave function of the nucleus for some prescribed value of the  gauge field $A$,  or as the exponential  $\sim \rme^{-S[A]}$ of some effective action $S[A]$. 

The quantity $W[A]$ is in general difficult to calculate. Its variation with energy can be controlled, to some extent, as we shall see in the next section. In this section we shall use for it the simple Gaussian ansatz of the   McLerran Venugopalan (MV) model \cite{McLerran:1993ni,McLerran:1993ka}. The main virtue of this  model is to allow the  explicit calculation of the  average over the target field, and this independently of the strength of the field. (Another instance where averages of Wilson loops can be calculated non-perturbatively is at strong coupling, within the AdS/CFT framework \cite{Liu:2006he}. The techniques to be discussed in this section bears strong similarities with those exposed in Ref.~\cite{Nachtmann:1996kt} in the context of the QCD stochastic vacuum picture.)  The  MV model offers also a simple picture of saturation, as resulting from self-interactions of the gauge fields. 

In the next subsection, we recall the basic idea of the MV model,  which is to assimilate  the color field of a nucleus to the Weizs\"{a}ker-Williams (WW) field created by color charges carried by the valence quarks, assumed to be uncorrelated. 
Then we illustrate the techniques of calculating observables according to Eq.~(\ref{observableaverage}) with a Gaussian $W[A]$, and we recover simply results  obtained in the previous section in a somewhat broader context. Finally we analyze the gluonic content of the WW field, and comment on the picture of saturation that  the model suggests. 

\subsection{The McLerran Venugopalan model}

 Let us start by recalling  that the electric and magnetic fields created by an electric charge $e$ moving at the speed of light in the negative $z$ direction, and at the origin of the transverse coordinate system, are given by
\beq
E^i(x^+,\x)= \frac{e}{2\pi} \frac{x_\perp^i}{\x^2}\delta(t+z),\qquad B^i=\frac{e}{2\pi} \frac{\epsilon^{ij}x_\perp^j}{\x^2}\delta(t+z).
\eeq
In a gauge where $A^+=0$,   these can be expressed in terms of the following gauge potentials
\beq\label{gaugeAi0}
A^i=0, \quad\del^+ A^-=0,\quad A^-(x^+,\x)\equiv\alpha(x^+,\x),
\eeq
or, equivalently, 
\beq\label{gaugeA-0}
A^-=0,\quad A^i(x^+,\x)=-\int_{-\infty}^{x^+} \rmd y^+\del^i\alpha(y^+,\x),
\eeq
where $\alpha(x^+,\x)$ obeys Poisson's equation
\beq
-\del_\perp^2 \alpha(x^+,\x)=e\delta(x^+)\delta^{(2)}(\x).
\eeq
The two sets of gauge potentials are related by a gauge transformation that is independent of $x^-$, viz., $A^\mu\to A^\mu+(1/e) \del^\mu\theta(x^+, \x)$. The choice $\del^-\theta=-e\alpha$ preserves $A^+=0$, and enforces $A^-=0$, leaving $A^i=(1/e) \del^i\theta$ as the only non trivial component.
The fields $\E$ and $\B$ are transverse, and orthogonal to each others, and are analogous to the fields of  an electromagnetic wave. This is the origin of the method of equivalent photons of Weizs\"{a}ker and Williams.\\

The MV model generalizes this notion to the non abelian case. 
More precisely,  the field $A_\mu$ is obtained as the solution of  the classical Yang-Mills equations:
\beq\label{YMequations}
\left[  D_\mu, F^{\mu\nu} \right] =J^\nu,\qquad [D_\mu,J^\mu]=0,\qquad D_\mu=\del_\mu-igA_\mu,
\eeq
in the presence of a given color source $J^\nu$. This source is carried by the valence quarks of the nucleus moving near the light cone $x^+\approx 0$. It is the analog of the electric current discussed above, with however complications due to the non abelian gauge symmetry: Because of the presence of the gauge field in the covariant derivative, the equation  $[D_\mu,J^\mu]=0$  implies that $J^\mu$ depends on the choice of a gauge. Assume, as before, that the nucleus is moving near the light cone $x^+=0$ (i.e., in the negative $z$ direction). Then,  we define, in the covariant gauge $\del_\mu A^\mu=0$, $J^\mu(x^+,\x)=\delta^{\mu-}\rho(x^+,\x)$, where $\rho(x^+,\x)$ is a color charge density (in the covariant gauge). Because of Lorentz contraction, this density  is sharply peaked along the light cone axis $x^+=0$ (i.e., $\rho(x^+,\x)\sim \delta(x^+)$), and is independent of $x^-$. Viewed as the density of valence quarks, it can be written as  $\rho^a(x^+,\x))=g\sum_{i=1}^{N^{\rm val}} \delta(x^+_i)\delta(\x-\x_i) t^a(i)$, with $t^a$ a matrix in the fundamental representation of the gauge group, and $N^{\rm val}=A N_c$ is the number of valence quarks in a nucleus containing $A$ nucleons.

 The question of determining the probability distribution of the field configurations, $W[A] $, is reduced to that of finding the probability distribution of the color charges $\rho$.     The crucial ansatz of the  MV model is that the color charges carried  by the valence quarks are uncorrelated, their distribution being a Gaussian entirely characterized by the density-density correlation function
 \beq\label{rhorhoMV}
\langle   \rho^a(x^+,\x)\rho^b(y^+,\y)\rangle =\delta_{ab}\delta(x^+-y^+)\, \delta^{(2)}(\x-\y) \mu(x^+,\b),
\eeq
where $\mu(x^+,\b)$ is the density of valence quarks (and $\b=(\x+\y)/2$). 
 Arguments that motivate such an ansatz can be formulated in the case of a heavy nucleus \cite{McLerran:1993ni,Kovchegov:1996ty,Kovchegov:1997pc}: they rest on the idea that if the nucleus is large enough, it is likely that two valence quarks seen at a given point in the transverse plane belong to distinct nucleons, hence are uncorrelated. In other words, we ignore the correlations arising from the grouping of valence quarks into color singlet nucleons \cite{Lam:1999wu}.
By integrating both sides of Eq.~(\ref{rhorhoMV}) over $x$ and $y$, taking into account that the average projects onto color singlet states, one  finds
 \beq\label{mudef}
 \mu=\frac{g^2}{2N_c} \frac{N^{\rm val}}{\pi R^2} =\frac{g^2}{2} \frac{A}{\pi R^2}, 
 \eeq
 where the factor $1/(2N_c)=(1/N_c) {\rm Tr} t^a t^b $ originates from the color singlet projection.
 We have set $\mu=\int \rmd x^+ \mu(x^+,\b)$, ignoring the impact parameter dependence (we assume a constant density in the transverse plane, as we did earlier  in getting Eq.~(\ref{Sonegluonexch})). 
 
The color WW field that accompanies the valence quark is localized in the transverse plane, as is the QED case. We shall be using systematically a  gauge where $A^+=0$. This leaves a freedom analogous to that observed earlier for QED (see eqs.~(\ref{gaugeAi0}) and (\ref{gaugeA-0})). The choice $A^i=0$ and $\del^+ A^-=0$, which is easily seen to be equivalent to the covariant gauge $\del_\mu A^\mu=0$, plays a special role. In this gauge the field strength tensor has a simple Abelian structure, $F^{i-}_a=\del^i \alpha_a$, where $\alpha_a$ ($\alpha_a=A^-_a$ in this gauge), is given by the Poisson equation
\beq\label{Poisson}
-\nabla_\perp^2 \alpha_a(x^+,\x)=\rho^a(x^+,\x). 
\eeq
It is easily verified that the current $J^-$ is conserved, $D^+ J^-=(\del^+-igA^+)J^-=0$, since $A^+=0$ and $\rho(x^+,\x)$ does not depend on $x^-$. 
The linear relation (\ref{Poisson}) between field and charge in the covariant gauge entails therefore
\beq\label{2pointAA}
\langle  \alpha_a(x^+,\x)\alpha_b(y^+,\y)\rangle=\delta^{ab}\delta (x^+ - y^+)\gamma(x^+, \x-\y),
\eeq
where
\beq\label{correlatorgamma}
\gamma(x^+, \x-\y)=  \mu(x^+)\int_\k \frac{  \rme^{i\k\cdot (\x-\y)} }{k_\perp^4}.
\eeq
This infrared divergent integral should be cutoff at a scale characterizing confinement effects, that is $k_\perp\lesssim \Lambda_{\rm QCD}$ (reflecting the fact that  color fields should be  screened beyond a distance $r_0\gtrsim 1/\Lambda_{QCD}$).

\subsection{The dipole $S$-matrix, momentum broadening and saturation momentum}\label{momentumbroadMV}

The Gaussian ansatz, together with Eq.~(\ref{rhorhoMV}) for the correlation function of the random field,  allows the easy calculation of expectation values of Wilson lines according to Eq.~(\ref{observableaverage}). In this subsection we shall revisit some of the results that were obtained in Sect.~\ref{sec:eikonal}, and complete these with more explicit expressions. The calculations to be presented, although to a large extent equivalent to those of Sects.~\ref{sec:twogluonexch} and \ref{sec:momentumboradening1},  differ  in that they look more systematic. For instance, the random field calculation automatically generates the correct multiple scattering series, as required when the field is strong: there is no need to consider scattering on individual nucleons as we did in Sect.~\ref{sec:eikonal}, and then perform the multiple scattering over individual nucleons.   However, the present calculation relies heavily on the ansatz (\ref{2pointAA}), which assumes that the local field fluctuations are uncorrelated, a strong assumption indeed. It certainly covers the assumption that successive scattering on nucleons are independent, but is also blind to any type of correlations (such as for instance nucleon-nucleon correlations).
\begin{figure}[htbp]
\begin{center}
\includegraphics[scale=0.40]{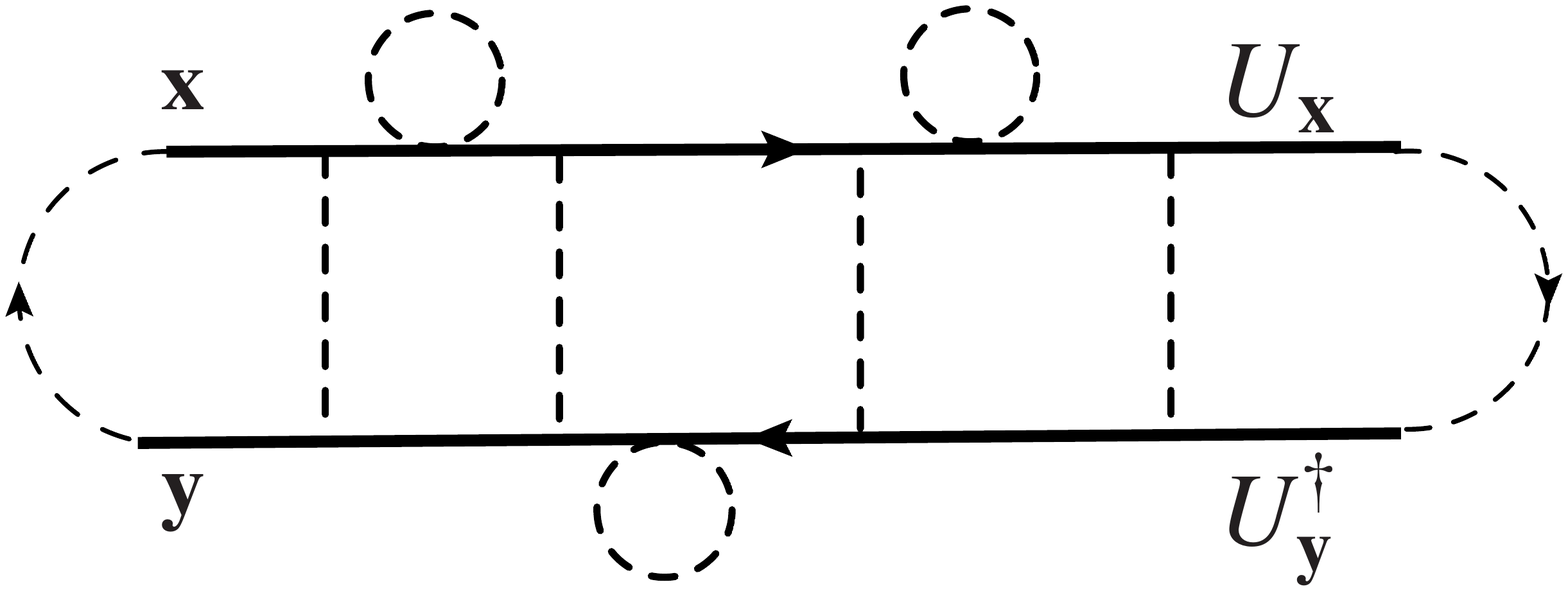}
\vspace{-4cm}
\caption{\label{fig:Wick_theorem} Illustration of the Gaussian averaging over the random field. Light-cone time $x^+$ runs from left to right. The horizontal line oriented to the right (left) represents the Wilson line $U$ ($U^\dagger$).  The vertical dashed line (at a given $x^+$) corresponds to the correlator $\gamma(x^+,\x-\y)$. The dashed circle represents the correlator $\gamma(x^+,0)$. The final result, given by Wick's theorem as the exponential of the leading order result, $g^2C_R \left[ \gamma(x^+,0)-\gamma(x^+,\x-\y)\right]$, finds its origin in  the fact that the interactions (field correlations) are instantaneous (in the language of multiple scattering this is equivalent to having independent successive collisions) and that the color singlet structure of the dipole is unaffected by each interaction. The dashed semi-circles at the left and right ends  indicate the color flow.  More details on this graphical notation can be found in Ref.~\cite{Blaizot:2012fh}.    }
\end{center}
\end{figure}

\begin{figure}[htbp]
\begin{center}
\includegraphics[scale=0.5]{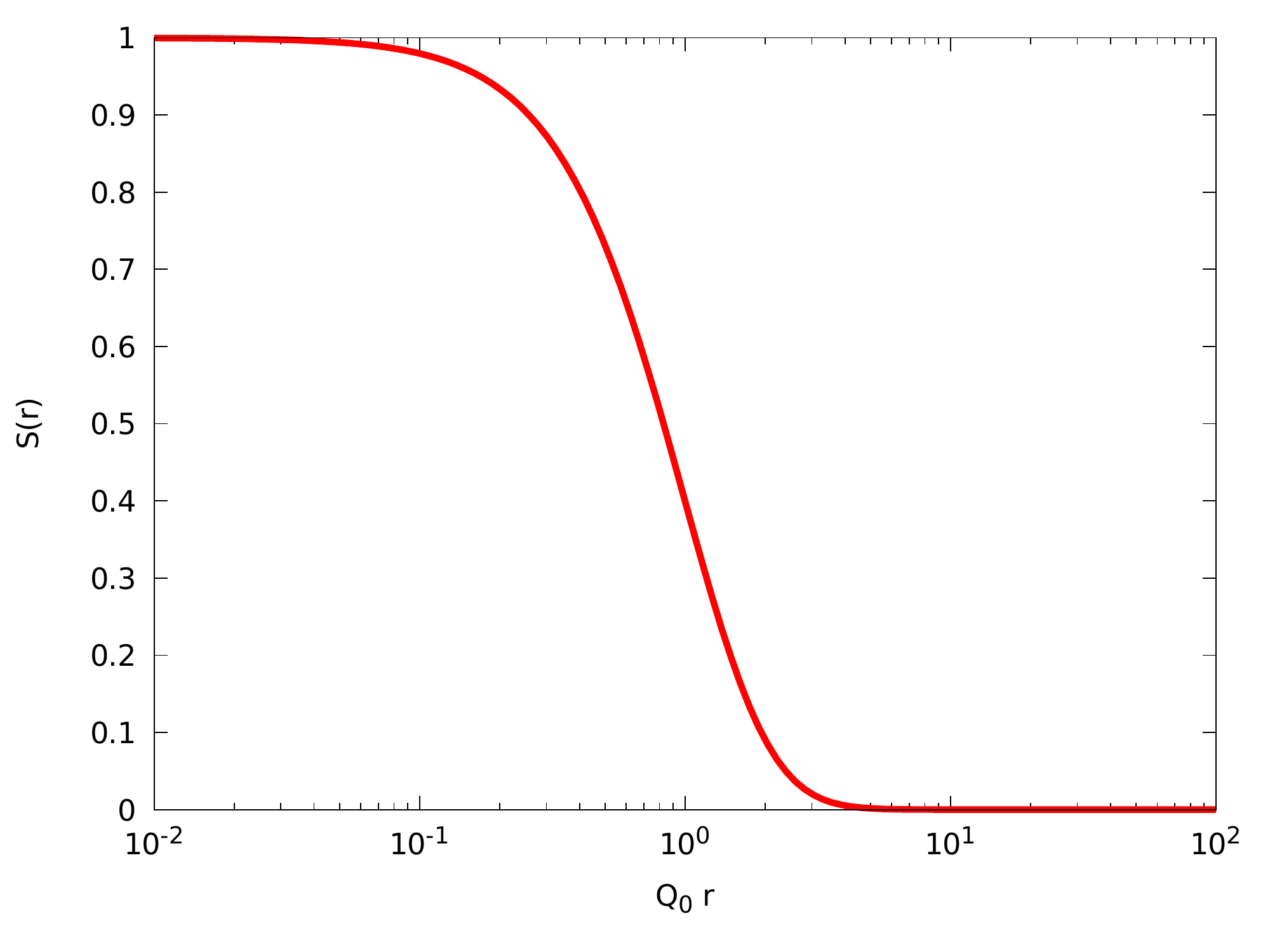}
\end{center}
\caption{\label{fig:Sr} The behavior of the dipole S-matrix as a function of the dipole size $r$ (measured in units of $Q_0^{-1}$, with $Q_0$ defined in Eq.~(\ref{Sr1})). The small size dipoles are little affected by the target field ($S\lesssim 1$), a reflexion of  ``color transparency". Large dipoles however have large cross sections,  they undergo multiple scattering, and the $S$-matrix eventually reaches the black disk limit ($S\approx 0$) for large sizes.  }
\end{figure}

Consider then  the forward $S$-matrix element for a color dipole in representation R (with $d_R$ the dimension of the representation)
\beq\label{SR2}
S_R(\b,\r)=\frac{1}{d_R} \langle{\rm Tr}\left(U_\x U^\dagger_\y\right)\rangle
\eeq
with as before,  $\r=\x-\y$, $\b=(\x+\y)/2$, and (see Eq.~(\ref{Ueikonal}))
\begin{equation}\label{Ueikonal2}
U_\x= {\rm T}\,{\rm exp}\left({\rm i}g\int_{-\infty}^{\infty} dz^+
A^-_a(z^+,{\x}) t^a\right). 
\end{equation}
Note that $S$, as written, depends on the choice of a gauge. It can be made gauge invariant, if desired, by closing the two Wilson lines into a (gauge invariant) loop, that is,  by adding two gauge links connecting $\x$ and $\y$ at $z^+=\pm\infty$. In the covariant gauge, where $A^i=0$,  these contribute unit factors. This is the gauge in which we shall pursue the calculation. In this gauge, $A^-=\alpha$, and the Gaussian average over $\alpha$ is easily done using Wick's theorem and Eq.~(\ref{rhorhoMV}). This  automatically leads to the exponentiation of the leading order result (see for instance \cite{Gelis:2001da,Blaizot:2004wu,Fukushima:2007dy} for examples of such calculations). An illustration of the procedure, together with the relevant diagrams, is provided in Fig.~\ref{fig:Wick_theorem}.
One finds \cite{JalilianMarian:1996xn} 
\beq\label{dipoleR}
S_R(\r)=\exp\left\{ -\mu C_R g^2 I(\r)   \right\},\quad I(\r)\equiv \int_\k \frac{1-\rme^{i\k\cdot\r}}{k_\perp^4} \approx  \frac{r_\perp^2}{16\pi} \ln \frac{r_0^2}{r_\perp^2},
\eeq
with $\mu $ given in Eq.~(\ref{mudef}). 
The integral $I(\r)$ is logarithmically divergent. As observed already for the integral (\ref{correlatorgamma}), it should be cut off at the lower end at  $k_\perp\gtrsim \Lambda_{QCD}\sim 1/r_0$. At the upper end, the natural cutoff is provided by the dipole size, $k_\perp\lesssim 1/r_\perp$. Hence our final estimate in Eq.~(\ref{dipoleR}). 
We may then write 
\beq\label{Sr1}
S_R(r_\perp)\approx \exp\left\{ - \frac{1}{4} r_\perp^2 Q_s^2\right\},\qquad Q_s^2=\alpha_s C_R \mu \ln \frac{r_0^2}{r_\perp^2}\equiv Q_0^2 \ln \frac{r_0^2}{r_\perp^2},
\eeq 
a form reminiscent of the phenomenological dipole cross section, Eq.~(\ref{GBW0}), with the saturation momentum estimated in the MV model. Note that $Q_s$ is not a constant, but retains a weak (logarithmic) dependence on a scale, that of the largest momenta ($\sim 1/r_\perp$) that contributes to it. 
An illustration of the $r_\perp$ dependence of  $S(r_\perp)$ is given in Fig.~\ref{fig:Sr}. 

We can express $Q_s$ in terms of the gluon density of the target, thereby recovering Eq.~(\ref{barQs}).  Recalling the expression (\ref{Gvalencequark}) for the gluon density of a single valence quark, we see that 
\beq\label{Qs2gluon}
Q_s^2=\frac{4\pi^2\alpha C_R}{N_c^2-1}\, L\rho \,xG_N(x,1/r_\perp^2),
\eeq
where we have set $xG_N(x,Q^2)=N_c xG_q(x,Q^2)$ and, in Eq.~(\ref{Qs2gluon}), we may use $Q_s^2$ in place of  $1/r_\perp^2$  in the argument of the gluon density. (The conventional notation  $xG(x,Q^2)$ suggests an energy dependence which is however not present in the MV model.) This result indicates that the integrated gluon density is additive in the MV model. 
It is also easily verified that  Eq.~(\ref{Sr1}) is compatible with the multiple scattering calculation that leads to Eq.~(\ref{Sexpon}). Finally, after Fourier transform, and assuming $Q_s$ to be constant, one  recovers Eq.~(\ref{momentumbroadening1}). Ignoring the mild $r_\perp$ dependence of $Q_s$ is a good approximation, except at large momentum where the true decay is a power law rather than a Gaussian. The correct estimate (obtained by properly keeping the $r_\perp$ dependence of $Q_s$) yields 
\beq\label{singlescat}
S_R(k_\perp\gg Q_s)\approx\frac{C_R}{2}\frac{g^4}{k_\perp^4}L\rho\approx \frac{4\pi}{Q_0^2}\left(\frac{Q_0^2}{k_\perp^2}\right)^2,
\eeq
as expected from perturbation theory.

\subsection{Gluon distributions in the MV model}

The picture that we have developed so far for the field of the nucleus, as seen by an elementary color dipole, is that of  a random field with uncorrelated fluctuations. We would like now to analyze the partonic content of this field. To do so, we need to change to the light cone gauge of the target, i.e. to  the gauge $A^-=0$ analogous to the second gauge choice in QED, Eq.~(\ref{gaugeA-0}).  It is indeed in this gauge that counting the number of gluons in the wavefunction of the target naturally leads to the expression  (\ref{xGxFF})  for the integrated gluon distribution in terms of the correlator of the color electric field.

In this gauge ($A^+=0=A^-$), the non trivial components of the gauge field are the transverse components, 
given in terms of the field $\alpha$, solution of Eq.~(\ref{Poisson}), by 
 \beq\label{AiLCg}
A^i(x)=-\int_{-\infty}^{x^+}dy^+\,U^\dag(y^+,\x)\;\del^i\alpha(y^+,\x), \label{Afieldvec}
\eeq
where
\beq\label{WilsonU1} 
U(x^+,\x)\equiv {\mathcal
P}\exp{\left[ig\int_{-\infty}^{x^+}dz^+\alpha(z^+,\x)\right]}.
\eeq 
We have chosen $A^i(x^+\to -\infty)=0$, but could have made the equivalent choice $A^i(x^+\to +\infty)=0$. In this gauge, $F^{i-}=-\del^- A^i = U^\dagger(x) \del^i\alpha(x)$. It is the presence of the Wilson line in Eq.~(\ref{AiLCg}) that makes the non abelian result distinct from the QED one, Eq.~(\ref{gaugeA-0}).

 There are a priori issues with Eq.~(\ref{xGxFF}): 
  i) It is not gauge invariant since it involves two field strength tensors at distinct points; ii) the average involves the  field $\alpha(z^+, \x)$ which is defined in covariant gauge. 
 It is however possible to render the expression of the gluon distribution function gauge invariant by inserting gauge links along a suitable path. A path exists for which these gauge links actually do not contribute in the gauge $A^-=0$ \cite{Iancu:2002xk}, so that finally all we need to evaluate is the correlator (\ref{FFweak}), with the field strength given in terms of $\alpha$ by the formulae above. (For more on this issue of gauge invariance in the definition of gluon distribution functions,  see \cite{Belitsky:2002sm,Dominguez:2011wm}.)

\begin{figure}[htbp]
\begin{center}
\includegraphics[scale=0.35 ]{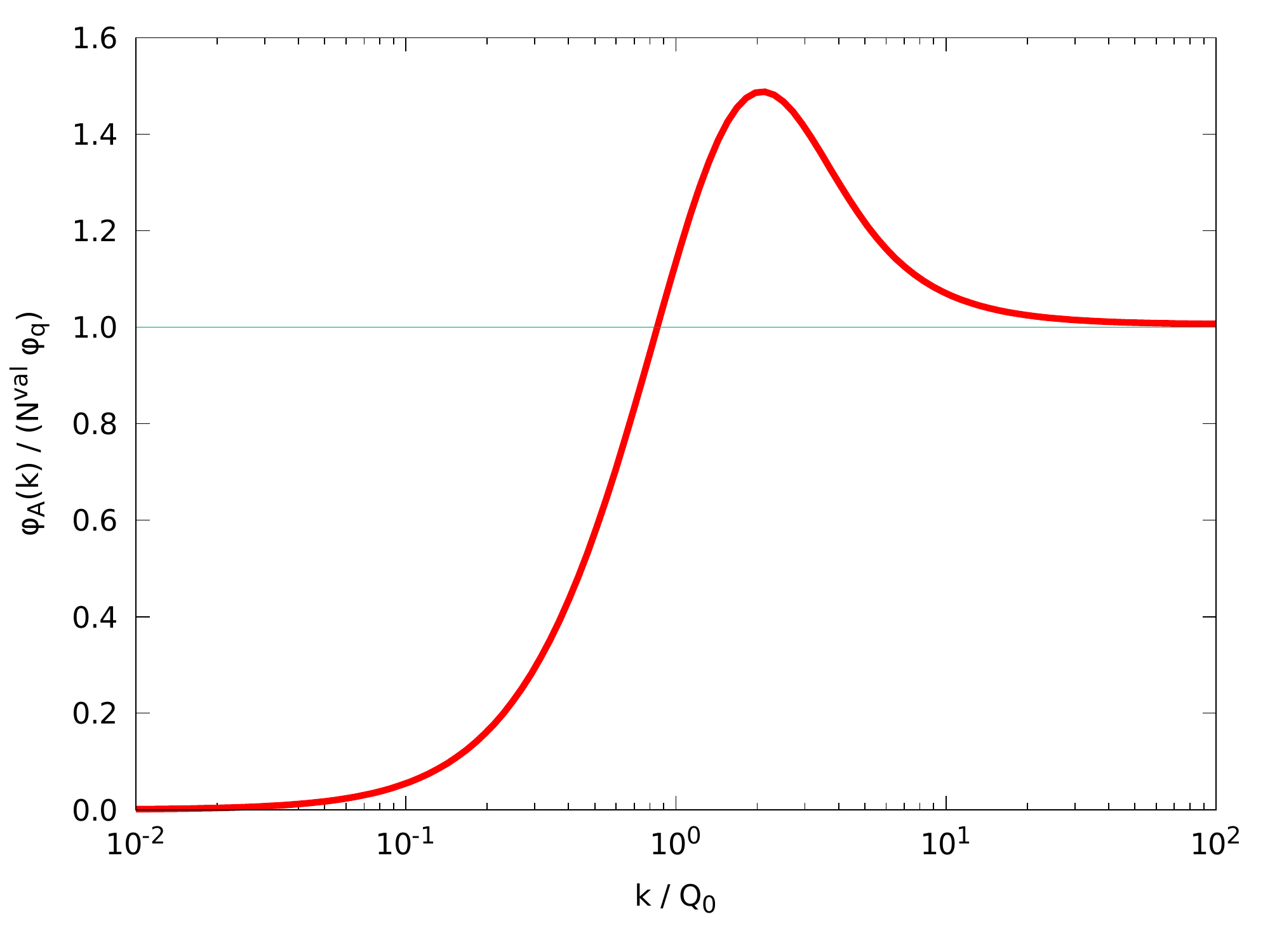}\includegraphics[scale=0.35]{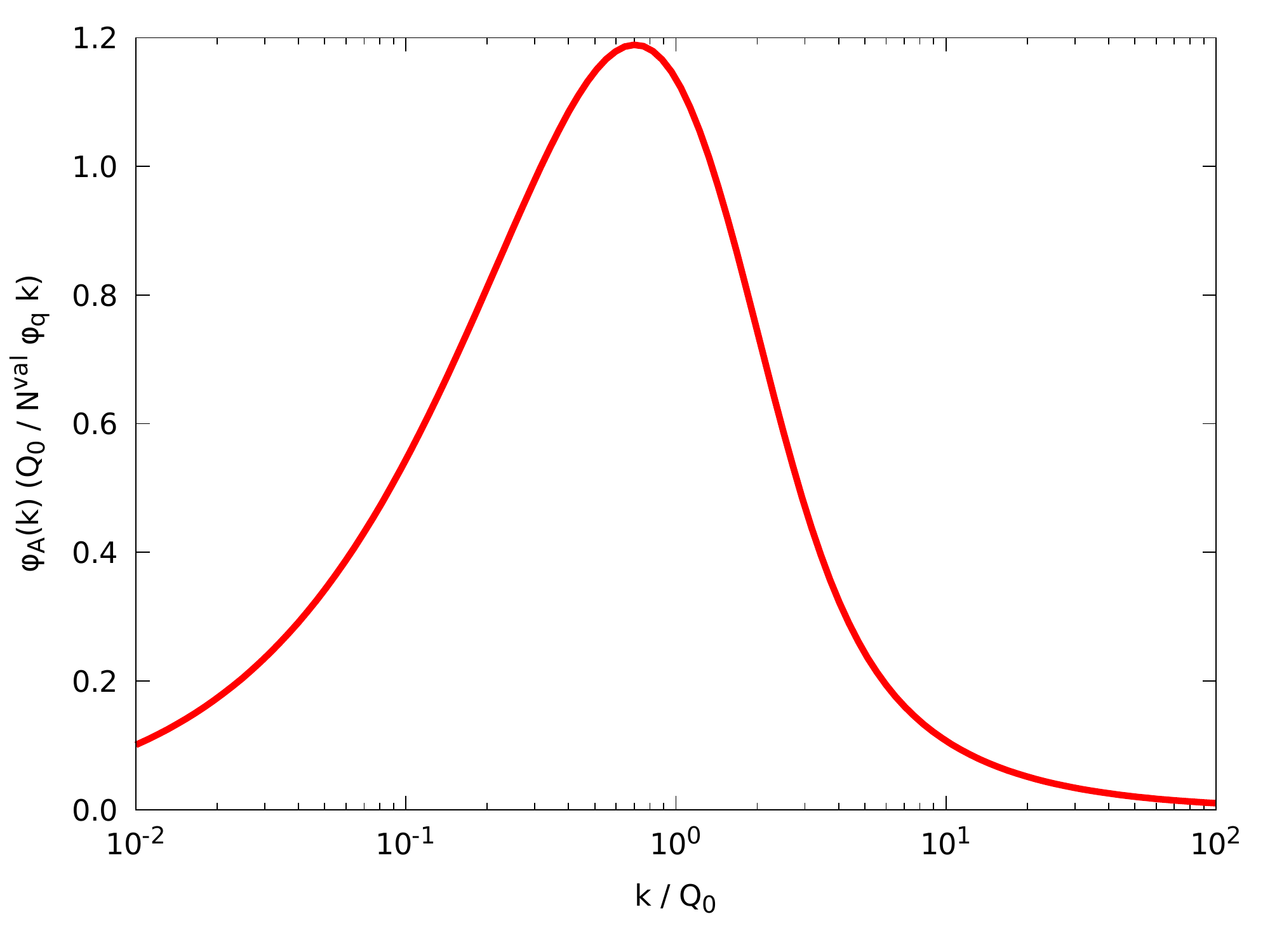}
\end{center}
\caption{\label{fig:phi} Left: The unintegrated gluon distribution $\varphi_A(\k)$ in the MV model, normalized to $N^{\rm val} \varphi_q$, with $\varphi_q$ the unintegrated distribution of a single valence quark. Multiple scattering effects cause a redistribution of the momenta, keeping the total number fixed. At large momenta, $\varphi_A(\k)$ is just the sum of the uncorrelated valence quark contributions.  The low momentum behavior can be understood as a correlation between effective color charges, on a scale of order $Q_s\gg \Lambda_{QCD}$. Right: The quantity  $\varphi_A(\k)/k$ representing the number of gluon in the interval $k,k+\rmd k$.}
\end{figure}

With the expression of the  field strength in terms of $\alpha$ obtained from  Eq.~(\ref{AiLCg}), 
and  using Wick's theorem as before, one gets 
\beq\label{FFMV1}
\langle F^{i-}_a(x^+,\x) F^{i-}_a(y^+,\y)\rangle&=&\langle \left(U^\dagger_{ab}\del^i\alpha^b\right)_{\x}\left(U^\dagger_{ac}\del^i\alpha^c\right)_{\y}\rangle\nn
&=&\langle\del^i\alpha^b(\x) \del^i\alpha^c(\y)\rangle\langle U_{ab}(\x)U_{ca}(\y)^\dagger\rangle\nn
& =& \delta(x^+ -y^+)\langle {\rm Tr}U(\x)U^\dagger(\y)\rangle\left(-\nabla^2_\perp \gamma_A(x^+,\x-\y)\right).\nn
\eeq
As we shall now verify, the main effect of the Wilson line correlator is to redistribute the gluons in momentum space, the total number of gluons remaining fixed \cite{Lam:1999wu}. From Eq.~(\ref{FFMV1}), a simple calculation yields $\varphi_A(x,\k)$, the unintegrated distribution function of a nucleus \cite{JalilianMarian:1996xn,Kovchegov:1998bi}
\beq\label{MVundf}
\varphi_A(x,\k)=\frac{k_\perp^2}{4\pi^2}\frac{4\pi R^2}{C_A g^2}(N_c^2-1)\int_{\r_\perp}\frac{\rme^{i\k\cdot\r}}{r_\perp^2}\left(  1-S(r_\perp) \right),
\eeq
where $S(r_\perp)=\exp\left\{ -\mu C_A g^2 I(r_\perp)   \right\}$ is  the dipole $S$-matrix in the adjoint representation, given in  Eq.~(\ref{dipoleR}). An illustration of the momentum dependence of $\varphi_A(x,\k)$ is given in Fig.~\ref{fig:phi}, left. The integrated gluon distribution function is obtained from $\varphi_A(x,\k)$ by integration over momentum, according to Eq.~(\ref{undf}).   When $Q^2\gg Q_s^2$, this is  easily done. One gets
\beq
\int_{\Lambda_{QCD}^2}^{Q^2}\rmd k_\perp^2 \frac{ \varphi_A(x,\k)}{k_\perp^2}\approx N^{\rm val} \frac{\alpha_s C_F}{\pi} \ln \frac{Q^2}{\Lambda_{QCD}^2}.
\eeq
where we have  set $r_\perp^2\sim 1/Q^2$ as the lower cutoff on $r_\perp$. The right hand side of this equation is nothing but $N^{\rm val} xG_q(x,Q^2)$ : as we anticipated, when $Q^2\gg Q_s^2$, the integrated gluon distribution  of the nucleus, $xG_A(x,Q^2)$, is just the sum of the corresponding distributions of its $N^{\rm val}$ valence quarks.

The limiting behaviors of $\varphi_A(x,\k)$ at small and large momenta are interesting (see \cite{Iancu:2004bx} for a detailed analysis). The large $k_\perp$ limit is the weak field limit, with the integral in Eq.~(\ref{MVundf}) dominated by small $r_\perp$. The leading behavior is captured by expanding $S(r_\perp)$ in  Eq.~(\ref{dipoleR}) to linear order, which is   equivalent to ignoring the effect of multiple scattering. One gets
\beq
\varphi_A(x,\k)&\approx&\frac{k_\perp^2}{4\pi^3}\frac{\pi R^2}{\alpha C_A }(N_c^2-1)\int_{\r}\frac{\rme^{i\k\cdot\r}}{r_\perp^2}\left(  \frac{C_Ag^2}{16\pi}\mu r_\perp^2\ln \frac{1}{r_\perp^2 \Lambda^2}  \right)\nn
&\approx& A \;\frac{(N_c^2-1)\alpha}{2\pi} =A\varphi_N(x,\k),
\eeq
where $\varphi_N(x,\k)=\frac{\alpha C_F}{\pi} N_c$ is  the unintegrated gluon distribution for a nucleon, treated here as a collection of three independent valence quarks.  We recover the additive property of the weak field limit.

At low momenta ($k_\perp\lesssim Q_s$), the dominant contribution is coming from the large sizes, $r_\perp\gtrsim 1/Q_s$. Then, in Eq.~(\ref{MVundf}), one can ignore $S(r_\perp)$ which is small. One gets
\beq
\varphi_A(x,\k)
\approx\frac{k_\perp^2}{4\pi^3}\frac{\pi R^2}{\alpha_s C_A }(N_c^2-1)\, \pi\ln\frac{Q_s^2}{k_\perp^2},
\eeq
that is,
\beq\label{occupation}
\frac{1}{\pi R^2} \frac{\varphi_A(x,\k)}{k_\perp^2}=\frac{(N_c^2-1)}{4\pi^2 \alpha_s C_A }\ln\frac{Q_s^2}{k_\perp^2}.
\eeq
In estimating the integral, we have used the fact that it is now  limited not by $r_0$, but rather by $1/Q_s\ll r_0$ since we assume that $Q_s\gg \Lambda_{\rm QCD}$.
One sees that in the saturated regime, the quantity in Eq.~(\ref{occupation}), akin to an occupation factor, is of order $1/\alpha_s$.


It is tempting to relate the depletion in  $\varphi_A(k_\perp\ll Q_s )$, as seen in Fig.~\ref{fig:phi}, left, to a general property of saturation that we shall discuss more in the next section. 
This connection is based on the general relation between the charge correlator and the unintegrated distribution function 
\beq\label{phiandcorr}
\langle \rho^a(k^-,\k) \rho^a(-k^-,-\k)\rangle=4\pi^2 N^{\rm val}\,\varphi(x,\k).
\eeq
According to Eq.~(\ref{occupation}), $\langle \rho^a(\k) \rho^b(-\k)\rangle\sim k_\perp^2$ for $k_\perp\lesssim Q_s$ (to within a logarithmic correction), which implies correlations between color charges resulting in a complete screening on a distance scale of order $1/Q_s$.  This provides a new interpretation of the saturation momentum, with $1/Q_s$ playing the role of a correlation length. Of course the charges in question are not those of the valence quarks, but rather those of an effective distribution that reproduces the effect of the non linearities of the gauge fields (or said differently, of the effect of multiple scattering encoded in the Wilson lines). 

As a final remark, the quantity $\varphi_A/k$ which represents the number of gluons between $k$ and $k+\rmd k$, is plotted in Fig.~\ref{fig:phi}, right . As one can see, this is maximum for $k\sim Q_s$. As we shall see later this has important consequences for the phenomenology, since most gluons that are freed in a heavy collison have momenta near this maximum. 
In contrast, the color dipole that we discussed earlier in this section is blind to the redistribution of gluons in momentum space. This is because its interaction is determined by the integrated gluon distribution function, and this is not affected. 
Thus, viewed by a  dipole, the MV nucleus appears as a collection of independent valence quarks accompanied by their respective clouds of virtual gluons.

\section{Non-linear evolution equations and statistical aspects of saturation}\label{sec:evolution}

The MV model  provides us with a way to calculate the interaction of an elementary dipole with the strong, random,  color field of a nucleus. The calculation uses  a strict eikonal approximation for the dipole,  a Gaussian probability for the random field, and the result is  independent of the rapidity interval between the dipole and the nucleus. A dependence on rapidity, hence on the energy of the collision,   enters when one takes into account radiative corrections, that is the possibility of soft gluon emissions during the interaction process. This is what this section will be about. 
As outcome of the analysis we shall obtain non  linear evolution equations,  a determination of the energy dependence of the saturation momentum $ Q_s$, a new perspective on  the phenomenon of saturation, a microscopic understanding of the phenomenon of geometrical scaling, as well as a new vision on the wave function of a hadron at high energy.

We shall gain insight into the role of these radiative corrections by considering the radiated gluons as either corrections to the wave function of the dipole, or as fluctuations of the color field of the target. Both point of views are a priori equivalent, but lead to different descriptions. When boosting the projectile, we shall be led to  picture  the evolution of the dipole wave function as a  cascade of dipoles, representing successive soft gluon emissions. On the other hand, boosting the target is a complicated operation, which is a priori difficult, if not impossible, to implement. We shall not attempt to do it directly. In line with the  previous section, we shall view the target as   a random color field, with some probability $W[A]$ that a given field configuration $A(x)$ be realized in a given event.  We shall then assume that the change of the target under a boost is entirely captured by the change in $W[A]$, and determine that change of $W[A]$ by a self-consistency argument.

\subsection{Evolution of the dipole operator in a fixed background}\label{Sec:BKop}

We start by considering the forward $S$-matrix element of a right moving color dipole 
\beq\label{dipolexy}
S(\x,\y)=\frac{1}{N_c}{\rm Tr} \left(U_\x U^\dagger_\y    \right),
\eeq
where 
\beq\label{WilsonJIMWLK}
U_\x={\rm P}\exp\left(   ig\int_{-\infty}^{\infty} \rmd z^+ \alpha_a(z^+,\x)t^a\right),\quad \alpha_a(z^+,\x)=A^-_a(z^+,\x) ,
\eeq
and the field of the target, $A^-_a(z^+,\x) $,  is at this point considered as a fixed background field (i.e., no average over the field of the target is performed yet).  
Recall that $\alpha_a(z^+,\x)$ has a finite extension in $z^+$, and  is (nearly) independent of $x^-$. We use a gauge with $A^+=0$ and $A^-\ne 0$.

We assume that the nucleus is moving with rapidity $Y_0<0$, and call $Y$ the rapidity interval between the dipole and the nucleus, so that the rapidity of the dipole is $Y+Y_0$. 
We assume, to simplify the discussion, that the dipole rapidity is large, but not so large (typically $\alpha_s Y\ll 1$), so that we can describe the interaction as that of a  ``bare dipole" with the target. We want to study the effects of radiative corrections that occur when  we increase the rapidity interval: $S$ will then depart from its simple expression in terms of Wilson lines given by Eq.~(\ref{dipolexy}), in particular it will acquire a dependence on rapidity. To emphasize this dependence it  will be denoted by $ S_Y(\x,\y)$, or by $S_{\x\y}^Y$, with $Y$ the rapidity of the dipole relative to that of the target. 

\begin{figure}[htbp]
\begin{center}
\vspace{-0.5cm}
\hspace{4cm}\includegraphics[scale=0.5]{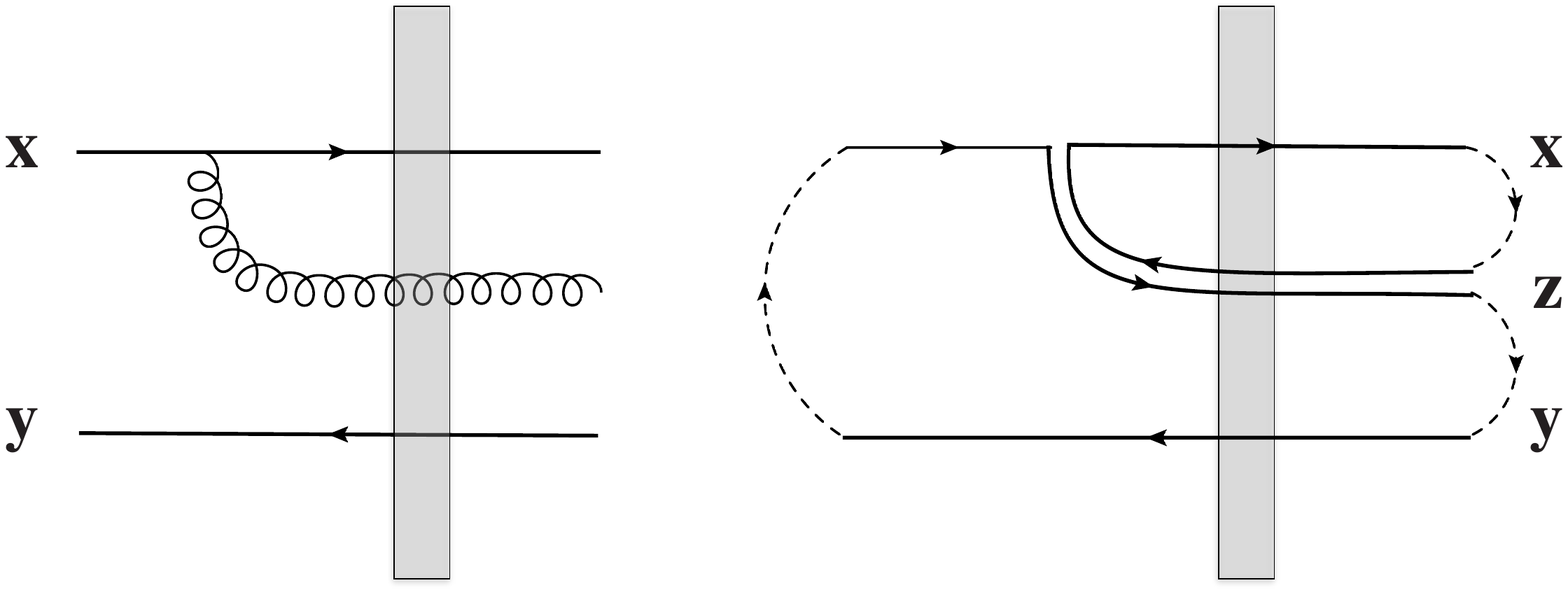}
\vspace{-5cm}
\end{center}
\caption{\label{fig:dipole_splitting} Basic process leading to the ``evolution''  of the interaction of a dipole with the field of a nucleus (represented by the ``Lorentz contracted'' shaded area). There is another diagram, not drawn, where the gluon is emitted from the antiquark at $\y$.  An increase of the rapidity gap between the dipole and the target by an amount $\rmd Y$ increases the phase space for the emission of one gluon. This extra gluon modifies the effective interaction of the dipole as it traverses the color field of the nucleus.  Right: In the large $N_c$ limit, a gluon can be represented by an overlapping quark-antiquark pair, indicated by the double line. The interaction with the target of  the dipole-gluon system is then equivalent to that of a pair of independent dipoles, with endpoints $\x,\z$ and $\z,\y$ respectively.  Note that the three coordinates $\x,\y,\z$ do not change during the interaction (eikonal approximation).  The dashed lines indicate the color flows, exhibiting the two color singlets that emerge from the emission of a gluon.}
\end{figure}

When we boost  the dipole by a certain rapidity amount $\rmd Y$, we increase, by an amount $\rmd P\sim\alpha\rmd Y$, the probability that a gluon be emitted in the interaction with the target. The basic process  is depicted in Fig.~\ref{fig:dipole_splitting}.  After the emission of a gluon, the original dipole turns into a dipole-gluon system, whose propagation  in the field of the hadron differs from that of the original dipole.  
\begin{figure}[htbp]
\begin{center}
\vspace{-0.5cm}
\includegraphics[scale=0.6]{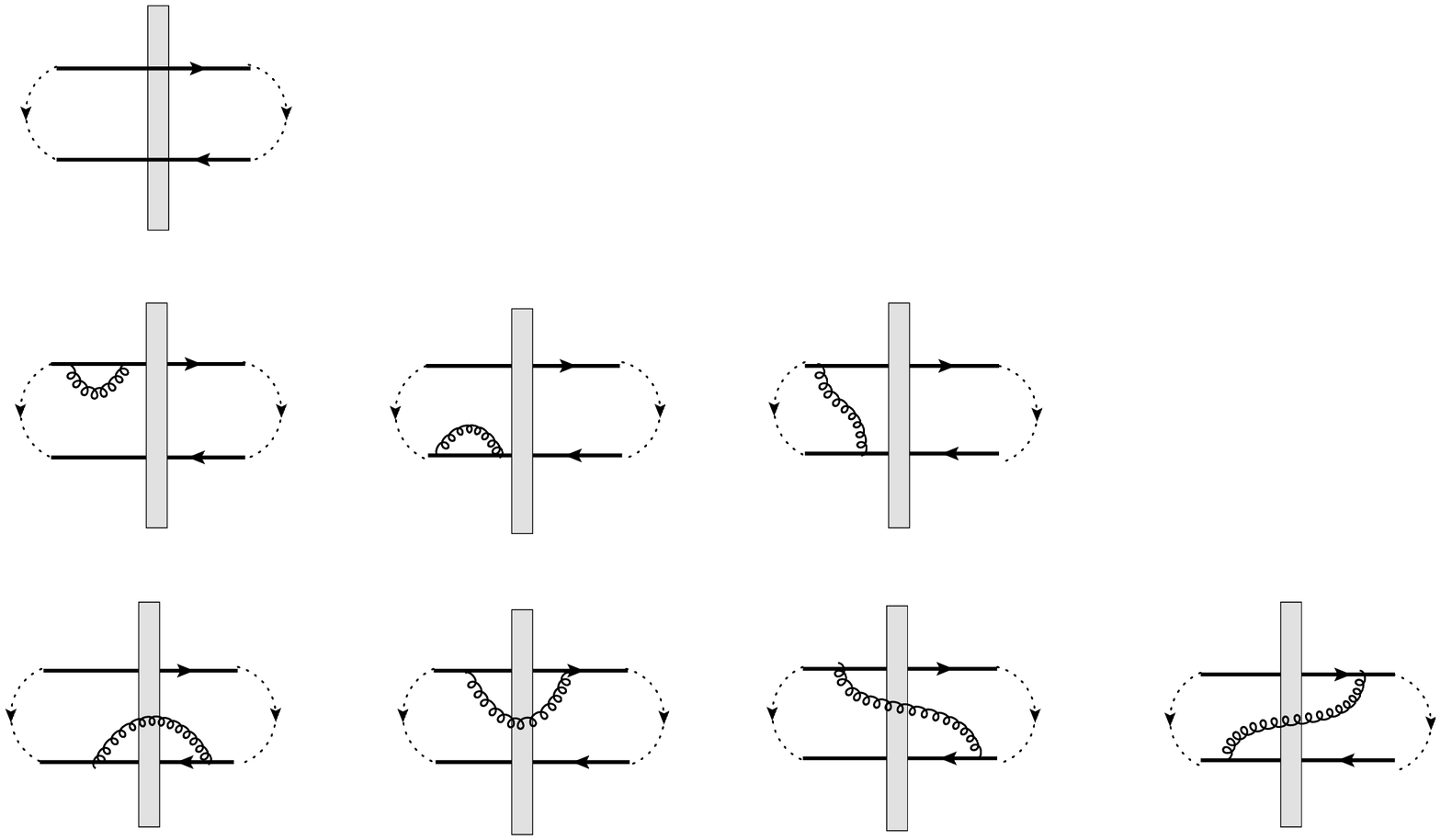}
\vspace{-3cm}
\end{center}
\caption{\label{fig:dipole_evolution} Evolution of a right moving dipole, following an increase $\rmd Y=\rmd k^+/k^+$ in rapidity, with $k^+\ll p^+$ the plus component of the emitted gluon, and $p^+$ that of the parent quark or antiquark. The diagram on top represents the bare dipole interacting with the field of the target, indicated by the shaded area. The diagrams in the second row represent ``virtual'' corrections, which may be interpreted here as corrections to the wave function of the dipole. Such corrections are essential to obtain a probabilistic interpretation of the process. The three diagrams where these corrections take place on the right of the target are not drawn. The diagrams in the last line represent genuine interactions of the dressed dipole with the target. The propagator of the soft gluon is an eikonal Wilson line. }
\end{figure}
The complete leading order calculation is illustrated by the diagrams in Fig.~\ref{fig:dipole_evolution}. 
 Since the emitted gluons are soft, the process can be treated in the  eikonal approximation (this includes both the vertex coupling the emitted gluon to the members of the dipole, as well as the propagation of the emitted gluon in the field of the target). 
The calculation of the diagrams in Fig.~\ref{fig:dipole_evolution} lead to the following equation (see e.g. \cite{Gelis:2012ri}  for an elementary derivation)
\beq\label{generaleqS}
\partial_Y S_Y(\x,\y)=-\frac{\alpha_sN_c}{2\pi^2}\int d^2\z\,{\cal K}_{\x \y \z}\,\left\{ S_Y(\x,\y)   - S_Y(\x,\z) S_Y(\z,\y)\right\},
\eeq
where $(\bar\alpha/2\pi) {\cal K}_{\x \y \z}\,d^2\z \,dY$, with
\beq\label{dipole_kernel}
 {\cal K}_{\x \y \z}\equiv \frac{(\x-\y)^2}{(\x-\z)^2(\y-\z)^2}, 
\eeq
 gives the probability that the dipole with  two color charges located at positions $\x$ and $\y$ emits a gluon located at $\z$ \cite{Mueller:1993rr}\footnote{The kernel $ {\cal K}_{\x \y \z}$,  sometimes denoted ${\cal M}_{\x \y \z}$ \cite{Triantafyllopoulos:2005cn}, differs from the corresponding kernel that apperas in the JIMWLK equation (see Eq.~(\ref{chidef}) below), which is of the factorized form
 $$
 \tilde {\cal K}_{\x \y \z}\equiv \frac{(\x-\z)\cdot (\z-\y)}{(\x-\z)^2(\y-\z)^2}.
  $$
  Note that ${\cal K}_{\x \y \z}= 2 \tilde{\cal K}_{\x \y \z}-\tilde{\cal K}_{\x \x \z}-\tilde{\cal K}_{\y \y \z}$ . }
 . It is the coordinate space version of the kernel of the BFKL equation (\ref{BFKL0}). 
 
 It is important to understand where the rapidity dependence of $S_Y$ comes from. The direct calculation of the diagrams displayed in Fig.~\ref{fig:dipole_evolution} yields divergent expressions, the divergence coming from the integration over the longitudinal momentum ($k^+)$ of the emitted gluon, and is generically of the form $\int \rmd k^+/k^+$. Cutoffs need to be introduced at both ends of this logarithmically divergent integral. What is assumed in writing Eq.~(\ref{generaleqS}) is that this cutoff on the phase-space for soft emission grows as $\rmd Y$. (Recall that $k^+=k_\perp \rme^y/\sqrt{2}$, and the rapidity $y$ of the emitted gluon lies somewhere between that of the projectile and that of the target.) The presence of this logarithmic divergence in the calculation of radiative corrections plays a crucial role in this entire section.\\

  While Eq.~(\ref{generaleqS}) has been obtained as a result of a bona fide quantum field theoretical calculation (and there are indeed subtle aspects in this calculation), the right hand side of this equation can be given  a probabilistic interpretation. The basic stochastic mechanism is the splitting of the dipole, which occurs, as we have seen,  with a probability $\rmd P=(\bar\alpha/2\pi) {\cal K}\,\rmd Y$.  It is convenient, at least in the limit of large number of colors, i.e, large $N_c$,  to look at this splitting as producing two new dipoles (see Fig.~\ref{fig:dipole_splitting}), thus promoting color dipoles to basic degrees of freedom  \cite{Andersson:1990dp,Mueller:1993rr}.  The equation ~(\ref{generaleqS}) can be written as $S_{Y+\rmd Y}=(1-\rmd P) S_Y+\rmd P S^2_Y$. The change  $\rmd S_Y=S_{Y+\rmd Y}-S_Y$ has two contributions: either the dipole  splits, with probability $\rmd P$, and then the $S$-matrix is that of the two resulting dipoles scattering independently  off the
colour field
$A^-$ of the target; this is the origine of the term $\sim S^2$ in the rhs of Eq.~(\ref{generaleqS}).  Or it does not split, which occurs with probability $1-\rmd P$, in which case the $S$-matrix is that of the original dipole.  

Based on this probabilistic interpretation, one may develop a picture of the projectile wave function as a cascade of dipoles \cite{Mueller:1993rr} that interact independently of each other with the target (see also \cite{Salam:1995uy,Mueller:1996te}).   Thus, for instance, the evolution of the  inclusive one body distribution (dipole density) is given by \cite{Iancu:2003uh}
 \beq
 \del_Y n^{_{\! Y}}_{\x\y} =\frac{\bar\alpha_s }{2\pi}\int_\z \left(  {\cal K}_{\x \z \y} n^{_{\! Y}}_{\x\z}+{\cal K}_{\z \y \x} n^{_{\! Y}}_{\z\y}-{\cal K}_{\x \y \z}n^{_{\! Y}}_{\x\y} \right),
 \eeq
 where $n_{\x\y}^{_{\! Y}}$ is  the density of dipoles with end points $\x$ and $\y$, at rapidity $Y$.  
 This equation has an obvious probabilistic interpretation: in the step $\rmd Y$, the dipole $\{\x\z\}$ has a chance to split into two dipoles
  $\{\x\y\}$ and $\{\z\y\}$, thereby increasing the density $n_{\x\y}$ (the dipole $\{\z\y\}$ is not counted in the inclusive density). The same reasoning applies to  $n_{\z\y}$. The last term is the loss term corresponding to the splitting of the dipole $\{\x\y\}$. This equation is formally identical to the BFKL equation. It predicts in particular an exponential growth with increasing rapidity of the number of dipoles in the projectile wave function.

 The equation (\ref{generaleqS}) for $S$ (a complex quantity) is a closed equation, and one coud imagine solving it  for a given realization of the random field.  But perhaps this is not very useful, for a number or reasons (see however \cite{Lappi:2016gqe}).  We have just remarked that  this equation describes the interaction of a cascade of independent dipoles with a given color field. It ignores possible interactions among the dipoles. Such interactions would be visible if we were to consider the evolution of a pair of dipoles, initially of the form $S_{\x_1\y_1}S_{\x_2\y_2}$. The evolution of this object would involve, in addition to the evolution of each independent dipoles, also contributions coming from interactions due to gluon exchanges between the two dipoles. In short, $[S_{\x_1\y_1}S_{\x_2\y_2}]^Y\ne S_{\x_1\y_1}^YS_{\x_2\y_2}^Y$. These interactions lead to more complex color structures than dipoles or product of dipoles (commonly referred to as quadrupoles, or sextupoles -- see for instance \cite{Triantafyllopoulos:2005cn,Iancu:2011ns}
for the explicit form of the corresponding equations).
 
 Another issue is that the physical interpretation of $S$ is easier after one takes the expectation value over the target field (in particular this turns $S$ into a real quantity).  In doing so, one introduces correlations that are due in particular to the fact that the two offsprings propagate in the same color field, with additional  correlations coming possibly from radiative corrections.  It follows that after taking the average over the target field, Eq.~(\ref{generaleqS})  is not closed anymore since $\langle S^2\rangle\ne \langle S\rangle^2$. It becomes in fact the first equation of an infinite hierarchy of equations that couple  correlators of arbitray numbers of Wilson lines \cite{Balitsky:1995ub}. We shall return to this hierarchy later in this section. In the next two subsections, we focus on the approximate form of the first equation obtained by forcing factorization.

\subsection{The BK equation} 
The Balitsky-Kovchegov (BK) equation \cite{Balitsky:1995ub,Kovchegov:1999ua} is the equation that results when taking the expectation value of Eq.~(\ref{generaleqS}) over the field of the target (at a fixed rapidity $Y_0$) and assuming factorization of the term quadratic in $S$ (see also \cite{Braun:2000wr}). That is 
\beq\label{BK}
\partial_Y \langle S_{\x,\y}^Y\rangle_{_{Y_0}} =-\frac{\bar\alpha_s}{2\pi}\int_\z\,{\cal K}_{\x \y \z}\,\left\{ \langle S^Y_{\x,\y}\rangle_{_{Y_0}}   - \langle S_{\x,\z}^Y\rangle_{_{Y_0}} \langle S_{\z,\y}^Y\rangle_{_{Y_0}}\right\}.
\eeq
The factorization, viz.  
$
\langle S_{\x,\z}^YS_{\z,\y}^Y\rangle=\langle S^Y_{\x,\z}\rangle \langle S^Y_{\z,\y}\rangle$, neglects the correlations that arise in particular from the fact that the two dipoles that emerge from a splitting  propagate in the same background field. Such correlations are generically suppressed at large $N_c$, by powers of $1/N_c$.  Note that the average done here does not necessarily involve correlations in the target, beyond those taken into account, for instance, in the MV model averaging.
As we have already indicated, Eq.~(\ref{BK}), prior to factorization, is  the first in an infinite hierarchy of equations for the correlators of an arbitrary number of Wilson lines.  The factorization of the non linear terms  allows to close that hierarchy at the level of the 2-point function. Since from now on we shall mostly deal with average $S$-matrix elements, we shall drop the angular brakets in order to alleviate the notation (they will be reinstated later when they will be explicitly needed to avoid confusion).

Equation~(\ref{BK}) is a non linear equation with two easily identified fixed points. Writing schematically this equation as $\del_Y S=-S(1-S)$, the two fixed points appear at $S=0$ and $S=1$. Small deviations  $\delta S$ from these fixed points are controlled by the linearized equation, $\del_Y (\delta S)=-(1-S)\delta S  +S \delta S$, which suggests that the fixed point $S=1$ is unstable, while $S=0$ is stable.  This observation will guide our forthcoming qualitative discussion, deferring a more quantitative analysis to the next subsection.

For analyzing the vicinity of the fixed point $S=1$, it is convenient to write $S =1-T$. Here $T$ denotes the imaginary part of the scattering amplitude (the scattering amplitude is purely imaginary, as already mentioned). The equation for $T $ reads
\beq\label{BKT}
\partial_Y  T_{\x \y} =\frac{\bar\alpha_s }{2\pi}\int_\z{\cal K}_{\x \y \z}\,\left[    T_{\x\z} +   T_{\z\y}-   T_{\x\y} -    T_{\x\z}  T_{\z\y}  \right].
\eeq
In the vicinity of the fixed point, we can drop the last term, quadratic in $T$. The resulting equation is the coordinate space version of the BFKL equation, written here as an equation for the scattering amplitude instead of an equation for the unintegrated distribution function, as in Eq.~(\ref{BFKL0}).  We have already seen that the solution of the BFKL equation grows exponentially as one increases the rapidity interval. A natural interpretation  is to relate this growth of the scattering amplitude  to the growth of the dipole density in the projectile, which effectively increases the probability of interaction of the initial dipole. But, anticipating on a reasoning that we shall make more systematic later, one may also attribute the growth of the scattering amplitude to the  growth of the saturation momentum, that is, to a change in the properties of the target. To see that, imagine that the only averaging done so far is that corresponding to the MV model, which gives the target a saturation momentum $Q_{s0}$. Consider then a dipole of initial size  $r_\perp$. Define the effective saturation momentum $Q_s$ as  the momentum for which, say, $T(r_\perp)\simeq 1/2$ (in terms of the MV model expression (\ref{Sr1}) for instance, this entails $Q_{s0} r_\perp\simeq 1.5$). If, from that point, we increase the rapidity interval, $T(r_\perp)$ increases, due to the BFKL evolution. Since $T(r_\perp)$ is an increasing function of $r_\perp$,  it follows that  $T(r_\perp')\simeq 1/2$ will be reached for a smaller value $r_\perp'<r_\perp$, that is, for   a larger value of  the effective saturation momentum. In other words, the initial dipole of size $r_\perp$ ``sees'' effectively a denser target after evolution. It is worth keeping in mind the interpretational aspect of the reasoning: we have not \emph{calculated} here any modification of the target. In fact the projectile knows very little about the target, except for the existence of a scale that necessarily enters the dipole cross section that characterizes its interaction with the target.

Consider now the fixed point $S=0, T=1$, which corresponds to the saturation regime, or the black disk limit. The way it is reached is already suggested by the previous discussion. As the evolution proceeds, the number of dipoles increases, and at some point their density is so high that the probability that two dipoles interact simultaneously with the field becomes of order unity. This  occurs when the density of dipoles in the projectile is of order $1/\alpha_s^2$ (to see that, recall that $T$ is $\sim \alpha_s^2 n$, in leading order, two gluon exchange approximation, and observe that the non linear terms in Eq.~(\ref{BK}) are comparable with the linear ones when $ T\sim 1$). Here, saturation comes in not because the gluon density of the target is increasing (the target is left unchanged by the evolution), nor because the dipole density in the projectile is so large that the dipoles start to recombine (they do not interact among themselves), but because the density of dipoles in the projectile is so high that they start to interact simultaneously with the target, causing a rapid drop in the probability for the original dipole not to interact with the target. 

We shall return later to this subtle   interplay between the properties of the projectile and those of the target in the evolution in rapidity. 

\begin{figure}[htbp]
\begin{center}
\includegraphics[scale=0.45,angle=0]{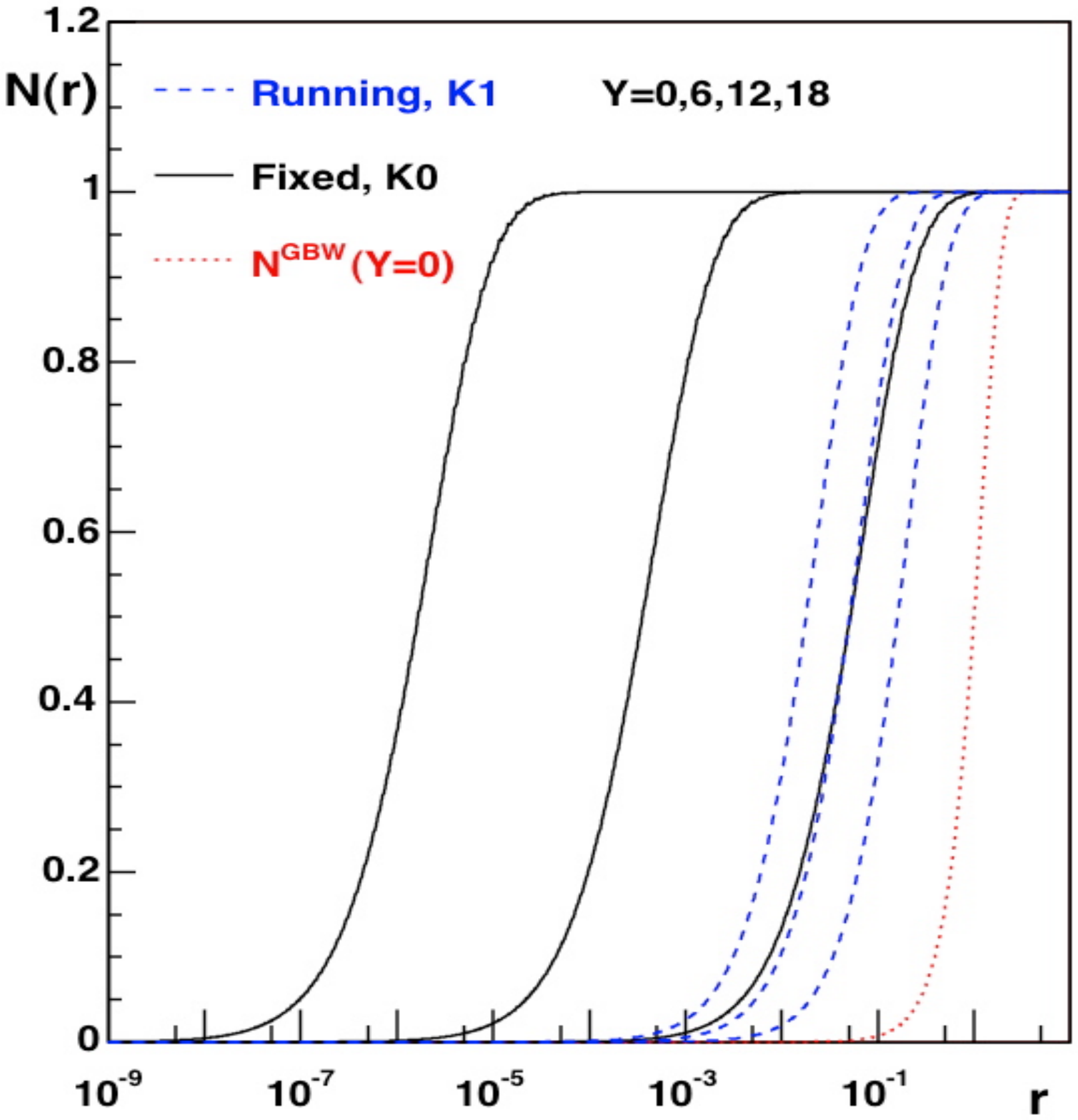}
\end{center}
\caption{\label{fig:trav} The dipole scattering amplitude $T(r)$  as a function of the dipole size for various  rapidities $Y$ (figure taken from Ref.~\cite{Albacete:2004gw}).  The curves  are obtained by solving the BK equation with fixed coupling (full lines) or running coupling (dashed lines). The travelling waves are clearly visible, with a `saturation front' joining small and large dipoles at a scale $r_s(Y)$ that decreases with increasing $Y$. Note that the evolution is slower for running coupling than for fixed coupling, a feature that seems generic in all these types of calculations.}
\end{figure}

\subsection{General properties of the solution of the  BK equation}\label{sec:solutionBK}

We turn now to a more quantitative discussion of the solution of the BK equation. While no explicit solution is known, the equation has been studied numerically, and the general properties of its solutions are well understood.  As an illustration, we show in Fig.~\ref{fig:trav} solutions of the BK equation for the scattering amplitude with simple initial conditions \cite{GolecBiernat:2001if,Albacete:2004gw}.   After some transitory regime, the striking feature of the solution is that the shape of the scattering amplitude as a function of the logarithm of the dipole size reproduces itself as one increases the rapidity. We shall later interpret this behavior as characterizing a travelling wave. We now proceed to a brief quantitative analysis of the solution near the two fixed points, where analytical studies are possible.

If one ignores the non linear term in Eq.~(\ref{BKT}), which is appropriate for small dipole sizes, one gets the BFKL equation
\beq\label{BFKLBK}
\partial_Y T_{\x \y} =\frac{\bar\alpha_s }{2\pi}\int_\z\,{\cal K}_{\x \z \y} \,\left(    T_{\x\z}+T_{\z\y}-T_{\x\y} \right).
\eeq
The exact solution of this equation is known, but we shall be satisfied here with the semi-quantitative behavior at high energy.   Eq.~(\ref{BFKLBK}) is a linear equation,  and because the kernel is  invariant under scale transformations, one may look for eigenstates that are power laws, i.e., of the form $T(r)\sim r^{2(1-\gamma)}$, with $r\equiv |\x-\y|$. The BFKL equation takes then the form 
\beq\label{BFKLsol}
\del_Y T(r)=\bar\alpha\chi(\gamma) T(r), \qquad \chi(\gamma)=2\psi(1)-\psi(\gamma)-\psi(1-\gamma),
\eeq
where $\psi(z)=\rmd \ln \Gamma(z)/\rmd z$ is the digamma function.
 The function $\chi(\gamma)$ has a minimum at $\gamma=1/2$, with $\chi(1/2)=4\ln 2\equiv \omega$, and this determines  the dominant behavior at high energy. Note that the $r$-dependence of the dipole amplitude at very high energy differs from the quadratic dependence expected from color transparency, i.e. $T(r)\sim r^2$. This  would  correspond to $\gamma=0$,  whereas the solution is now $T(r)\sim r$, corresponding to an ``anomalous'' dimension $\gamma=1/2$. In addition, one finds an exponential growth with rapidity, i.e., $T(r,Y)\sim \rme^{\omega \bar\alpha Y}$. It follows that the effective saturation momentum, defined for instance from the dipole size $r_s\sim 1/Q_s$ at which $T(r_s)$ has a given value (say $\sim 1/2$ - see previous subsection), exhibits also an exponential growth with rapidity. The growth of the saturation momentum is obtained by solving the BFKL equation with the constraint $r_s Q_s={\rm cste}$, which shifts the saddle point from the minimum of $\chi(\gamma)$ to $\gamma_0$ solution of  $\gamma_0\chi'(\gamma_0)=\chi(\gamma_0)$ (see also Eq.~(\ref{frontposition}) below). One then gets \cite{Gribov:1984tu,Iancu:2002tr} $\gamma_0\simeq 0.63$, and  
\beq\label{QsBFKL}
Q_s^2(Y)\approx Q_0^2 \,\rme^{\lambda Y},\qquad \lambda\simeq 4.88 \bar\alpha_s.
\eeq
This value of $\lambda$ ($\lambda\simeq 1$ for $\bar\alpha_s\simeq 0.2$) is larger than that obtained from phenomenological analysis of DIS, $\lambda\simeq 0.3$ (see Sect.~\ref{sec:dipolemodel}).  More refined analysis \cite{Triantafyllopoulos:2002nz,Mueller:2002zm}, that include  NLO corrections to the BFKL kernel, in particular running coupling corrections  (\cite{Fadin:1998py}, and more recently \cite{Iancu:2015vea}), yield values  in good  agreement with the phenomenology. In line with a remark made in the previous subsection, it is interesting to observe that the energy dependence of the saturation momentum is essentially driven by the BFKL evolution: saturation is of course essential, but only as a constraint on the evolution, the details of the saturation mechanism being seemingly irrelevant. 
\\

Turning now to large dipole sizes, we note that $S$ is then small and one can ignore the quadratic term in Eq.~(\ref{BK}). The resulting equation is simply
\beq\label{generaleqS1}
\partial_Y \langle S^Y_{\x,\y}\rangle =-\frac{\bar\alpha_s}{2\pi}\int_\z\,{\cal K}_{\x \y \z}\, \langle S^Y_{\x,\y}\rangle .
\eeq
This equation is valid in the deep saturated region,  when all three dipoles $\{\x,\y\}$, $\{\x,\z\}$ and $\{\z,\y\}$, are large compared to $1/Q_s$, that is, $1/Q_s^2\ll (\x-\z)^2, (\y-\z)^2\lesssim r_\perp^2$, where $r_\perp=|\x-\y|$ is the size of the original dipole, and $Q_s $ the effective saturation momentum. It is then easy to extract the leading logarithm of the integral coming from integrating all sizes from $1/Q_s$ to $r_\perp$ (see e.g. \cite{Mueller:Cargese}). One gets
\beq
\partial_Y \langle S^Y(r_\perp)\rangle \approx -\bar\alpha_s \ln(Q_s^2 r_\perp^2)\, \langle S^Y(r_\perp)\rangle,
\eeq
whose solution can be written as \cite{Levin:1999mw}
\beq\label{LT1}
S^Y(r_\perp)&=&S^{Y_s}(r_\perp) \exp \{ -\frac{\bar\alpha \lambda}{2} (Y-Y_s)^2 \}\nn &=& S^{Y_s}(r_\perp)\exp \left\{ -\frac{\bar\alpha}{2\lambda} \ln^2(Q_s^2(Y) r_\perp^2 )\right\}.
\eeq
The  rapidity $Y_s$ is chosen such that $Q^2_s(Y_s) r_\perp^2\simeq 1$, that is $Y_s$ is the rapidity at which the dipole of size $r_\perp$ starts to feel the saturation, and we have used  rapidity dependence of $Q_s$ obtained earlier, Eq.~(\ref{QsBFKL}), 
$Q_s^2(y)=Q_s^2(Y_s)\exp \{ \lambda (y-Y_s) \}.$ The equation (\ref{LT1}) describes the evolution of the scattering matrix as the rapidity grow beyond saturation and the black disck limit is approached. 

Perhaps more interesting is the transverse momentum dependence. This can be analyzed in terms of the correlator of the target color charges in Fourier space \cite{Mueller:2002pi,Iancu:2001md}. This again requires an interpretation, based on the general relation (\ref{phiandcorr}) between the unintegrated gluon distribution and the correlator of color charges, as well as a relation such as Eq.~(\ref{MVundf}) between the unintegrated gluon distribution and the scattering amplitude. Then, knowing the scattering amplitude from the BK equation, one can infer the resulting correlator. To proceed, let us define for any given $k_\perp$, the rapidity $Y_s$ as the rapidity corresponding to the onset of saturation:
\beq
Q_s^2(Y_s)=k_\perp^2=Q_0^2 \rme^{\lambda Y_s}, 
\eeq
where we have used Eq.~(\ref{QsBFKL}) in the last equality. 
One can show that in the saturation region, $k_\perp^2\ll Q_s^2(Y)$, 
\beq
\langle \rho_a(\k)\rho_a(-\k)\rangle_Y\approx \frac{k_\perp^2}{\pi} \left( Y-Y_s(k_\perp)\right)= \frac{k_\perp^2}{\lambda \pi} \ln\frac{Q_s^2(Y)}{k_\perp^2}.
\eeq
One recovers a behavior that was already identified in the unintegrated distribution function of  the MV model (Eq.~(\ref{occupation})), the main difference being here that the saturation momentum depends explicitly on the rapidity. As was already commented, the behavior in $k_\perp^2$ reflects strong correlations on distances of order $Q_s^{-1}$.

\subsection{Statistical features of saturation}

At this point we shall digress about  mathematical properties of the non linear BK equation that are common to  analogous equations in other branches of physics. In fact, it can be argued that the BK equation belongs to a large universality class of non linear equations whose asymptotic solutions have universal behaviors. As we have seen earlier, the dynamics described by the BK equation is that of a cascade of dipoles. These dipoles split, with some probability given by QCD, and then interact independently of each other with the color field of the target. When a dipole splits, it produces dipoles of different sizes. There are therefore two probabilistic elements in the process: the splitting proper, and a modification of the dipole sizes which, except for rare fluctuations, can be viewed as a random walk in the space of dipole sizes. This makes the cascade of dipoles akin to what are known as ``branching random walks''. 

The Fisher-Kolmogorov-Petrovsky-Piscounov (FKPP) equation is a central equation in the study of such branching random walks (see Ref.~\cite{Munier:2014bba}, and references therein, for an excellent introduction in the context of QCD. See also \cite{Marchesini:2015ica} for the occurence of the FKPP equation in QCD in a  context different than the present one.). An elementary example of a branching random walk is that of a particle on a line undergoing a random walk, and allowed also to split locally into two identical particles. The FKPP equation is typically written as an equation for the probability $Q(x,t)$  that the particle, initially located at $x=0$ stays on the left of a given position $x$ at time $t$. It takes the form
\beq\label{FKPP}
\frac{\del Q}{\del t}=\frac{\del^2 Q}{\del x^2}-Q+Q^2.
\eeq
The first term on the right hand side  is a diffusion term which  reflects the ordinary random walk. The non linear term originates from the branching of the random walk and the fact that the offsprings also undergo random walks. 

A mapping between this equation and the BK equation can be made at high energy \cite{Munier:2003vc,Munier:2003sj}, by transforming the linear part of the BK equation which controls the vicinity of the unstable fixed point. This involves using a diffusion approximation to the BFKL kernel.  
The resulting FKPP equation \cite{Munier:2014bba} reads  
\beq\label{FKPP_BK}
\del_Y S(Y,x)=\del^2_x S(Y,x) -S(Y,x)+S^2(Y,x),\qquad  x\equiv \ln (r^2/r_0^2),
\eeq
where the rapidity $Y$ plays the role of time and $x=\ln (r^2/r_0^2)$ is a logarithmic measure of the dipole size.  It is in term of this variable that the diffusion term in the right hand side acquires a simple form. 

There has been much studies of this equation (see for a recent discussion see  \cite{Munier:2009pc,Munier:2014bba}, as well as \cite{Mueller:2014fba} and references therein).
We shall here only focus on one specific property that illuminates one aspect of the physics encoded in the BK equation. This is the existence of  travelling wave solutions, that exist for rather generic initial conditions \cite{Munier:2003vc}. Recall that a travelling wave is a function 
\beq
S(Y,x)=S_0(x-X(Y)),
\eeq
with $S_0(x)$ the initial profile. 
 We have already encountered such behavior when discussing numerical solutions of the BK equation (see Fig.~\ref{fig:trav}). The property of geometric scaling  naturally emerges from such solutions. Recall that geometrical scaling is the property that the dipole cross section, a function a priori of $k_\perp$ and the rapidity $Y$, is in fact a function of only the ratio $k_\perp^2/Q_s^2$, with all the rapidity dependence contained in $Q_s(Y)$. The travelling wave connects the unstable fixed point $S\simeq 1$ to the stable fixed point $S\approx 0$ via a wave front whose shape is independent of the rapidity. The rapidity specifying the location of the front, which can be identified with the saturation momentum, is  determined by properties of the solution near the unstable fixed point $S=1$, that is by the solution of the BFKL equation (see \cite{Mueller:2014fba} for a recent study). These properties are in fact universal, and independent of the details of the nonlinearity of the equation.  One gets, to leading order, (with $\chi(\gamma)$ given in  Eq.~(\ref{BFKLsol}))
\beq\label{frontposition}
\frac{\del \ln Q_s^2}{\del Y}=\bar\alpha_s \chi'(\gamma_0)-\frac{3}{2\gamma_0 Y},\qquad \gamma_0\chi'(\gamma_0)=\chi(\gamma_0).
\eeq
To within a power correction, on recovers the growth of the saturation momentum that we have already mentioned, $Q_s^2=Q_0^2\rme^{\lambda Y}$ with $\lambda=\bar\alpha_s\chi'(\gamma_0)$ (which yields $\gamma_0=0.63$ and $\lambda\simeq 4.88 \bar\alpha_s$). Note that these results are for fixed coupling. The results for running coupling are rather different, in fact the resulting equation does not belong to the same universality class \cite{Mueller:2014fba}. As already mentioned, one important effect of the running coupling is to slow down the evolution, as illustrated in Fig.~\ref{fig:trav}.

 One can go beyond the FKPP equation and include fluctuations that result from the merging, or recombination,  of particles. This leads to  the so-called reaction diffusion processes, described by the stochastic FKPP equation (see \cite{Iancu:2004es} and the review \cite{Munier:2009pc}). In terms of QCD, the physics described by such processes is akin to that involving Pomeron loops, which we shall not discuss in this paper (see e.g. \cite{Triantafyllopoulos:2005cn}), and references therein). 
 
 What emerges from such studies is a feature already alluded to earlier, that non linearities are important for saturation, but not their details. The energy dependence of the saturation momentum is mostly driven by the dynamics near the unstable fixed point, i.e., by the BFKL dynamics. This is a remarkable, and puzzling, feature. Indeed,  in the present context of a collision of a color dipole with a nucleus, the BFKL dynamics controls the evolution of the projectile, and knows very little about the target!  

\subsection{Non linear evolution equations: B-JIMWLK}

We now return to the calculation which led to Eq.~(\ref{generaleqS}) and is illustrated in Fig.~\ref{fig:dipole_evolution}. In Sect.~\ref{Sec:BKop}, it was natural to consider the emitted gluon as a correction to the wave function of the projectile. But, in principle, the increase in the rapidity gap could be achieved by keeping the dipole intact, and boosting the target. Since the boost just corresponds to a change of frame, the resulting value of the $S$-matrix should be the same. This is not so easy to verify, however, because calculating the effect of the boost on the target is a non trivial task. Our strategy will therefore be more modest: we shall  postulate a general form for the effect of the boost on the target, and fix the unknown elements by exploiting the aforementioned  equivalence between the two points of view. 

To be more specific, let us recall that the equation (\ref{generaleqS}) for  $S^Y$ involves the propagation of a dipole in a given background field, that of the target in a given event. In other words, $S^Y$ in Eq.~(\ref{generaleqS}) may be considered as a functional of the gauge field of the target. However, it is clearly impossible to express the change in $S^Y$ due to the dipole splitting as a change in the background field in which the unperturbed dipole would propagate. This  is clear from a simple examination of the Feynman diagrams in Fig.~\ref{fig:dipole_evolution}. However, the situation becomes simpler if we consider not the $S$ matrix, but its average over the field of the target, and we remember that the background field itself is a random field: the equivalence between the two descriptions will then be expressed not as a change in the background field proper, but in its probability distribution. All that we shall be able to say about the target concerns the evolution with rapidity of the probability $W[A]$ that a given field configuration be realized in an event.  But this is also all we need to know in order to calculate observables in situations where the field of the target can be considered as classical, i.e., non fluctuating, during the interaction.  \\

We proceed now with a simple observation.  Let us define  functional derivatives ${\delta/}{\delta\alpha_\u^a}$ by their actions  on the end points of the Wilson lines
\beq\label{fctalderivatives}
\frac{\delta}{\delta\alpha^a_\u} U_\x =ig\delta(\x-\u) t^a U_\x,\qquad  \frac{\delta}{\delta\alpha_\u^a} U^\dagger_\x =-ig\delta(\x-\u) U^\dagger_\x t^a.
\eeq
Recall that the gauge field that enters the Wilson line in Eq.~(\ref{WilsonJIMWLK}) has limited extent in $x^+$, so that the derivatives may be considered as acting on the field at the largest $x^+$. 
These definitions can be used to rewrite Eq.~(\ref{generaleqS}) as
\beq\label{generaleqS2}
\partial_Y S^{_{Y}}(\x,\y)&=&-\frac{\alpha_sN_c}{2\pi^2}\int d^2\z\,{\cal K}_{\x \y \z}\,\left\{ S^{_{Y}}(\x,\y)   - S^{_{Y}}(\x,\z) S^{_{Y}}(\z,\y)\right\},\nn
&=& {\cal H}[\alpha, \del_\alpha] S_{\x\y}[\alpha]
\eeq
where ${\cal H}$ is the following functional differential operator
\beq\label{JIMWLK_H}
{\cal H}=-\frac{1}{(2\pi)^3} \int_{\u\v\z} {\cal K}_{\x\y\z} \frac{\delta}{\delta\alpha^a_\x}\left[ 1+U_\x U^\dagger_\y -U_\x U^\dagger_\z-U_\z U^\dagger_\y \right]^{ab}\frac{\delta}{\delta\alpha^b_\y},
\eeq
and  
${\cal K}_{\x \y \z}$ is the dipole kernel given in Eq.~(\ref{dipole_kernel}).  Equation~(\ref{generaleqS2}) is an algebraic identity, a simple rewriting of Eq.~(\ref{generaleqS}). The equation is still valid for the $S$-matrix in a given background, but the right hand side of Eq.~(\ref{generaleqS2}) emerges now as a differential operator that involves directly the target field $\alpha(x^+,\x)$ (the role of the functional differential operator is essentially to bring down the color matrices at the appropriate places). 
It can be verified that ${\cal H}$ is Hermitian, a property  that will be useful shortly.

At this point, we can take the average over the target field. Recall that $Y_0<0$ is the rapidity of the target, and that of the projectile is $Y+Y_0$. We label $S^Y$ by the rapidity $Y$ of the projectile measured from that of the target. Then we take the average over the target field, with a probability distribution that we call $W_{Y_0}[\alpha]$, very much like in the MV model.  That is
\beq\label{avwithWY0}
\langle S^{Y}_{\x\y}\rangle_{Y_0}=\int [\rmd \alpha] W_{Y_0}[\alpha] S^{Y}_{\x\y}[\alpha].
\eeq
Now, when $Y\to Y+\rmd Y$, the average $S$ matrix changes, and this change can be expressed in two ways. One may focus on the projectile, and modify $S^Y$ according to Eq.~(\ref{generaleqS2}), i.e.,  $S^Y\to S^{Y+\rmd Y}$, the average being taken afterwards with the distribution $W_{Y_0}$.  The corresponding change reads
\beq
\langle  S^{Y+\rmd Y}_{\x,\y}-  S^Y_{\x,\y}\rangle_{Y_0}=\langle  \rmd  S^Y_{\x,\y}\rangle_{Y_0}= \int [\rmd \alpha]W_{Y_0}[\alpha]\, \rmd S^Y_{\x,\y},
\eeq
where $\rmd S_{\x,\y}=\left( \partial_Y S_{\x,\y}\right)\rmd Y$ is given by Eq.~(\ref{generaleqS2}). Or, we may decide to interpret the change as resulting from the  evolution of the target, that is, from a modification of the distribution $W[\alpha]$, $W_{Y_0}\to W_{Y_0-\rmd Y}$. The corresponding change reads
\beq
\langle  S^Y_{\x,\y}\rangle_{Y_0-\rmd Y}-\langle  S^Y_{\x,\y}\rangle_{Y_0}=\int [\rmd \alpha]\left(W_{Y_0-\rmd Y}[\alpha]-W_{Y_0}[\alpha]\right)\, S^Y_{\x,\y}.
\eeq
By equating the two changes, and this is the major step in the argument, one gets
\beq\label{majoridentification}
0&=&\int [\rmd \alpha] \left\{   W_{Y_0}[\alpha]{\cal H}[\alpha, \del_\alpha] S^Y_{\x\y}+\del_{Y_0} W_{Y_0}[\alpha]\, S_{\x,\y}^Y  \right\}\nn
&=&\int [\rmd \alpha] \left\{ {\cal H} [\alpha, \del_\alpha] W_{Y_0}[\alpha] +\del_{Y_0} W_{Y_0}[\alpha]\right\} S_{\x,\y}^Y,
\eeq
where we have used the hermiticity of ${\cal H}$ in order to perform an integration by part. This equation is satisfied if $W$ obeys the equation 
\beq\label{firstBal}
\del_{Y_0}W_{Y_0}[\alpha]=-{\cal H}[\alpha, \del_\alpha] W_{Y_0}[\alpha].
\eeq
This equation yields the change to be made in the field distribution of the target when the rapidity interval increases by $\rmd Y$, so as to reproduce the original calculation of the dipole modification. We have assumed, of course, that all what happens to the target when one makes a boost is the change in $W$, and it is not a priori obvious that we could be able to do so. By construction, this equation is equivalent to the first Balitsky equation  \cite{Balitsky:1995ub}, that is the average of Eq.~(\ref{generaleqS}), without the factorization that yields the BK equation. This is natural, since the manipulations that lead to Eq.~(\ref{firstBal}) take into account the  specific correlations that are generated when we average over the field of the target.

  Extending the approach that we have just outlined in order  to recover Balitsky's  infinite hierarchy of coupled evolution
equations \cite{Balitsky:1995ub} does not require any new conceptual element.  All one needs to do is to consider averages of arbitrary numbers of Wilson lines with open color indices (i.e., not restricting oneself to color dipoles) \cite{Mueller:2001uk}.  It is also convenient to consider these Wilson lines as the stochastic elements (rather than the field of the target), the probability that a 
 field
$U_{\x}
$ be realized in a collision with rapidity gap $Y$ being given by a functional $Z_{Y}[U]$. Thus the averages of
products of Wilson lines are written as \cite{Weigert:2000gi}:
\begin{equation}
   \label{eq:expect-val}
   \langle U_{\x_1}^{(\dagger)} \ldots U_{\x_n}^{(\dagger)}
\rangle_Y
   = \int [{\rm d}\mu(U)]\ U_{\x_1}^{(\dagger)}
\ldots U_{\x_n}^{(\dagger)}\ Z_Y[U] \,.
\end{equation}
This formula generalizes Eq.~(\ref{avwithWY0}) above. 
Here,
$U^{(\dagger)}$ is a generic notation for either $U$ or $U^\dagger$ 
(in any representation of
SU$(N)$), and
${\rm d}\mu(U)$ denotes the group invariant
measure.
Then, as recognized in  \cite{Weigert:2000gi},  
Eq.~(\ref{eq:expect-val}) is consistent with the entire hierarchy of Balitsky's equations if
$Z_Y[U]$ obeys the following equation:
\begin{equation}
\label{eq:RGdef}
  \partial_Y Z_Y[U]\,=\,\frac{1}{2}\int_{\x,\y}
\,\nabla_{\x}^a \, \chi^{a b}_{\x \y}[U] \,\nabla_{\y}^b\,
Z_Y[U] ,
\end{equation}
where 
$\nabla_{\x}^a$ is a Lie derivative which generates translations on the
group manifold (the proper generalizations of the derivatives that we introduced in Eq.~(\ref{fctalderivatives})). The kernel $\chi^{ab}_{\x\y}[U]$ has the following factorized form
\begin{equation}
\label{chidef}
\chi^{ab}_{\x\y}[U]\,=\,  \int_\z {\bf
e}^{ac}({\x,\z})\cdot
           {\bf e}^{bc}({\y,\z}), 
\end{equation}
 where ${\bf e}^{ab}({\x,\z})$ is a matrix in  color indices
and a vector in transverse coordinates, given by
\begin{equation}
\label{e}
{\bf e}^{ab}({\x,\z})
= \frac{1}{\sqrt{4 \pi^3}}\,\frac{(\x-\z)}{(\x-\z)^2}
\bigl(1-U_{\x}U^\dagger_{\z}\bigr)^{ab}.
\end{equation}

Equation~(\ref{eq:RGdef}) is the generalization of Eq.~(\ref{firstBal}) for the probability distribution $W_Y[A]$, and it reduces to it when restricted to the evolution of a color dipole. It is a particular writing of the so-called Jalilan-Marian, Iancu, McLerran, Wiegert, Leonidov,  Kovner (JIMWLK) equation \cite{JalilianMarian:1997jx,JalilianMarian:1997gr,JalilianMarian:1998cb,Kovner:2000pt,Weigert:2000gi,Iancu:2000hn,Iancu:2001ad,Ferreiro:2001qy}. The JIMWLK equation is a compact rewriting of the infinite hierarchy of Balitsky equations, to which it is therefore equivalent. It allows the calculation of the evolution of arbitrary correlators of Wilson lines. It has the  form of a (functional) Fokker-Planck
equation, involving fields in the transverse plane, $U_\x$, that take values on the  (curved) $SU(N_c)$ 
manifold. The functional $Z[U]$ can indeed be interpreted as a probability distribution, in particular, $\del_Y \int [\rmd \mu[U]] z[U]=0$. The evolution from rapidity $Y_0$ to rapidity $Y$ can be written as 
\beq
Z_Y[U]=\exp^{-(Y-Y_0){\cal H}} Z_{Y_0}[U],
\eeq
where ${\cal H}$ is the JIMWLK hamiltonian, whose explicit expression can be deduced from Eq.~(\ref{eq:RGdef}) (it differs slightly from that appropriate for the dipole evolution in Eq.~(\ref{JIMWLK_H}); see footnote after Eq.~(\ref{dipole_kernel})). It  is positive (semi)definite, so that the evolution drives the system towards a trivial fixed point $Z[U]=1$ at large rapidity, a point at which $Q_s\to \infty$. We recover properties already encountered when discussing the BK equation.

The form (\ref{eq:RGdef}) of the equation  places the Wilson lines at the heart of the description, and treats them as the basic stochastic elements. For this reason it is mathematically the most elegant version of the JIMWLK equation.  Although it  may look formal,  the associated Langevin equation turned out to be the most convenient way to proceed in numerical calculations \cite{Weigert:2005us}. The JIMWLK equation is often written in a form that is closer to that given in Eq.~(\ref{firstBal}), where the basic degree of freedom is the field $\alpha$ rather than the Wilson line $U$.  The connection between the two formulations was clarified in \cite{Blaizot:2002np}.

\subsection{The CGC  effective theory}

The non linear evolution equations that we have discussed in the previous subsection are often interpreted as a renormalization group equation, and the overall scheme regarded as an effective theory,  the ``CGC effective theory'' \cite{Gelis:2010nm}. While this offers an interesting new perspective on the non linear equations that we have just discussed, it should be clear that there is no  new physical ingredients involved,  beyond those that we have already mentioned. In this approach the focus is put entirely on the target, and an attempt is made to ``construct" the probability $W_Y[A]$ of fields that can be considered as frozen during the interaction, and its evolution with rapidity. 
The renormalization group aspects arise from the   (logarithmic) divergence of the $k^+$ integration, and the observation that a boost operates as a cutoff,  namely it increases (or decreases, depending on how one wants to view things), the logarithmic phase space for radiative corrections. The effective theory comes in as we interpret the probability distribution as the exponential of some effective action, and the radiative corrections as fluctuations that are gradually integrated into the classical background field as the phase space increases. 

To make this more explicit, let us recall how we treated the probability distribution $W_{Y_0}[A]$ that allows us to calculate the interaction of a dipole with a nucleus. Let us assume for definitness  that at the rapidity $Y_0$, the MV model is a good description. The probability $W_{Y_0}[A]$ is the probability that a particular classical field, or equivalently a particular distribution of color sources, be realized in a given event.  Then, make a small boost, $Y_0\to Y_0-\rmd Y$, allowing for soft gluon radiation. As we have seen, this soft radiation can be absorbed into a modification of the probability distribution, $W_{Y_0}\to W_{Y_0-\rmd Y}$. In the spirit of the MV model, one could interpret the modification of $W$ as accounting for the additional color charges carried by the radiated gluons. Because of the specific kinematics (that leading to a large logarithmic contribution), these new charges may be considered as ``frozen" during the interaction, and they add up statistically to the original charges of the valence quarks of the MV model: these new charges modify the probability to find a given distribution of charges in a given event.  And indeed the purpose of the new distribution $W_{Y_0-\rmd Y}$ is to allow the calculation of the interaction of the dipole with the nucleus as that of a bare dipole with this new distribution  of charges. The  original formulation  \cite{Iancu:2000hn,Iancu:2001ad,Ferreiro:2001qy} proceeded through a one-loop calculation of the fluctuations of the gluon field in the existing background  at rapidity $Y_0$, with only the 2-point function being needed in leading (logarithmic) order. The use of correlators of color charges, rather than the correlators of Wilson lines,  introduces some particular, and subtle, features. However,   although the calculations differ in details, conceptually the manipulations are identical to those indicated earlier. This is the case in particular of  the crucial step corresponding to the identification (\ref{majoridentification}).  

A similar one-loop calculation of small fluctuations around a strong color background is performed in  \cite{Gelis:2008rw}, in the context of nucleus-nucleus collisions, with an analogous conclusion:  in the leading logarithmic approximation, the one loop correction is driven by the JIMWLK hamiltonian. We return to the role of  the JIMWLK hamiltonian in nucleus-nucleus collisions in the next section.

\subsection{Summary and further developments}

In this section, we have discussed how gluon densities evolve with  the energy of the collision, or equivalently the rapidity gap between the projectile and the target. We have seen that the evolution is governed by non linear equations,  saturation emerging naturally from the non linearity of these equations.  The study of these equations illustrated various facets of the saturation phenomenon.

Our discussion was limited to the case where the projectile is a simple object, a color dipole, while the target is a nucleus. This entails automatically an asymmetry in the treatment of the collisions, and invites for some caution in the interpretation of these equations, as we have repeatedly emphasized along this section. The basic calculation of radiative corrections, at the source of the evolution, is a well defined perturbative calculation which has a direct interpretation in terms of dressing the dipole wave function, usually referred to as ``projectile evolution". The radiation of gluon provides a stochastic element, at the source of fluctuations in the projectile properties, while the target is treated at a mean field level. This target  classical field is a random field though, and it is possible to transfer all stochastic properties of the interaction, in particular those arising from radiative corrections, to the probability distribution $W$ of this random field. This is the co-called ``target evolution". Note that the   distribution $W$  is not calculated (except in the MV model), only its evolution with rapidity is constrained. In that respect, the probability $W$ plays a role somewhat similar to that of the parton distribution in the collinear factorization scheme. 

Establishing these non linear equations represents a major progress in the field during the last two decades. However many conceptual issues remain: the tools at our disposal to describe high density gluon systems provide neither a complete nor a systematic description. This remark concerns as well the CGC effective theory briefly discussed in the previous subsection. 

We shall now list a few directions where progress is being made, or should be made. 
The non linear equations that describe the interaction of an elementary object with a nucleus are under reasonable control, with  however recent works still clarifying some of their formal features 
\cite{Jeon:2013zga,Binosi:2014xua,Caron-Huot:2013fea}. Study of higher order corrections have been undertaken \cite{Kovner:2013ona}. This includes in particular running coupling  corrections \cite{Balitsky:2006wa,Kovchegov:2006vj} (see also \cite{Albacete:2007yr} for numerical solution). 

The JIMWLK and BK equations are closely related, with the JIMWLK equation capturing finite $N_c$ contributions that are left out of the BK equations. 
Numerical solutions of JIMWLK and BK reveal that these two non linear equations provide in fact very similar evolutions, at least for  the  simplest correlators (see the discussion in \cite{Weigert:2005us} and references therein, see also \cite{Kovchegov:2008mk}, \cite{Lappi:2015fma} and \cite{Dumitru:2011vk}). 

More recently much progress has been achieved in the determination of the NLO evolution of the BK equation \cite{Balitsky:2008zza}, identifying large double logarithmic corrections that after resummation lead  to  a simple modification of the kernel of the BK equation \cite{Beuf:2014uia,Iancu:2015vea}. Generically, it appears that the main effect of these various next to leading corrections, as well as those mentioned above,  is to slow down the evolution (as illustrated for instance in Fig.~\ref{fig:trav}).

Finally, a major open  issue is related to the asymmetric treatment of the projectile and the target in all these approaches. Most discussions involve a simple projectile for which radiative corrections represent important fluctuations.   On the other hand the target is treated rather in a mean field approximation, in which fluctuations are absent. A lot of efforts have been put in trying to overcome this asymmetry, with partial success in limited cases.  For a representative work on this issue see \cite{Kovner:2005en} (and also \cite{Blaizot:2005vf} as well as \cite{Triantafyllopoulos:2005cn}).

\section{Particle production}

As we have already argued, most particles  produced in heavy ion collisions originate from processes involving partons that carry a relatively small fraction of the energy of the colliding nucleons. Such small $x$ partons are likely to come from the end point of a cascade of smaller and smaller $x$ partons (see Sect.~\ref{sec:partons}). It appears therefore that evolution plays an important role in the production process. On the other hand coherence is also potentially  important since, near saturation, small $x$ partons overlap both longitudinally and transversely, thereby motivating a description of the interactions in terms of classical fields. The classical fields, obtained as solutions of the Yang-Mills equations, capture  effects of saturation (see Sect.~\ref{WWfields}), but not of evolution.  However, these classical fields are actually random fields, and the effect of the evolution can be encoded, to some extent,  in the probability distributions of these random fields (see Sect.~\ref{sec:evolution}). How this is justified, and implemented in practice, will be briefly discussed in this section. 

Until now, the only interactions that we have considered have been those of an elementary probe, such as a color dipole, with a complicated target, such as a nucleus.  Our main concern was to understand the structure of the nucleus, and the role of the dipole there has been that of a test particle, whose presence does not alter the field of the nucleus in which it propagates. This field is then obtained as the solution of Yang-Mills equations in the absence of the test particle, which, in  a covariant gauge, reduces to solving the Poisson equation. With the study of particle production we come to a different situation. Here the solution of the Yang-Mills equation represents a non trivial step: we need indeed the explicit  time-dependence of the total field generated during the collision in order to analyze its particle content and deduce the spectrum of produced particles. 

Because it concerns quantities that can be measured, this discussion of particle production will naturally lead us to a (brief)  connection to experimental data, and a rapid survey of a few selected phenomena where high gluon densities could be playing a visible role.

\subsection{Interaction of classical fields}

We start by recalling how  a nucleus-nucleus collision is viewed  in the MV model. 
The two nuclei, say A and B, are described as two collections of valence quarks moving in opposite directions near their respective light cones (see Fig.~\ref{fig:light_cone}). The valence quarks of the respective nuclei do not interact directly with each other (except in rare, hard processes): they are too far apart in rapidity. Rather, the interaction takes places through their classical WW fields. These are obtained by solving  the classical Yang-Mills equations, with appropriate (retarded \footnote{The retarded conditions are connected to the fact that we calculate inclusive quantities.}) boundary conditions. These conditions can be fixed on the light cone where the  field just after the collision can in fact be calculated exactly, given the distributions of color charges. At later time the field can be analyzed in terms of normal modes, using  standard reduction formulae, or Fourier analysis,  and one obtains, after averaging, the inclusive distributions of particles. 

In more precise terms,  the first step consists therefore in solving the Yang-Mills equations (\ref{YMequations}). 
The color current, before the nuclei interact, is that carried by the two independent nuclei, 
\beq
J^+(x)=\rho_B(x^-,\x),\qquad  J^-(x)=\rho_A(x^+,\x),
\eeq
where the color charges $\rho_{A,B}(x^\pm ,\x)$ (defined in covariant gauge) are localized  near the light-cone directions $x^\pm=0$. 
 Given the classical solution, the average number of gluons produced in a collision is given by
\beq
\langle N_g\rangle =\int_\q \frac{1}{2E_q} \sum_\lambda \langle  |{\cal M}_\lambda(\q)|^2\rangle,
\eeq
where ${\cal M}_\lambda(\q)=q^2 A^\mu(q) \epsilon_\mu^{(\lambda)}(\q)$ is the amplitude to find a gluon with momentum $\q$ and polarization $\lambda$ in the classical field $A^\mu(q)$. Similar formulae hold for other inclusive distributions. The average denotes an average over events, i.e. the MV average over the (independent) color charge distributions in the two nuclei, using the correlator (\ref{rhorhoMV}) in each nuclei.  

The classical approximation of the MV model is most accurate if both sources are strong, i.e., $J^\nu\sim 1/g$ (in the range of transverse momenta that are being probed, typically $k_\perp\lesssim Q_s$). Then the solution of the Yang-Mills equations is $A^\mu\sim 1/g$, and all terms in the classical equation of motion are of the same order of magnitude. From the point of view of perturbation theory, solving the classical equations of motion amounts to resumming the tree graphs: When the sources are strong, $\rho\sim 1/g$,  all such tree graphs contribute indeed at the same order, $\sim 1/g$. This is because, adding a source $\sim 1/g$ entails adding a vertex $\sim g$, resulting in a contribution of order unity. In contrast, adding a loop implies adding two vertices, without any compensation, making the graph of relative order $g^2$ compared to the initial graph. 

There are situations where both sources cannot be considered as large.  This is the case in particular of  pA collisions, where the source of the nucleus may be large, of order $1/g$, while that of the proton is small, of order $g$. In that case, one can linearize the Yang-Mills equations with respect to the weak source, and obtain semi-analytical solutions.  This case,  often referred to as the ``dilute-dense'' case,   is discussed in more details in the next subsection. The case of nucleus-nucleus collisions, referred to as ``dense-dense'', is more difficult and the solution  of the full non linear equations has only been obtained numerically.  Finally, the case where all sources are weak corresponds to usual perturbation theory, for which the classical approximation does not provide useful guidance. There is here an aspect of the CGC effective theory worth emphasizing: the effort to organize an expansion around the classical, i.e. strong field, solution \cite{Gelis:2008rw}.

As just mentioned, the complete solution of the Yang-Mills equations for nucleus-nucleus collisions in the MV model can only be obtained via numerical simulations, and effort initiated in Ref.~\cite{Krasnitz:1998ns} (see also  \cite{Krasnitz:2003jw,Lappi:2003bi}). However a number of results concerning the main characteristics of the gauge fields after the collision can be obtained analytically. 
Thus, the field just after the collision can be calculated in terms of the WW fields of each nuclei. 
 One can estimate the field at short times, that is at times $\tau\lesssim 1/Q_s$ using an expansion in $\tau Q_s$ \cite{Chen:2015wia}. The  non linear evolution beyond this early time is not known analytically,  but a linearization of the equations of motion has proven to be rather accurate \cite{Blaizot:2008yb}. Corrections to the linear approximation can, in principle, be estimated, but this has not been done yet. 
An interesting feature of the field immediately after the collision is that it becomes longitudinal: at $\tau=0^+$, the transverse color electric and magnetic WW fields vanish. This is a non abelian effect, as shown by the fact that, just after the collision,  and in a suitable gauge, $E^z(\x)=ig[A^i_{_{\! A}}(\x), A^i_{_{\! B}}(\x)]$, where $A^i_{_{\! A}}$ and $A^i_{_{\! B}}$ are the transverse component of the gauge fields created by the nuclei A and B, respectively. This commutator expresses the color precession of the color charges that go through each other during the collision. That particular configuration of fields has been dubbed 
 ``glasma'' \cite{Lappi:2006fp}. This feature is  reminiscent of the Lund string model \cite{Andersson:1983ia}, although the Lund model includes only a color electric field while the glasma contains both electric and magnetic fields. Also, the size of the flux tubes is of order  is $Q_s^{-1}$ in the glasma, whereas in the  Lund model, it is related to a correlation length in the vacuum The presence of these longitudinal fields entails a particular structure of the energy momentum tensor, $T^{\mu\nu}={\rm diag}(\epsilon,\epsilon,\epsilon, -\epsilon)$. The negative longitudinal pressure  reflects the existence of a restoring force (string stretching) in the longitudinal direction. This particular form of the initial energy-momentum tensor is used as initial condition in studies of the approach to hydrodynamics (see e.g. \cite{Gelis:2013rba}, and the recent reviews \cite{Gelis:2016upa,Fukushima:2016xgg}.

\subsection{Particle production in proton-nucleus and nucleus-nucleus collisions}

Further analytical progress can be made in the dilute-dense system, typically in the case of proton-nucleus collisions.  
To keep the discussion simple, we view the proton as a collection of three valence quarks, interacting independently of one another with the color field of the nucleus.  The calculation of the resulting field, and its Fourier analysis in terms of normal modes, according to the procedure outlines in the previous subsection, yields  (see  \cite{Blaizot:2004wu} and references therein)
\beq\label{prodpA}
\frac{\rm d N}{\rmd y \rmd^2 \q}=\frac{\alpha_s }{2\pi^4C_F}\frac{1}{q_\perp^2}\frac{{\cal S}_{\rm pA}}{\pi R_A^2 \pi R_p^2}\int _{\k}  \phi_{\rm p}(\k)\, \phi_{\rm A}(\q-\k).
\eeq
This expression gives the average number of gluons produced with rapidity $y$ and transverse momentum $\q$ in a pA collision (at a given impact parameter). To obtain Eq.~(\ref{prodpA}) from the more general formula given in Ref.~\cite{Blaizot:2004wu},  we have, as in other places in this paper, made the crude assumption that the densities are uniform in the transverse plane. We denote by $R_A$ and $R_p$  the radii of the nucleus and the proton, respectively, and 
 ${\cal S}_{\rm pA}$ is the overlap area at the given impact parameter of the pA collision. The distribution $\phi_{\rm A}$ is given by the Fourier transform of the correlator of two Wilson lines in the adjoint representation:
 \beq\label{phiGrperp0}
\frac{\phi_{\rm A}(x,\k)}{\pi R_A^2}=\frac{k_\perp^2 (N_c^2-1)}{4\alpha_s N_c}\int_{\r}\rme^{i\k\cdot\r}S(\r),
\eeq
where $S(\r)$ is the average (over the field of the nucleus) matrix element of a gluon dipole. We can rewrite this equation in terms of the scattering amplitude $T=1-S$
\beq\label{phiGrperp}
\frac{\phi_{\rm A}(x,\k)}{\pi R_A^2}= \frac{C_F}{2\alpha_s }\int_{\r}\rme^{i\k\cdot\r}\nabla^2_{\r} T(\b,\r).
\eeq

In the weak field limit, we may  expand $S$ in Eq.~(\ref{phiGrperp0}) to second order in $\alpha_s$ and, using  Eq.~(\ref{singlescat}), get 
\beq\label{wflim}
\phi_A(x,\k)\approx \frac{4\pi^3}{k_\perp^2}\frac{C_F\alpha_s}{\pi} {AN_c}.
\eeq
The quantity $\phi_A(x,\k_\perp)$ plays the role of an unintegrated gluon distribution (note that it differs from the gluon distribution of the WW field, as given by Eq.~(\ref{MVundf})). In the weak field limit, the distribution  denoted  here $\phi(x,\k)$ is related to the unintegrated distribution defined in Eq.~(\ref{undf}) by $\phi(x,\k)=\frac{4\pi^3}{k_\perp^2} \varphi(x,\k)$,
where $\varphi(x,\k)$ is dimensionless.  One recovers in Eq.~(\ref{wflim})  the gluon distribution in weak field (or large $k_\perp$), and its additivity property. 
Note also that the total multiplicity obtained by integrating the spectrum (\ref{prodpA}) over $p_\perp$ is plagued with a logarithmic infrared divergence whose origin can be traced back to our very crude picture of the proton, as three independent valence quarks. This is usually taken into account by  introducing an IR cutoff of order $p_\perp\lesssim \Lambda_{QCD}$. 

The formula (\ref{prodpA}) exhibits a factorization property, as an integral over $k_T$ of two unintegrated distribution functions. This is the so-calle  $k_T$-factorization used in most phenomenological applications. While this property naturally emerges in the calculation of the inclusive  spectrum of particles produced in the collision of a dilute system on a dense system, it does not hold in the case of the collision of two dense, saturated objects, like a nucleus-nucleus collision.  Recent numerical studies within the MV model show that the violation of $k_T$ factorization can be large in the region of small $k_\perp$ \cite{Blaizot:2010kh}. In practice, a cut-off is used to control this low $k_\perp$ region, and this introduces uncertainties which are difficult to quantify (although one may argue that the effect of the cut-off may vary little with energy). Note that $k_T$ factorization is also violated in other processes, including dilute-dense situations, such as for instance  in quark-production  in pA collisions \cite{Blaizot:2004wv},  or in the  di-jet production at forward rapidity \cite{Marquet:2007vb}, a topic that we shall discuss later. Note also that dilute dense systems may be more accurately described via a hybrid formalism, mixing parton distributions for the projectile and Wilson lines for their interaction with the target \cite{Dumitru:2002qt} (see also \cite{Chirilli:2012jd}).

In the MV model, an average has to be done over the distributions of  color charges in the two nuclei, that is,  with a probability of the product form $W=W_A[\rho_A] W_B[\rho_B]$. These represent the usual Gaussian averages over the initial conditions, the factorization reflecting the fact that these initial conditions for nucleus A and nucleus B are uncorrelated. It has been argued \cite{Gelis:2008rw} that  this factorization property holds also for the calculation of inclusive distributions in the presence of radiative corrections (or equivalently small fluctuations around the background fields of the respective nuclei) at the leading logarithmic level. Then the distribution $W$ is to be replaced by $W_Y=W_{Y_0}[\rho_A] W_{Y+Y_0}[\rho_B]$,  where  $Y_0$ is the rapidity of nucleus A and $Y$ the rapidity interval between A and B, and  with $W_{Y_0}[\rho_A]$ and $W_{Y+Y_0}[\rho_B]$ evolved with the respective JIMWLK hamiltonians of A and B. For an example of such a calculation, see \cite{Lappi:2011ju}.\\

In the rest of this section, we discuss a selection of phenomena observed at RHIC and the LHC, where high parton density effects may play an important role. The three chosen examples emphasize various features of the regime of high parton density: the role of the saturation momentum and its energy dependence, the qualitative effects of high densities in particle production at forward rapidity, and the specific correlations coming from the presence of strong random background fields.  I emphasize that this is meant for illustration only, and I shall omit all the ``details''  that would be necessary in a decent comparison between theoretical predictions and experimental data. A more complete account can be found in the specialized reviews quoted in the Introduction.

\subsection{Particle multiplicities in hadronic collisions}

Our primary observation concerns the role of the saturation momentum $Q_s$ and its energy dependence. This is most clearly seen in bulk properties such as the inclusive distribution $\rmd N/\rmd y$ of particles  produced in a given rapidity bin. That the saturation momentum plays an important role here is clear from the previous discussion and the calculation of the inclusive distribution that has been sketched in the previous subsections:  
The relevant  unintegrated distribution functions are peaked at transverse momenta of the order of $Q_s$ (see e.g.  Eq.~(\ref{phiGrperp}), and also Fig.~\ref{fig:phi}). Thus, as a crude approximation, we may consider that all partons with momenta $k_\perp\lesssim Q_s$ are freed in the collision. This naturally leads to the following generic behavior for the multiplicity density
\cite{Kharzeev:2000ph,Kharzeev:2001yq} (see also \cite{Kharzeev:2002ei})
\begin{equation}\label{multiplicity}
\frac{1}{\pi R^2}\frac{dN}{dy}\sim 
\frac{Q_s^2}{\alpha_s(Q_s^2)},
\end{equation}
where all dependence on energy or centrality is contained in $Q_s$ (to simplify the discussion, we are considering central collisions of two identical nuclei, so there is a single saturation momentum).  In fact, Eq.~(\ref{multiplicity}) is nothing but a rewriting of  Eq.~(\ref{Qsaturation}) that relates  $Q_s$ to the integrated gluon  distribution function, $xG(x,Q_s^2)=\rmd N_g/\rmd y$, which illustrates the fact that the gluons that are freed at early times are those present in the wave function with transverse momenta $\lesssim Q_s$.

Now, as we have seen in Sects.~\ref{sec:dipolemodel} and \ref{sec:solutionBK}, one expects the saturation momentum to depend on energy. 
 Typically, 
\beq
Q_s^2=Q_0^2(\b) \left(\frac{x}{x_0}  \right)^\lambda, \qquad Q_0^2(\b)=Q_0^2(0) T_A(\b),
\eeq
where $T_A(\b)=\int dz \,n(\b,z)$, with  $n(\b,z)$ the nucleon density. This formula exhibits a factorization between geometrical effects (essentially the additive increase of the gluon density proportionally to the density of nucleons at a given point $\b$ in the transverse plane), and the energy dependence captured by the factor $x^\lambda$.

\begin{figure}[htbp]
\begin{center}
\includegraphics[scale=0.35]{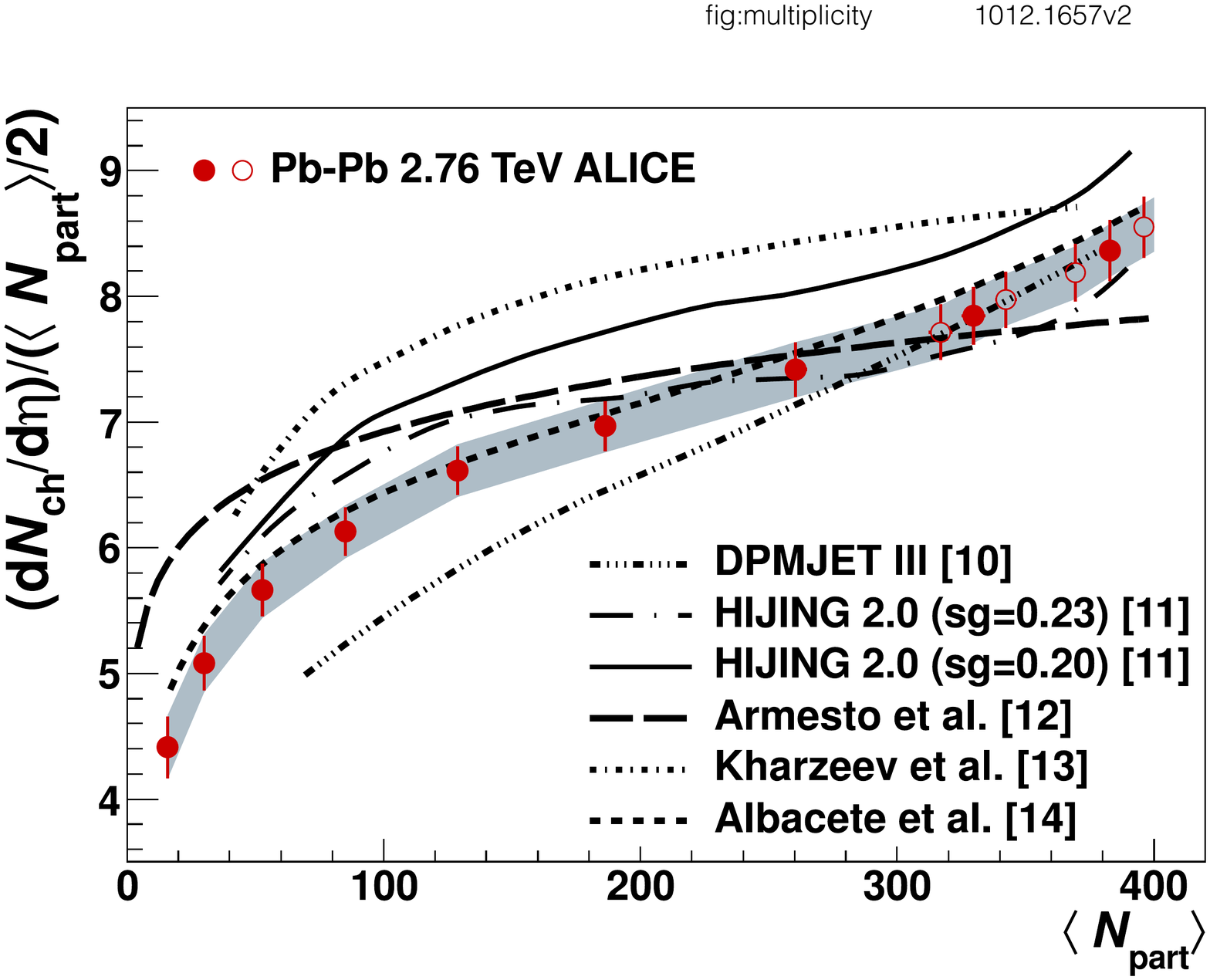}
\end{center}
\caption{\label{fig:multiplicity} The multiplicity $\rmd/\rmd \eta$, with $\eta$ the pseudo rapidity, observed by the ALICE collaboration for Pb-Pb collisions at $2.75$ TeV \cite{Aamodt:2010cz}
. The calculation by Albacete et al. provides the best fit to the data \cite{ALbacete:2010ad,Albacete:2012xq}. }
\end{figure}

The relations above provide of course only qualitative orientations. Detailed calculations have been performed, mainly based on the $k_\perp$ factorization \cite{Kharzeev:2000ph}, including a proper treatment of the geometry of the collisions. The energy dependence of the saturation scale has been determined by solving the non linear BK equation, with corrections taking into account the running of the coupling constant \cite{Albacete:2007sm,ALbacete:2010ad,Albacete:2012xq}. The initial conditions for the evolution are adjusted so that the resulting unintegrated distributions yield good fits to HERA data. As can be seen on  Fig.~\ref{fig:multiplicity}, a calculation along the lines just indicated, together with a reasonable adjustments of a few parameters,  is capable of accounting rather well for the ALICE data. In particular it provides a natural explanation for the similar centrality dependence of the multiplicity distribution  observed at RHIC and LHC. And the  power law dependence of the multiplicity density for central collisions \cite{Aamodt:2010pb}, $\rmd N/\rmd y\propto \sqrt{s}^\lambda$,  naturally follows  the energy dependence of the saturation momentum.  

Note that one 
only predicts here the distribution of `initial gluons', set free typically at
a proper time $\tau\sim Q_s^{-1}$. Between this early stage and the
 freeze-out, the system undergoes several non-trivial
steps: kinetic and chemical equilibration (possibly with additional
parton production), hadronization, etc. The fact that the predicted multiplicity accounts well for the data seems to imply that there is little room left for the late stages of the collision to contribute significantly to the multiplicity (or in the thermodynamical language to the entropy production after the initial period of particle production).

\subsection{Forward rapidity}

The presence of high density of gluons can be directly probed by measurements sensitive to the momentum broadening of the produced particles, and the ensuing redistributions of produced particles in momentum space (e.g. see Fig.~\ref{fig:phi}, left). By selecting particles that are produced at high rapidity, one favors situations where these particles propagate through high density regions, and hence suffer significant momentum broadening (recall that the production of a particle with a rapidity $y$ close to that of the projectile involves partons  in the target with momentum fraction $x\sim \rme^{-y}$, and hence probes the small $x$, high density part, of the target wave function ). 

\begin{figure}[htbp]
\begin{center}
\includegraphics[scale=0.50]{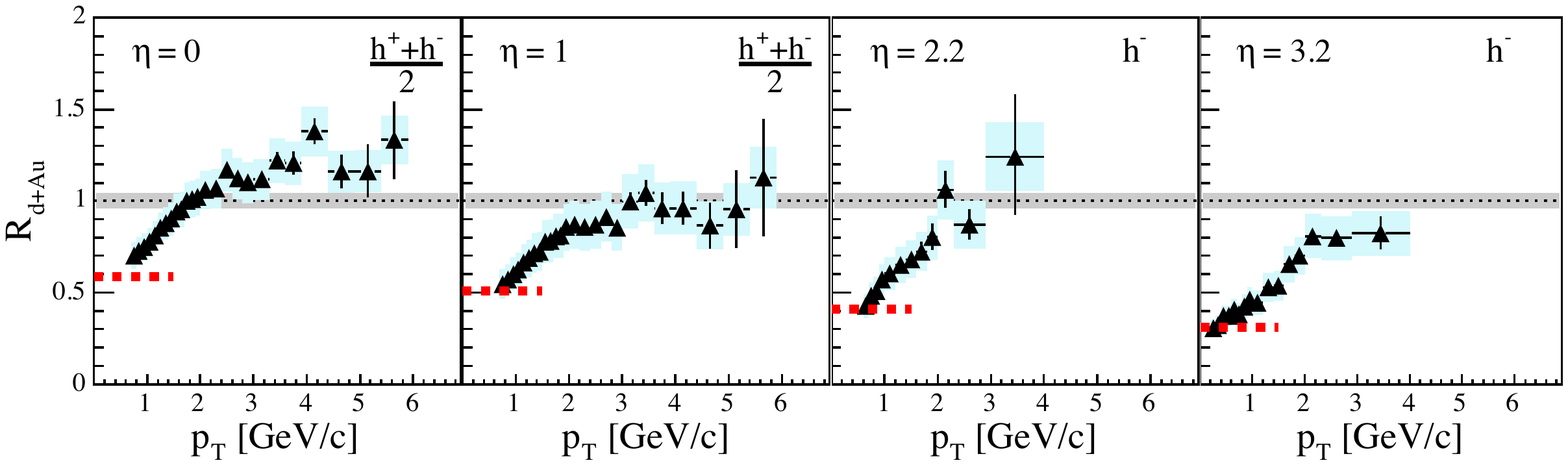}
       \end{center}
       \vspace{-5cm}
\caption{\label{fig:brahms_croning} The nuclear modification factor $R_{dA}$ showing the disappearance of the Cronin peak in d-Au collisions at RHIC  \cite{Arsene:2004ux}.}
\end{figure}

One of the early indications of a rapid evolution of particle production with rapidity was provided by BRAHMS data on d-AU collisions at RHIC, and the rapid disappearance of the Cronin peak with increasing rapidity \cite{Arsene:2004ux}, as illustrated in Fig.~\ref{fig:brahms_croning}. The quantity plotted there is the so-called nuclear modification factor, a ratio of particle yields normalized so that it is unity if the deuteron-gold interaction is a superposition of independent nucleon-nucleon collisions. Both the existence of the Cronin peak (the enhancement above unity for $p_\perp\gtrsim 2$ to 3 Gev), as well as its disappearance with increasing rapidity,  are natural consequences of momentum broadening in a dense gluon system and its evolution with energy \cite{Kharzeev:2003wz,Baier:2003hr,Kharzeev:2004yx,Albacete:2003iq,Iancu:2004bx}.  This phenomenon has been very much discussed as it was hoped that it could constitute the first glimpse of high gluon density effects.   However other competing explanations appear equally plausible, in particular those involving effects related to the proton constituents, namely large $x$ partons \cite{Kopeliovich:2005ym,Frankfurt:2007rn}.    

\begin{figure}[htbp]
\begin{center}
\includegraphics[scale=0.45]{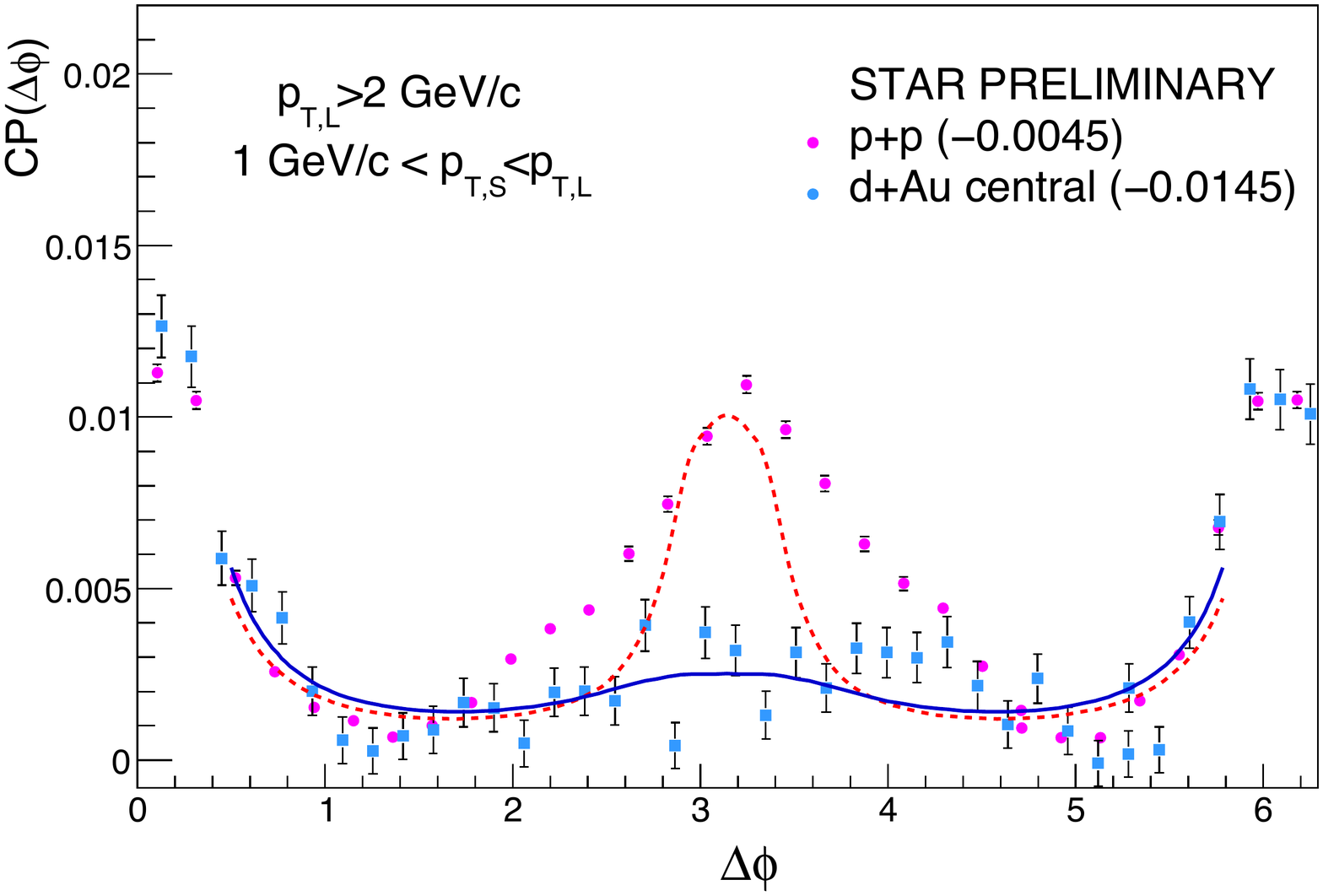}
       \end{center}
\caption{\label{fig:forward_correlation} The absence of the `away side' jet observed by the STAR experiment \cite{Braidot:2010zh}. The (blue) points represent the results of the calculation of Ref.~\cite{Albacete:2010pg}, from which the figure is taken.}
\end{figure}

Perhaps more conclusive evidence will come from 
the study of di-hadron production at forward rapidity. This is indeed a very interesting situation where one can  probe the very small $x$ component of the nucleus: with $y_1$ and $y_2$ denoting the rapidities of the produced hadrons, the $x$-values that are probed are given by $x\sqrt{s}=k_1{\rm e}^{-y_1}+k_2 {\rm e}^{-y_2}$. This can be very small if both $y_1$ and $y_2$ are large. The forward double inclusive pion production has been calculated  using a mixed formalism in which the wave function of the projectile  is described by a standard parton distribution function, with the propagation of these partons in the field of the nucleus  described by Wilson lines.  The calculation then reduces to the calculation of the average of some products of Wilson lines in the random field of the target. At present, approximations are used in order to evaluate the corresponding averages of  4 and 6 point functions. One  usually expresses those in terms of 2-point functions, whose evolution with energy is calculated using the BK equation (with running coupling) \cite{Albacete:2010pg}.   The physics one expects is somewhat similar to that of the Cronin effect, namely multiple scattering in the dense gluon system. Such multiple scattering of partons through the nucleus is expected to wash out the back-to-back correlations between the produced hadrons, leading eventually to  the disappearance of the away side jet, as was predicted first in \cite{Marquet:2007vb}.  Such an effect  has indeed been observed by STAR (see Fig.~\ref{fig:forward_correlation}). Although the final interpretation of the data is lacking,  
 the phenomenon is very suggestive of an initial state effect that can finds its natural explanation in terms of the large gluon density of the nucleus.

\subsection{Long range rapidity correlations}

An important feature of high density gluon systems is contained in various correlations. Some of these correlations are generated by the average over the charge distributions, with a specific pattern emerging when the charges are large \cite{Gelis:2009wh};  some correlations also result from the fact that the inclusive distributions are peaked at transverse momenta of order $Q_s$.  
\begin{figure}[htbp]
\begin{center}
  \includegraphics[scale=0.3]{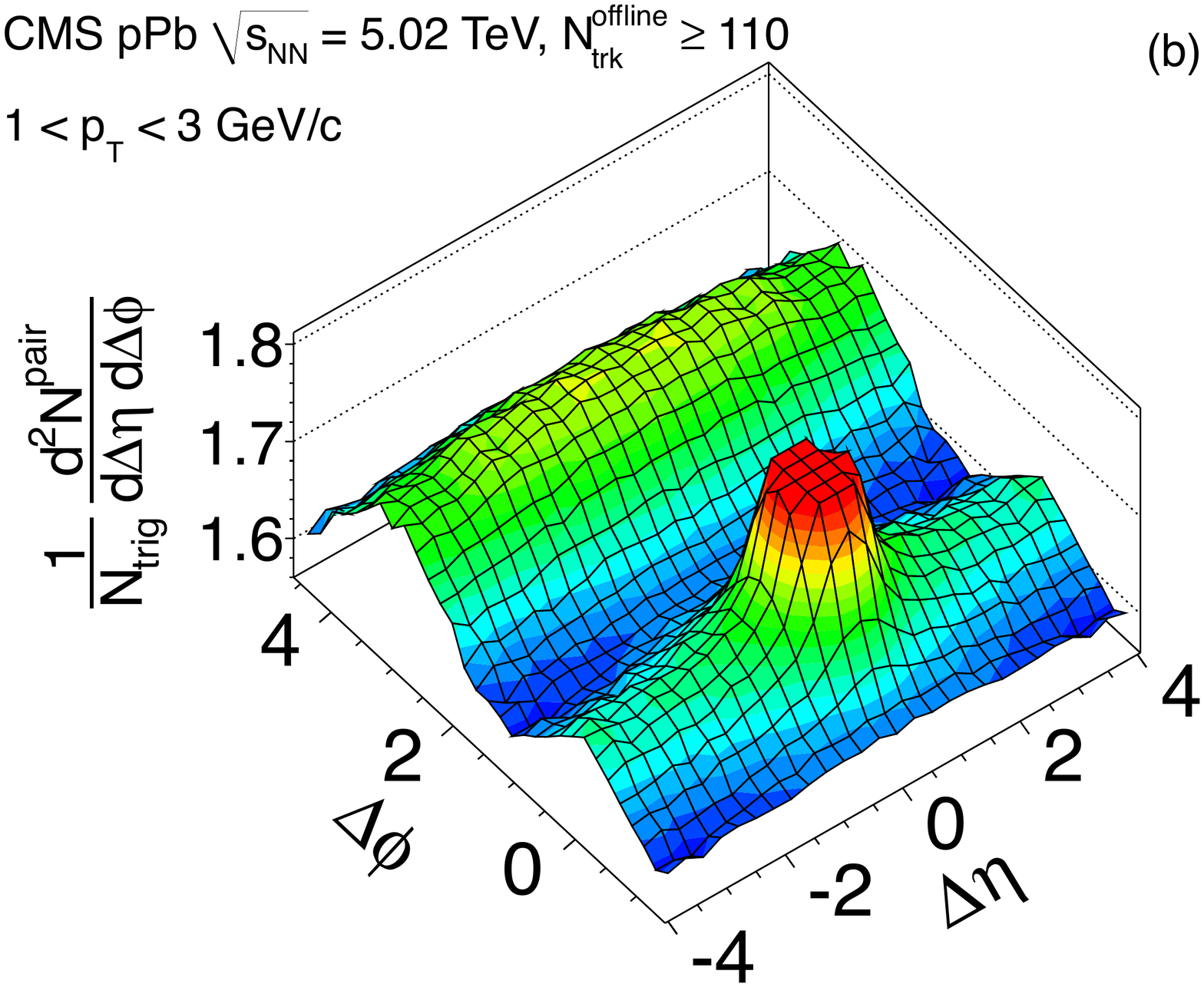}
   \end{center}
   \vspace{0cm}
\caption{\label{fig:ridge} Two particle correlations ($\eta$ is the pseudo rapidity and $\phi$ the azimuthal angle) measured by CMS in high multiplicity p-Pb collisions, for pairs of charged particles with $1<p_T<3$ GeV \cite{CMS:2012qk}.}
\end{figure}

The last example that I shall briefly mention may be reflecting some of these correlations. It concerns the observation of what has been called the ``ridge" phenomenon, and  is illustrated in Fig.~\ref{fig:ridge} in the case of p-Pb collisions at the LHC \cite{CMS:2012qk}. This figure suggests indeed the existence of correlations among pairs of not too soft particles ($1<p_\perp< 3$ GeV), the correlation extending over several units of (pseudo) rapidity and being collimated in azimuthal angles around 0 and $\pi$. Analogous observations were made for nucleus-nucleus collisions both at RHIC and the LHC, while at the LHC the ridge phenomenon has been observed also in pp collisions. 
There is a simple causality argument that indicates that in the rapidly expanding quark-gluon plasma 
long range rapidity correlations can only be produced at very short time \cite{Dumitru:2008wn}, where the system is the densest. This, combined with the observation that the phenomenon occurs in the range of momenta that are of the order of the saturation momentum, suggests indeed that the observed correlations could be due to the presence of high gluon densities.

The long range rapidity correlations are perhaps not too surprising. They simply reflect the boost invariance of the particle production mechanism, a generic feature at high energy. 
The azimuthal correlations however, may be more subtle.  In nucleus-nucleus collisions, a natural mechanism involves the collective flow that boosts the produced particles in the same direction, thereby extending in azimuth the correlations already present in rapidity. But in small systems, the phenomenon may reflects correlations that are already present in the initial wave functions. For instance,  specific correlations arise when averaging over strong color sources ($\rho\sim 1/g$). Then, as already mentioned,  a new ordering of the contributions of various diagrams emerges, which differs from the natural ordering in powers of the coupling constant \cite{Gelis:2009wh,Dusling:2012iga}. Some azimutal correlations may also arise from the fact that the transverse momenta of the produced particles are bound to be close to $Q_s$. Explicit calculations along those lines are indeed capable of providing a systematic account for the data (see \cite{Dusling:2012wy} and references therein). However, the  topic is presently very much debated. Its theoretical aspects are critically reviewed in \cite{Kovner:2012jm}.

\section{Conclusion and outlook}

We have seen that a striking feature of high gluon density is the phenomenon of saturation, which manifests itself in various ways: it characterizes the  slowing down of the growth of the gluon distribution function with increasing energy, it ensures the unitarization of cross sections, it induces specific color correlations. Saturation sets in at a particular scale, the saturation momentum, which has, correspondingly, plural interpretations: separation scale between dilute and dense partons, threshold for the breakdown of perturbation theory, momentum broadening of partons propagating in matter, color correlation length. 
The saturation momentum emerges dynamically from the non linear evolution equations that control the growth with energy of the gluon densities.  Its value is typically such that  $Q_s\gg  \Lambda_{QCD}$, leading to a reasonably small coupling constant, thereby making dense systems of partons amenable to a weak coupling description, in spite of their ``strongly coupled" behavior seen in experiment.   

 Being the only important scale at high energy, the saturation momentum plays an important role in the phenomenology of DIS and heavy ion collisions. Perhaps the best evidence  for the existence of the saturation momentum comes indeed from DIS, and the phenomenon of geometrical scaling.   In the case of heavy ions, effort are made to identify  in the data some robust features that can be unambiguously attributed to the high density gluonic systems. The few selected examples that we have discussed in the last part of this paper illustrate the fact that, indeed, a consistent and coherent phenomenology is  emerging, encompassing both DIS and heavy ion experiments, although no compelling evidence of small $x$ evolution, nor of saturation, can be claimed at present. But possibilities exist for further progress in the study of very small $x$ parton systems, for instance at the LHC by  focussing on very forward particle production. One may also expect  much progress in the future from experiments at electron-ion colliders.  

Completing the picture implies also more theoretical work.  The last three sections illustrate the main concepts and techniques that are being used in the field.  The current theoretical framework puts emphasis on  Wilson lines, which emerge from the eikonal approximation, a natural approximation at high energy. These Wilson lines typically describe the propagation of the partons of a projectile through the color field of  a target. They are mandatory when the target field is strong, or equivalently when the parton propagation involves multiple scattering.  The average over the target random field  cannot be done in general, except by using a model such as the MV model. However the variation with energy of this average is controlled by non linear evolution equations that are known and being improved. A major open issue is that these evolution equations favor an asymmetric treatment of the projectile and the target, with some account of fluctuations in the projectile, but a mostly mean field treatment of the target. This is a major limitation of the present formalisms, in particular in view of applications to heavy ion collisions. Further connections need to be explored, among them the connection with the $t$-channel approach of hadronic collisions.  There are obviously many puzzles to be resolved.

 \ack
My  research is 
supported by the European Research Council under the
Advanced Investigator Grant ERC-AD-267258. Important parts of this review have been written as I was guest of several institutions: I would like to thank in particular the Yukawa Insitute for Theoretical Physics in Japan for hospitality during the summer 2015, as well as the Institute for Theoretical Physics of the University of Heidelberg where I staid as Jensen Professor in November 2015, and finally the European Center for Theoretical Studies in Nuclear Physics and Related Areas (ECT*) in Trento, where the paper was completed.   I have benefited from many discussions with colleagues on issues discussed in this review, as well as from comments on initial versions of the manuscript. I would like to thank in particular G. Beuf, F. Gelis, E. Iancu,  C. Marquet, Y. Mehtar-Tani, A. Mueller, L. McLerran, O. Nachtmann, J.-Y. Ollitrault, D. Triantafyllopoulos, R. Venugopalan. Figures 13 and 14 were prepared by F. Gelis, and  Feynman diagrams were drawn using Jaxodraw. Finally, I would like to address special thanks to G. Baym for his invitation to write this review... and for his patience in waiting for the manuscript.



\end{document}